\documentclass[nofootinbib,aps,11pt]{revtex4}
\usepackage{graphicx}
\usepackage{cancel}
\usepackage{amssymb}
\usepackage{textcomp}
\usepackage{amsmath}
\usepackage{bm}
\usepackage{times}
\usepackage{epsfig}
\usepackage{color}

\begin{document}
\begin{flushright}
NPAC-08-19\\
MADPH-08-1522
\end{flushright}
\title{\LARGE Leptoquarks and Neutrino Masses at the LHC}
\bigskip
\author{Pavel Fileviez P{\'e}rez$^1$, Tao Han$^{1}$, Tong Li$^{1,2}$, Michael J. Ramsey-Musolf$^{1,3}$}
\address{
$^1$Department of Physics, University of Wisconsin, Madison, WI 53706, USA\\
$^2$Department of Physics, Nankai University, Tianjin 300071, P.R.~China \\
$^3$Kellogg Radiation Laboratory, California Institute of Technology, Pasadena, CA 91125 USA}
\date{\today}
\begin{abstract}
The properties of light leptoquarks predicted in the context of a
simple grand unified theory and their observability at the LHC are 
investigated. The $SU(5)$ symmetry of the theory implies that the leptoquark 
couplings to matter are related to the neutrino mass matrix. We study 
the resulting connection between neutrino masses and mixing parameters and 
the leptoquark decays, and show that different light neutrino hierarchies 
imply distinctive leptoquark decay signatures. We also discuss low-energy 
constraints implied by searches for charged lepton flavour violation, 
studies of meson decays, and electroweak precision data. 
We perform a detailed parton-level study of the leptoquark signals and 
the Standard Model backgrounds at the LHC. With the clean final states 
containing a di-lepton plus two jets, the QCD production of the 
leptoquark pair can be observed for a leptoquark  
mass of one TeV and beyond.
By examining the lepton flavor structure of the observed events, 
one could further test the model predictions related to the neutrino mass spectrum. 
In particular, $b$-flavor tagging will be useful in distinguishing 
the neutrino mass pattern and possibly probing an unknown Majorana 
phase in the Inverted Hierarchy or the Quasi-Degenerate scenario.
Electroweak associated production of the leptoquark doublet can also be 
useful in identifying the quantum numbers of the leptoquarks and distinguishing 
between the neutrino mass spectra, even though the corresponding event 
rates are smaller than for QCD production. We find that with only the clean
channel of $\mu+\cancel{E}_T+$jets, one could expect an observable 
signal for a leptoquark masses of about 600 GeV or higher.
\end{abstract}
\maketitle
\section{Introduction}
\label{sec:intro}
The possibility of explaining  all fundamental interactions in nature in a unified framework
is one of the main motivations for considering physics beyond the Standard Model (SM) of particle physics.
In this respect, Grand Unified Theories (GUTs) are one of the most appealing SM extensions, as they may allow one
to understand the origin of the SM interactions
and  predict both the quantization of the electric charge and
the weak mixing angle~\cite{GUTReview}. In these theories the unification of
gauge couplings is realized at a scale $M_{GUT}\approx 10^{14-16}$ GeV.
The unification of the matter fields implies
the decay of the lightest baryon, the proton~\cite{review},
and the generic existence of scalar and vector leptoquarks (denoted by LQ unless
specified otherwise)~\cite{LQreview} that carry both lepton- and baryon-numbers,
since the quarks and leptons live in the same enlarged
representation of the gauge group.

Because of the important role played by LQs in GUTS and other SM extensions as well as their unique phenomenological features, the general properties and phenomenology of LQs have been  studied extensively\cite{Buchmuller,HewettRizzo}.
Recently, a simple GUT based on the SU(5) gauge symmetry that contains light LQs has been proposed in Ref.~\cite{paper1}. In this theory, the Higgs sector is composed of ${\bf 5_H}$, ${\bf 15_H}$ and ${\bf 24_H}$;  neutrino
masses are generated through the Type II seesaw mechanism~\cite{TypeII}; and there exist
light scalar leptoquarks consistent with the constraints from the gauge coupling unification
and experimental lower bounds on the proton lifetime.  A distinctive feature of this framework is that the LQ-matter Yukawa interactions are governed by the neutrino mass matrix. Consequently, different scenarios for the light neutrino mass hierarchy lead to broadly distinguishable patterns for LQ interactions with matter. In this paper, we explore this phenomenology in detail, concentrating in particular on its implications for studies at the Large Hadron Collider (LHC). 

While several phenomenological and cosmological aspects of this proposal
were studied in Refs.~\cite{paper2} and~\cite{paper3}, a comprehensive analysis of the corresponding LQ collider phenomenology has not appeared previously in the literature. Here, we report the results of such an analysis. In doing so, we consider the constraints on the model implied by the results of neutrino oscillation experiments, searches for charged lepton flavor violation, and studies of SM-suppressed meson decays. We find that although the constraints from low-energy flavor physics are severe, they do not preclude the possibility of observing a statistically significant LQ signal at the LHC if the LQ mass is sufficiently heavy. The constraints from electroweak precision observables (EWPO) have a marginal impact at best on the LHC discovery potential of the model. 

In searching for the LQ at the LHC, we identify the dominant production mechanisms and the 
experimentally fully reconstructable clean final states. 
We find that the ``golden signals" for LQ discovery are the QCD pair production of  
the charge $+2/3$ LQ (denoted $\Psi_1$), with its decays to a charged lepton plus a down-type
quark ($d_i$, $i=1-3$). After applying a series of judicious cuts to suppress SM backgrounds 
that would produce the same final state, we find that one could expect on the 
order of 1000 events with 100 fb$^{-1}$ integrated luminosity for a 400$-$600 GeV LQ.  
By examining the lepton flavor structure of the observed events, 
one could further test the model prediction for the neutrino mass spectrum. 
In particular, $b$-flavor tagging will be  useful in helping to  distinguish the neutrino mass
pattern and to probe an unknown Majorana phase in the ``Inverted Hierarchy" or the
``Quasi-Degenerate" scenario.
Assuming that one will have knowledge of the neutrino mass spectrum in the near future, 
obtained from either searches for neutrinoless double $\beta$-decay or future 
long baseline neutrino oscillation studies, the possible collider discovery analyzed 
here would not only provide a crucial consistency check for the neutrino mass pattern,
but also indicate a neutrino mass generation mechanism at a fundamental level.

Our study is organized in the remainder of the paper as follows: In Section II 
we present the properties of the LQs in the model. In Section III the constraints on
the physical couplings and parameters coming from neutrino oscillation experiments,
rare processes and EWPO are investigated. The predictions for leptoquarks decays are 
discussed in Section IV. Taking into account the effects of neutrino mass
and mixing we show the unique  predictions for the branching fractions
of all leptonic decays $\Psi_1\to d_i e^+_j$ and $\Psi_2\to d_i \bar{\nu}$.
The impact of the Majorana phases is also studied. The pair and associated
production mechanisms, cuts necessary for SM background suppression, 
and the signal observability at the LHC are investigated
in Section V. We summarize our results in Section VI.
\section{Light Leptoquarks and Grand Unification}
\label{sec:model}
A general classification of the properties of leptoquarks can be found
in Ref.~\cite{Buchmuller}. Here, we concentrate on the specific model 
realization of Ref.~\cite{paper1} that consists of a simple realistic 
extension of the Georgi-Glashow model~\cite{GG} in which the light neutrino masses
are generated through the Type II seesaw mechanism. 
The Higgs sector of this model is composed of ${\bf 5_H}$, ${\bf 24_H}$ and ${\bf 15_H}$, and
the matter fields live in the ${\bf \bar{5}}= (d^C, l)_L$ and ${\bf 10}=(u^C, Q, e^C)_L$
representations, where $l^T=(\nu, e)$ and $\ Q^T=(u, d)$.
As emphasized in Section \ref{sec:intro}, one of the main features of this GUT is the possibility of having
light leptoquarks consistent with all constraints coming from the proton lifetime and
the unification of gauge couplings~\cite{paper1,paper2,paper3}. Here,
we focus on the properties of these light scalar leptoquarks
\begin{equation}
\Phi_b^T = (\Psi_1, \Psi_2 ) \sim (\bm{3},\bm{2},1/6).
\end{equation}
The electric charges of the leptoquarks are $Q(\Psi_1)=I_3+Y=2/3$ and $Q(\Psi_2)=-1/3$, respectively.
The LQ $\Phi_b$ lives in the ${\bf 15_H}$ representation together with $\Delta \sim (1,3,1)$,
the scalar triplet field responsible for implementing the Type II seesaw mechanism. 
Using the interactions in the model~\cite{paper1} one finds
\begin{equation}
Y_\nu \ {\bf \bar{5}} \ {\bf \bar{5}} \ {\bf 15_H}  \ \supset \  Y_\nu \left[ \ l_L^T \ C \ i \sigma_2 \ \Delta \ l_L \ + \
\sqrt{2} \ \overline{d}_R \  \ l_L^\alpha \ \Phi_b^\beta \ \epsilon_{\alpha \beta}\right] \ \ \ ,
\end{equation}
where we have suppressed the generation indices on the lepton and quark fields and Yukawa coupling matrix $Y_\nu$. 

It is important to notice that the $SU(5)$ symmetry of the theory implies that the coupling of the
leptoquark to matter is defined by the mass matrix for neutrinos
\begin{equation}
 Y_\nu = {M_\nu \over \sqrt{2} \  v_{\Delta}} ,
\end{equation}
where $v_{\Delta}/\sqrt{2}$ is the vacuum expectation value of the neutral
component of the field $\Delta$. As shown in Ref.~\cite{paper1}
the leptoquark $\Phi_b$ has negative (positive) contribution to
$b_1 - b_2$ $(b_2 - b_3)$, where $b_i$ stands for the different beta function coefficients for the three SM gauge groups.
Therefore, the presence of this field helps to achieve unification without supersymmetry. If the LQ mass lies in the range of 
100 GeV$-$1 TeV, one obtains gauge coupling unification
in agreement with experimental observations at the low energies.

The Lagrangian relevant for our study is given by:
\begin{equation}
{\cal L} = (D_\mu \Phi_b)^{\dagger} (D^\mu \Phi_b) \ +{\cal L}_Y-V ,
\end{equation}
where
\begin{eqnarray}
{\cal L}_Y & = &  - \left( \sqrt{2} \ \overline{d}_R \ Y_\nu \ l_L^\alpha \ \Phi_b^\beta \ \epsilon_{\alpha \beta} \ + \ \text{h.c.} \right)\\
V & =&  M_{\Phi_b}^2 \Phi_b^\dagger \Phi_b \ + \ \lambda_1 (\Phi_b^\dagger \Phi_b) (H^\dagger H) \ +
\ \lambda_2 (\Phi_b^\dagger H) (H^\dagger \Phi_b) \ +\lambda_3 (\Phi_b^\dagger \Phi_b)^2\ \ \ ,
\end{eqnarray}
where $H^T=(H^+, H^0)$ is the SM Higgs doublet.
The tree-level masses of the leptoquarks, after the electroweak symmetry breaking
$\langle H^0 \rangle = v_0 / \sqrt{2}$, are given by:
\begin{equation}
M_{\Psi_1}^2 = M_{\Phi_b}^2 \ + \ \lambda_1 v_0^2/2 \qquad \text{and} \qquad M_{\Psi_2}^2 = M_{\Psi_1}^2 \ + \ \lambda_2 v_0^2/2.
\end{equation}
In principle, the dimensionless couplings $\lambda_1,\ \lambda_2$ can be either positive, 
negative, or zero\footnote{For negative $\lambda_1$ and/or $\lambda_2$, boundedness of the 
potential restricts the magnitude of coupling relative to that of $\lambda_3$ and the 
SM Higgs quartic coupling.}. Thus, we have no {\em a priori} prediction for which 
of the two LQs is the lightest\footnote{We note that SM radiative corrections will generate a mass 
splitting\cite{Cirelli:2005uq} $M_{\Psi_1} - M_{\Psi_2}\approx 106\ \mathrm{MeV}$ in the 
$\overline{\mathrm{MS}}$ scheme.}. In our study of the collider phenomenology, 
we will focus on the case in which $M_{\Psi_2} - M_{\Psi_1}\geq 0$, 
corresponding to $\lambda_2\geq 0$. 

Working in the  basis of physical fermions, 
the new LQ-matter Yukawa interactions read:
\begin{equation}
\overline{d}_R \ \Gamma_1 \ e_L \ \Psi_1\ \qquad
\text{and} \qquad
\overline{d}_R \ \Gamma_2 \ \nu_L \ \Psi_2 ,
\end{equation}
where
\begin{equation}
\Gamma_1 = \sqrt{2} \ D_R^\dagger Y_\nu E_L = \Gamma_2 \ V_{PMNS}^\dagger \ K_3^* \ \ {\rm and} \ \
\Gamma_2 = \sqrt{2} \ D_R^\dagger Y_\nu N_L = \sqrt{2} \ B \ K_3^* \ V_{PMNS}^* \ Y_\nu^{diag}.
\label{G12}
\end{equation}
Here, $V_{PMNS}$ is the leptonic (neutrino) mixing matrix; $D_R$, $E_L$, and $N_L$ transform the right-handed down quark, left-handed electron, and left-handed neutrino mass eigenstates into the corresponding flavor states; and 
$K_3$ is a diagonal matrix containing the unknown phases
\begin{eqnarray}
K_3 = \left(
\begin{array}{lll}
 e^{i\alpha_1} & 0 & 0 \\
 0 & e^{i \alpha_2} & 0 \\
 0 & 0 & e^{i\alpha_3}
\end{array}
\right).
\end{eqnarray}
The other mixing matrix, $B$, appearing in Eq. (\ref{G12}) involves the product of RH quark and LH charged lepton rotation matrices: is  $B=D_R^{\dagger} E_L^*$.
It has been pointed out \cite{upper1,upper2,paper3}
that in order to satisfy the constraints from proton decay,
the unitary matrix $B$ must be of the form
\begin{eqnarray}
B = \left(
\begin{array}{lll}
 0 & 0 & e^{ i \beta_1} \\
 0 & e^{i \beta_2} & 0 \\
 e^{i \beta_3} & 0 & 0
\end{array}
\right).
\end{eqnarray}
Once we impose the constraints  from the lower limit on the  proton lifetime, the LQ-matter couplings are dictated
 by the neutrino masses and mixing, as presented in detail in
Appendix A.
It is useful to notice that  the Yukawa couplings scale parametrically as
$\Gamma_{1,2}\propto m_\nu/ v^{}_\Delta$.
We summarize the interactions for the leptoquarks in Table.~\ref{int}.
\begin{table}[tb]
\begin{center}
\begin{tabular}[t]{|c|c|c|}
\hline
Fields & Vertices & Couplings
\\
\hline
$\Psi_i(i=1,2)$ & $\bar{d}\Psi_il_i$ & $i\Gamma_i P_L$\\
       Yukawa      & $l_1=e, \ l_2=\nu$ &  $P_L=(1- \gamma_5)/2$ \\
\hline\hline
QCD        & $\Psi_i(p_1)\Psi_i^\ast(p_2) G_\mu^a$ & $ig_s{\lambda_a\over 2}(p_1-p_2)_\mu$\\
        & $\Psi_i\Psi_i^\ast G_\mu^a G^{b\mu}$ & $ig_s^2{\lambda_a\over 2}{\lambda_b\over 2}$\\
\hline\hline
    3-point EW            & $\Psi_i(p_1)\Psi_i^\ast(p_2) A_\mu$ & $iQ_ie(p_1-p_2)_\mu$\\
                &                           & $Q_1=2/3, \ Q_2=-1/3$\\
\hline
        & $\Psi_i(p_1)\Psi_i^\ast(p_2) Z_\mu$ & $i{g_2\over 6}\cos\theta_WK_i(p_1-p_2)_\mu$\\
        &                           & $K_1=4-{1\over \cos^2\theta_W}, \ K_2=-2-{1\over \cos^2\theta_W}$\\
\hline
        & $\Psi_1(p_1)\Psi_2^\ast(p_2) W_\mu^-$ & $-i{g_2\over \sqrt{2}}(p_1-p_2)_\mu$ \\
\hline\hline
4-point EW & $Z_\mu Z_\nu\Psi_i\Psi_i^\ast$ & $iG_{ZZii}g_{\mu\nu}$\\
                &  & $G_{ZZ11}={(g_1^2-3g_2^2)^2\over 18(g_1^2+g_2^2)}, G_{ZZ22}={(g_1^2+3g_2^2)^2\over 18(g_1^2+g_2^2)}$\\
\hline
                & $A_\mu A_\nu\Psi_i\Psi_i^\ast$ & $iG_{AAii}g_{\mu\nu}$\\
                &                           & $G_{AA11}={8g_1^2g_2^2\over 9(g_1^2+g_2^2)}, G_{AA22}={2g_1^2g_2^2\over 9(g_1^2+g_2^2)}$\\
\hline
        & $Z_\mu A_\nu\Psi_i\Psi_i^\ast$ & $iG_{ZAii}g_{\mu\nu}$\\
        &                           & $G_{ZA11}={-2g_1g_2(g_1^2-3g_2^2)\over 9(g_1^2+g_2^2)}, G_{ZA22}={g_1g_2(g_1^2+3g_2^2)\over 9(g_1^2+g_2^2)}$\\
\hline
        & $W_\mu^+W_\nu^-\Psi_i\Psi_i^\ast$ & $iG_{WWii}g_{\mu\nu}$ \\
      & & $G_{WW11}=G_{WW22}=g_2^2/2$\\
      \hline
\end{tabular}
\end{center}
\caption{Feynman rules for the leptoquarks Yukawa and  gauge interactions. The momenta are all assumed to be incoming.}
\label{int}
\end{table}

\section{Constraints on the physical parameters}
In this section we discuss the constraints from neutrino oscillation experiments,
rare decays and collider experiments on the Yukawa couplings and
the physical parameters $M_{\Psi_1}, M_{\Psi_2}$ and $v_\Delta$
in this theory.
\subsection{Constraints From Neutrino Oscillation Experiments}
The physical Yukawa couplings of leptoquarks for the leptonic
decays are given by Eq.~(\ref{G12}). In order to
understand the constraints from neutrino physics we start
from the correlation between the neutrino masses and
mixing angles. The leptonic mixing matrix is given by
\begin{eqnarray}
V_{PMNS}=
\left(
\begin{array}{lll}
 c_{12} c_{13} & c_{13} s_{12} & e^{-\text{i$\delta $}} s_{13}
   \\
 -c_{12} s_{13} s_{23} e^{\text{i$\delta $}}-c_{23} s_{12} &
   c_{12} c_{23}-e^{\text{i$\delta $}} s_{12} s_{13} s_{23} &
   c_{13} s_{23} \\
 s_{12} s_{23}-e^{\text{i$\delta $}} c_{12} c_{23} s_{13} &
   -c_{23} s_{12} s_{13} e^{\text{i$\delta $}}-c_{12} s_{23} &
   c_{13} c_{23}
\end{array}
\right)\times \text{diag} (e^{i \Phi_1/2}, 1, e^{i \Phi_2/2}),
\end{eqnarray}
where $s_{ij}=\sin{\theta_{ij}}$, $c_{ij}=\cos{\theta_{ij}}$,
$0 \le \theta_{ij} \le \pi/2$ and $0 \le \delta \le 2\pi$.
The phase $\delta$ is the Dirac CP phase, and $\Phi_i$ are the Majorana phases.
The experimental constraints on the neutrino masses and mixing parameters,
at $2\sigma$ level are \cite{Schwetz}
\begin{eqnarray}
7.3 \times 10^{-5} {\rm eV}^2 \  < & \Delta m_{21}^2 & < \  8.1 \times 10^{-5} {\rm eV}^2, \\
2.1 \times 10^{-3} {\rm eV}^2 \  < & |\Delta m_{31}^2| & < \  2.7 \times 10^{-3} {\rm eV}^2, \\
                   0.28 \  < & \sin^2{\theta_{12}} & < \  0.37, \\
                   0.38 \  < & \sin^2{\theta_{23}} & <\  0.63, \\
                          & \sin^2{\theta_{13}} & <\  0.033,
\end{eqnarray}
and $\sum_{i} m_{i} < \ 1.2 \ {\rm eV}$. Following the convention, we denote the case
$\Delta m_{31}^2 \equiv m^2_3 - m_1^2 >0$ as the normal hierarchy (NH) and otherwise the
inverted hierarchy (IH).

\begin{figure}[tb]
\begin{center}
\begin{tabular}{cc}
\includegraphics[scale=1,width=8.5cm]{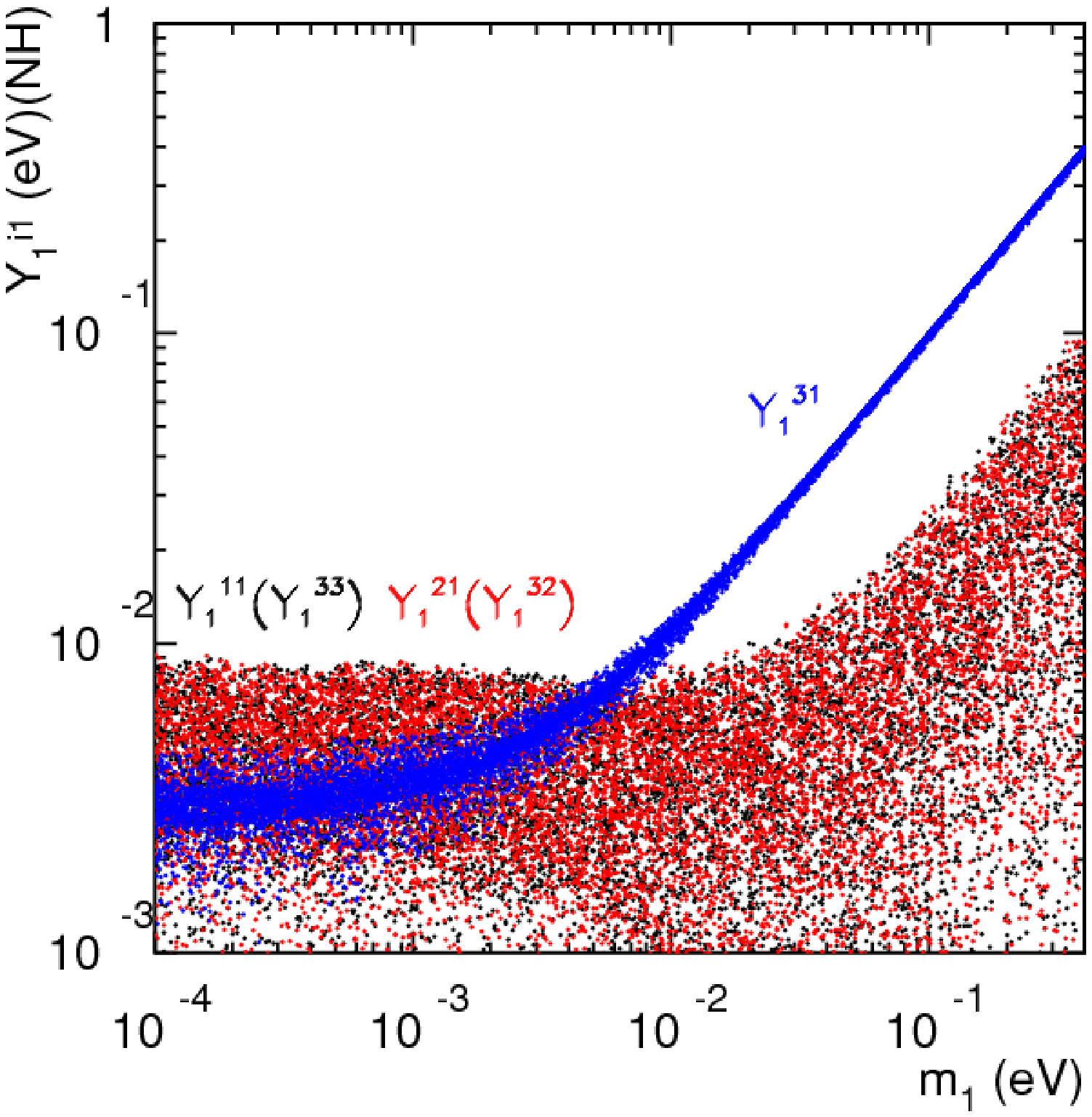}
\includegraphics[scale=1,width=8.5cm]{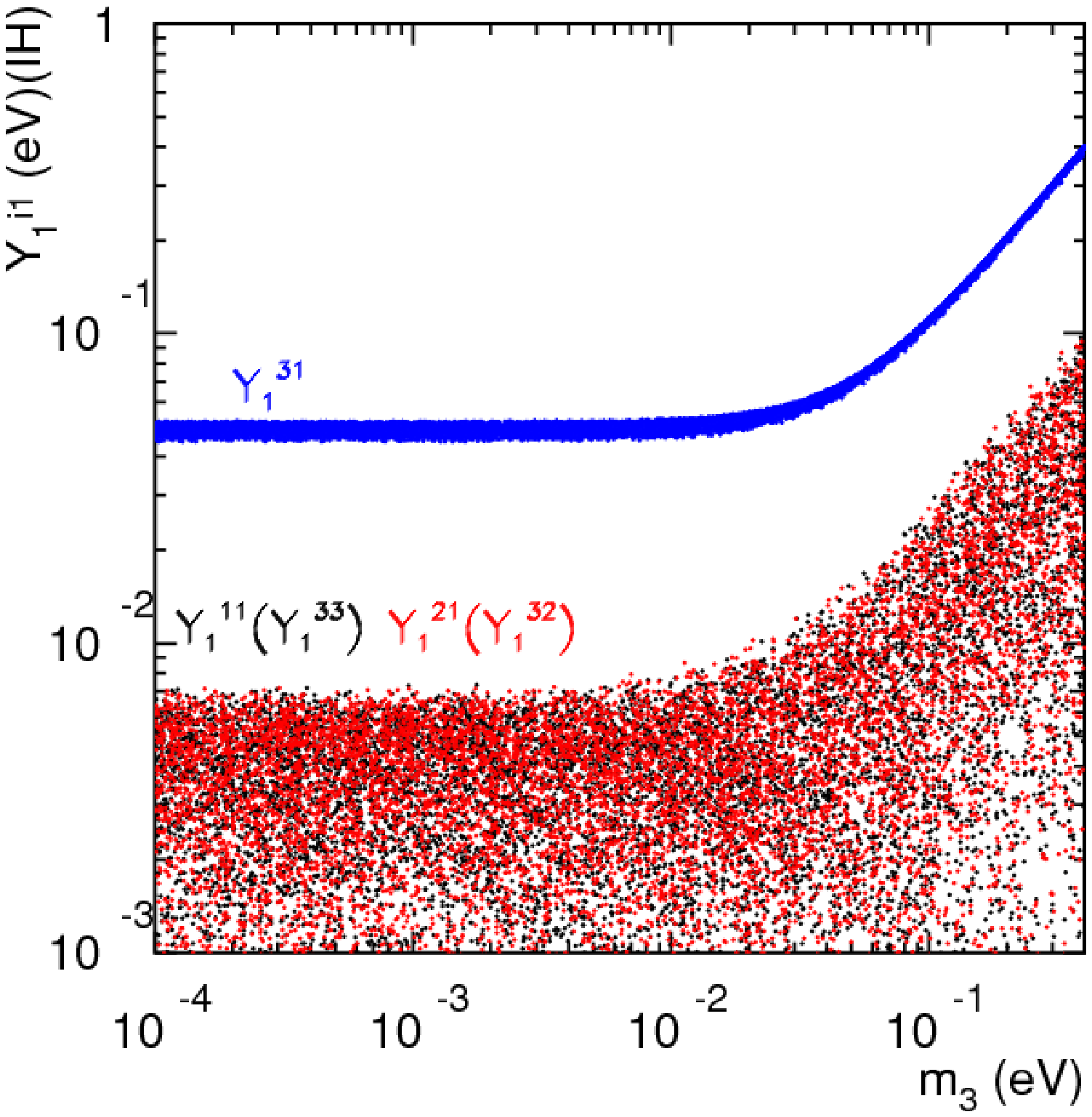}
\end{tabular}
\end{center}
\caption{Constraints on the leptoquark couplings 
$Y_1^{i1}=\Gamma_1^{i1}\times v_\Delta$ versus the lowest neutrino 
mass for NH (left) and IH (right) when all the phases vanish.
Due to the symmetry as in Eq.~(\ref{nmy}), the equal couplings 
are also indicated in the parenthesis.}
\label{y2ii}
\end{figure}
\begin{figure}[tb]
\begin{center}
\begin{tabular}{cc}
\includegraphics[scale=1,width=8.5cm]{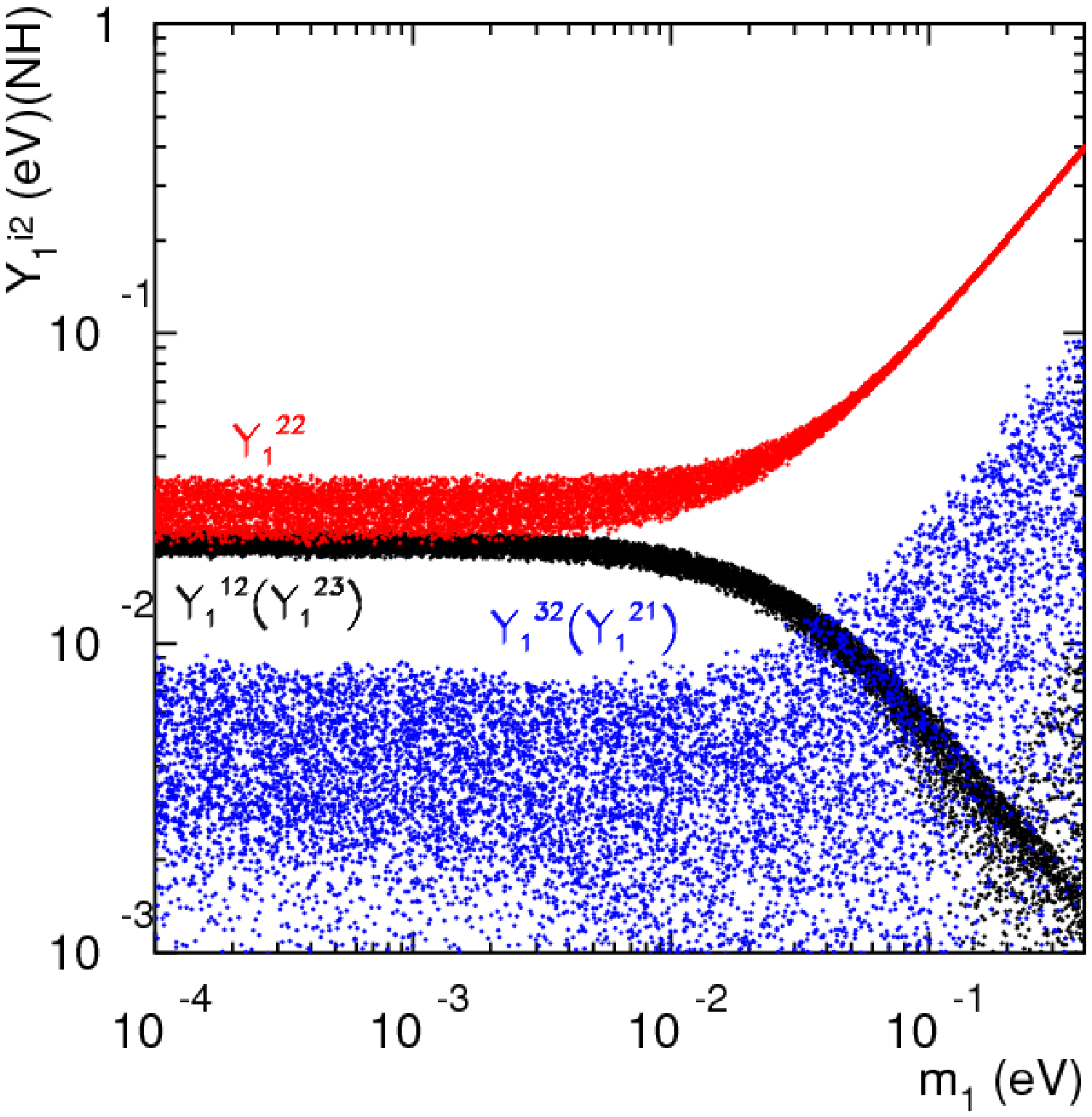}
\includegraphics[scale=1,width=8.5cm]{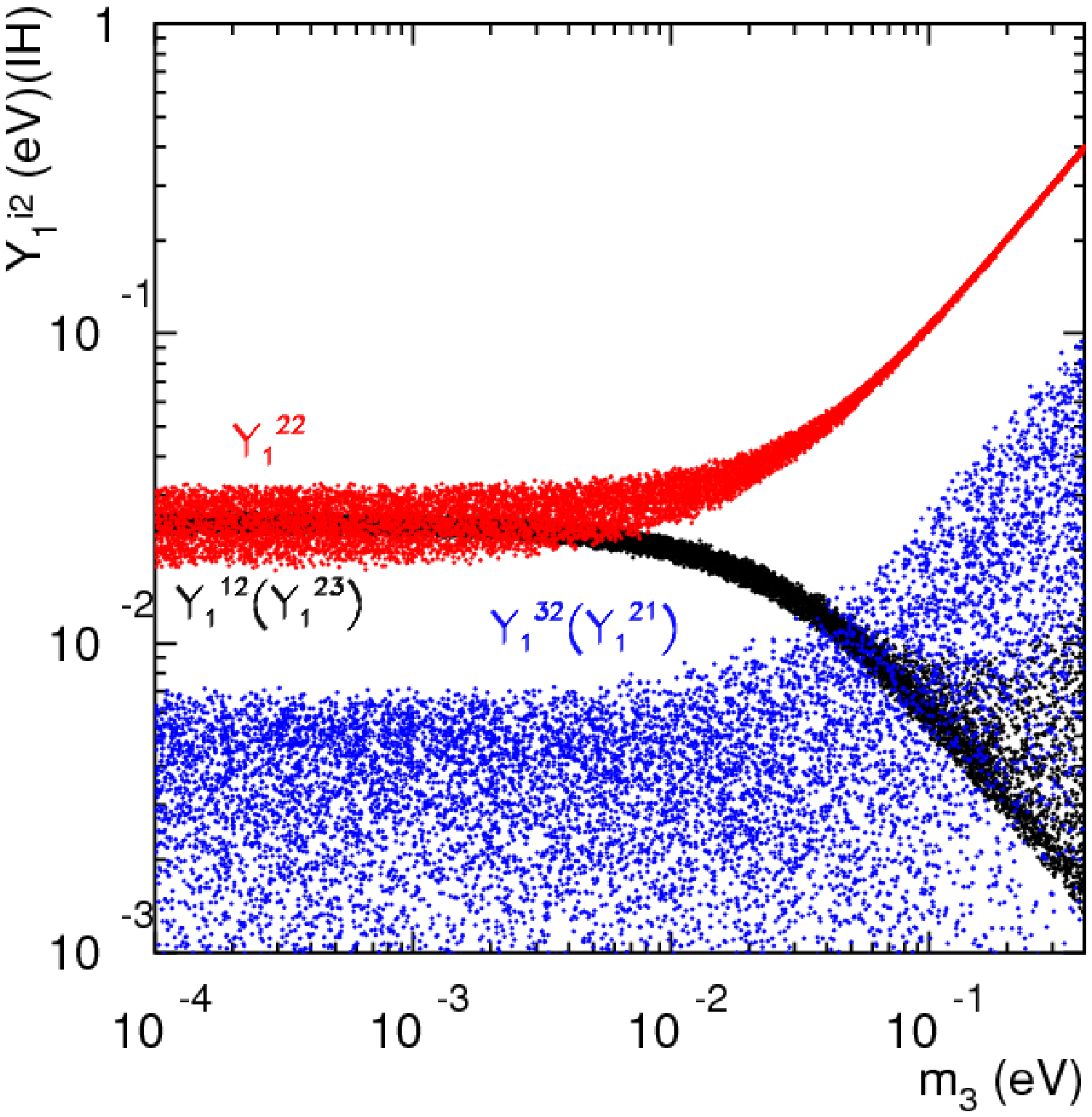}
\end{tabular}
\end{center}
\caption{Constraints on the leptoquark couplings $Y_1^{i2}=\Gamma_1^{i2}\times v_\Delta$ 
versus the lowest neutrino mass for NH (left) and IH (right) when all the phases vanish.
Due to the symmetry as in Eq.~(\ref{nmy}), the equal couplings are also indicated in the
parenthesis.}
\label{y2ij}
\end{figure}
In this section we neglect all the phases. The observed neutrino mass spectra indicate that
the neutrino mass matrix
\begin{eqnarray}
M_\nu=V_{PMNS}^\ast m_\nu^{diag}V^\dagger_{PMNS},
\end{eqnarray}
presents the following patterns
\begin{eqnarray}
\nonumber
&& M_\nu^{11}\ll M_\nu^{22}, M_\nu^{33} \quad {\rm for\ NH},\\
&& M_\nu^{11}> M_\nu^{22}, M_\nu^{33}\quad {\rm for\ IH},\\
{\rm and} && M_\nu^{23}> M_\nu^{12}, M_\nu^{13}\quad {\rm for\ both\ NH,\ IH}.
\nonumber
\end{eqnarray}
where the last relation for the off-diagonal elements is
due to the large atmospheric mixing angle $\theta_{23}$.
See for example Ref.~\cite{seesaw} for more detailed discussions.

In order to see the relationship between neutrino mass pattern and the
Yukawa couplings of the leptoquarks,
we first introduce a dimensionful coupling matrix relevant to
$\Psi_1$ Yukawa coupling
\begin{eqnarray}
&& Y_1=\Gamma_1\times v_\Delta=\left(
\begin{array}{lll}
 0 & 0 & 1 \\
 0 & 1 & 0 \\
 1 & 0 & 0
\end{array}
\right) M_\nu, \quad  {\rm or}\quad Y_1^{ij}=M_\nu^{(4-i),j}.
\label{nmy}
\end{eqnarray}
Here we have neglected the phases in the B matrix. Note that the 
neutrino mass matrix $M_\nu$ is symmetric but $Y_1$ is not,
with the first index $i$ for the down-type quarks and the 
second index $j$ for the charged leptons.

\begin{figure}[tb]
\begin{center}
\begin{tabular}{cc}
\includegraphics[scale=1,width=8.5cm]{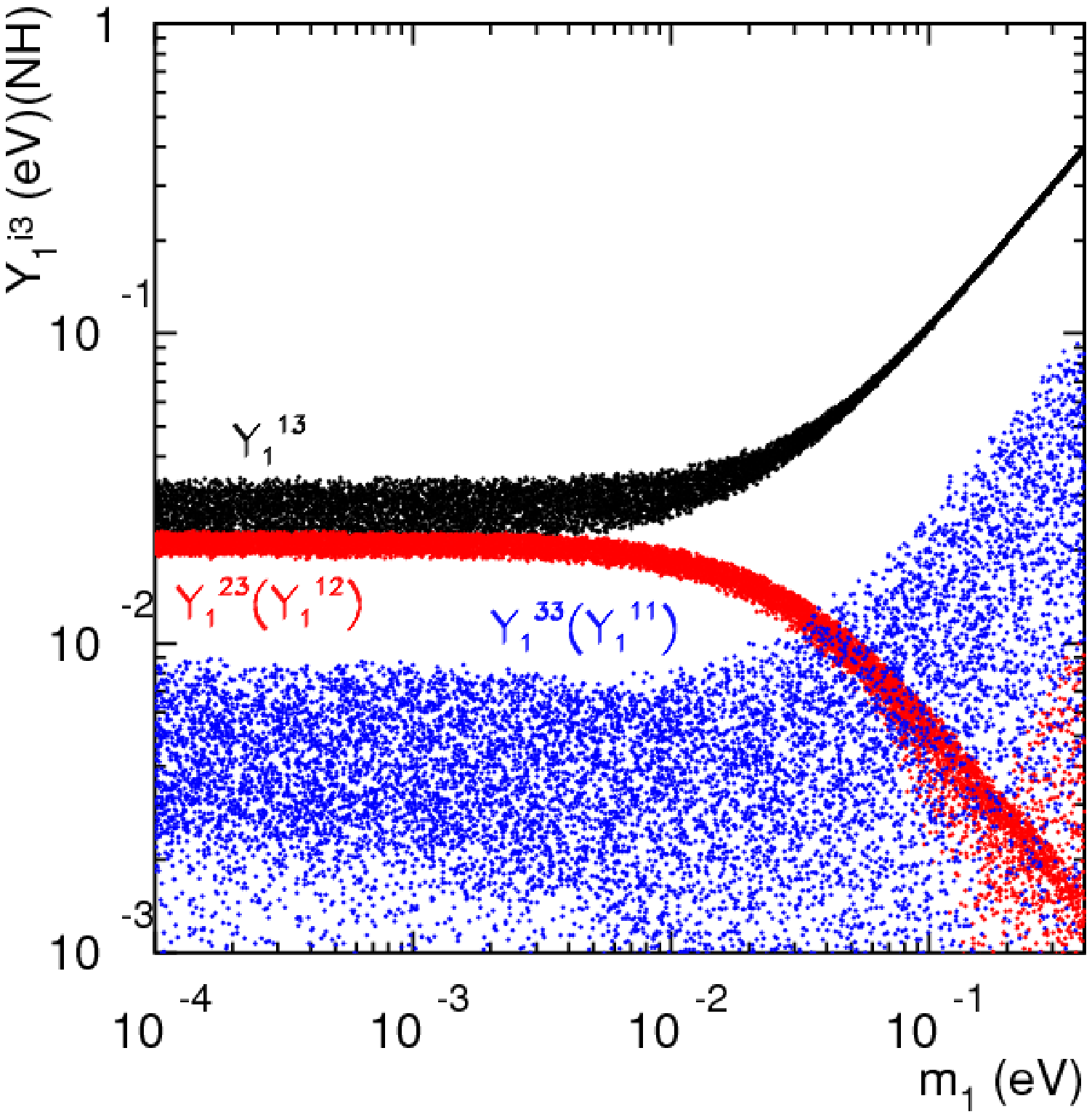}
\includegraphics[scale=1,width=8.5cm]{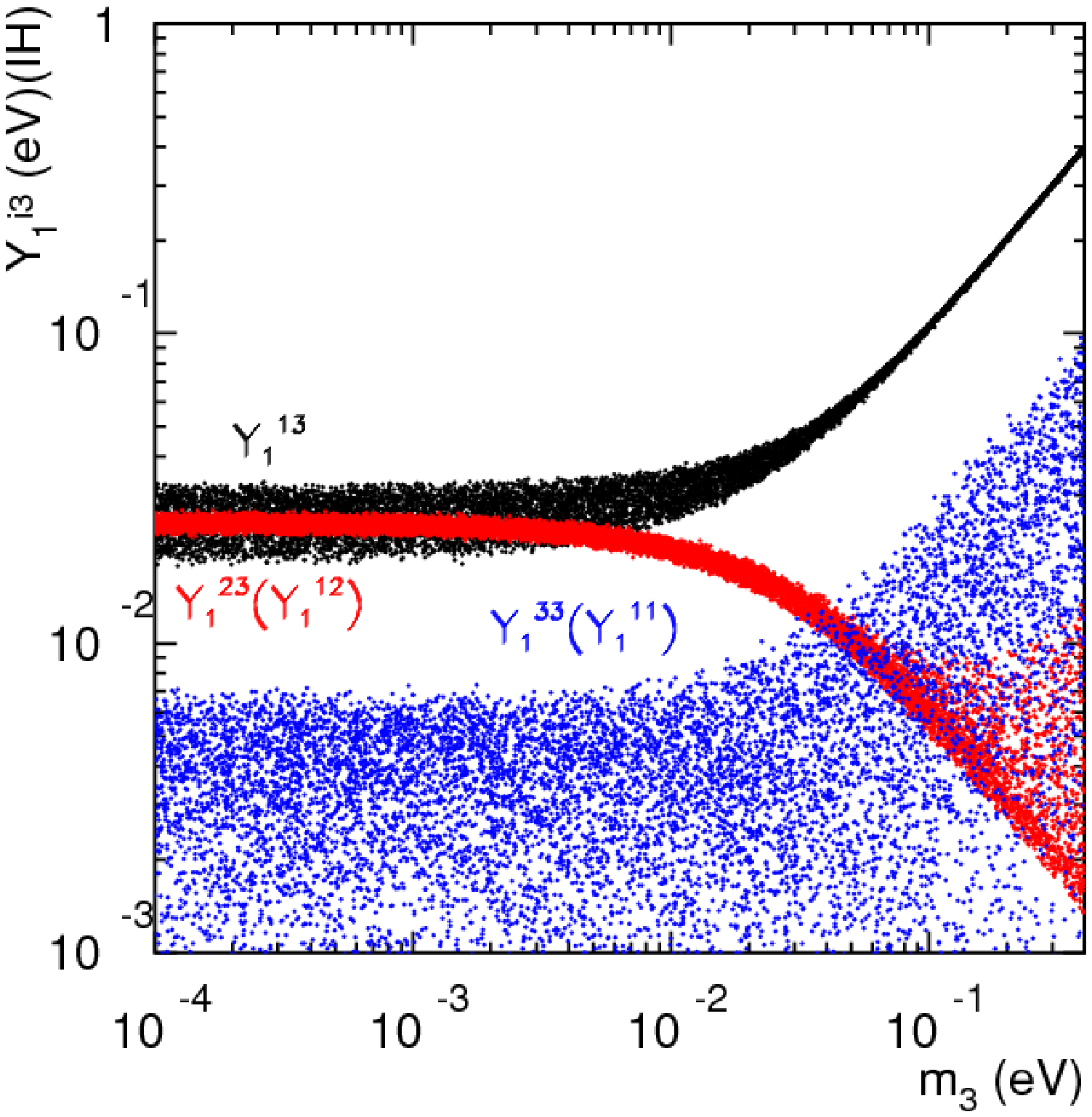}
\end{tabular}
\end{center}
\caption{Constraints on the leptoquark couplings $Y_1^{i3}=\Gamma_1^{i3}\times v_\Delta$ 
versus the lowest neutrino mass for NH (left) and IH (right) 
 when all the phases vanish.
Due to the symmetry as in Eq.~(\ref{nmy}), the equal couplings are also indicated in the
parentheses. }
\label{y2ji}
\end{figure}

We are thus able to determine all the Yukawa couplings that are consistent with a given set of neutrino masses
and oscillation parameters. To illustrate this relationship, we perform a numerical scan over the neutrino masses and mixing angles, using a uniform, random distribution while taking into account 
the above experimental constraints and neglecting all phases.
We show the scatter plots for the allowed values for the couplings of each lepton flavor
in Figs.~\ref{y2ii},~\ref{y2ij} and \ref{y2ji},  versus the lightest
neutrino mass in each spectrum, the normal hierarchy (left panels) and the inverted
hierarchy (right panels). We see two distinctive regions in terms of the lightest neutrino
mass. In the case $m_{1(3)} < 10^{-1}$ eV, 
\begin{eqnarray}
\nonumber
&& {\rm Fig.~\ref{y2ii}\ for}\ e:\quad 
 Y_1^{31} \simeq Y_1^{11}, Y_1^{21}\quad {\rm for\ NH},\\
 \nonumber
&&\qquad\qquad\qquad~~  Y_1^{31} \gg Y_1^{11}, Y_1^{21}\quad {\rm for\ IH},\\
\nonumber
&&  {\rm Fig.~\ref{y2ij}\ for}\ \mu :\quad Y_1^{22} \gtrsim Y_1^{12} >  Y_1^{32}
\quad {\rm for\ both\ NH\ and\ IH},\\
\nonumber
&&  {\rm Fig.~\ref{y2ji}\ for}\ \tau :\quad Y_1^{13} \gtrsim Y_1^{23} >  Y_1^{33}
\quad {\rm for\ both\ NH\ and\ IH}.
\end{eqnarray}
On the other hand, for  $m_{1(3)} > 10^{-1}$ eV, we have the quasi-degenerate spectrum 
$
M_\nu^{11} \approx M_\nu^{22} \approx M_\nu^{33},
$
and the leading couplings in each lepton flavor are 
\begin{eqnarray}
Y_1^{31} \approx Y_1^{22} \approx Y_1^{13} .
\end{eqnarray}

\begin{figure}[tb]
\begin{center}
\begin{tabular}{cc}
\includegraphics[scale=1,width=8.5cm]{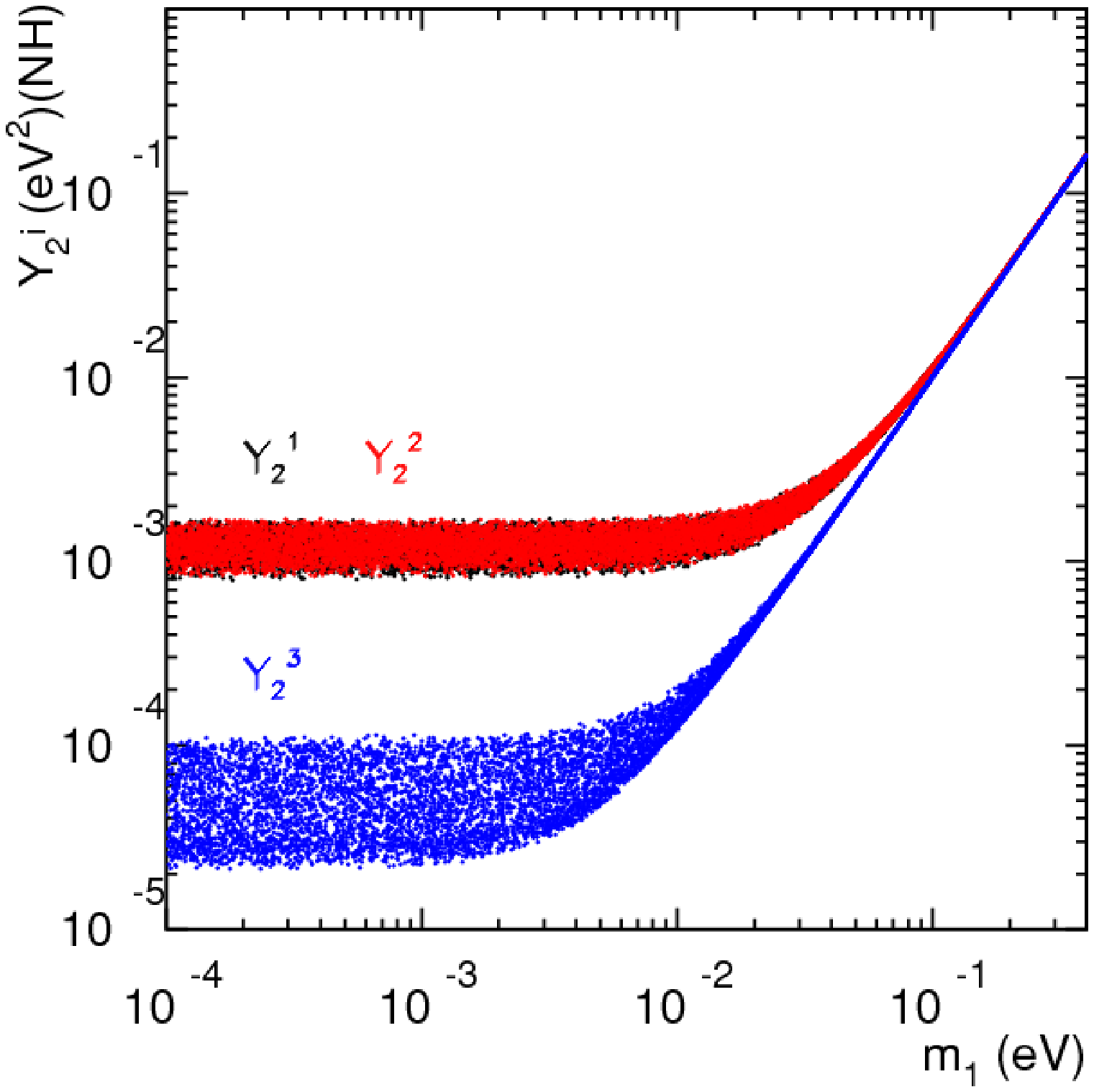}
\includegraphics[scale=1,width=8.5cm]{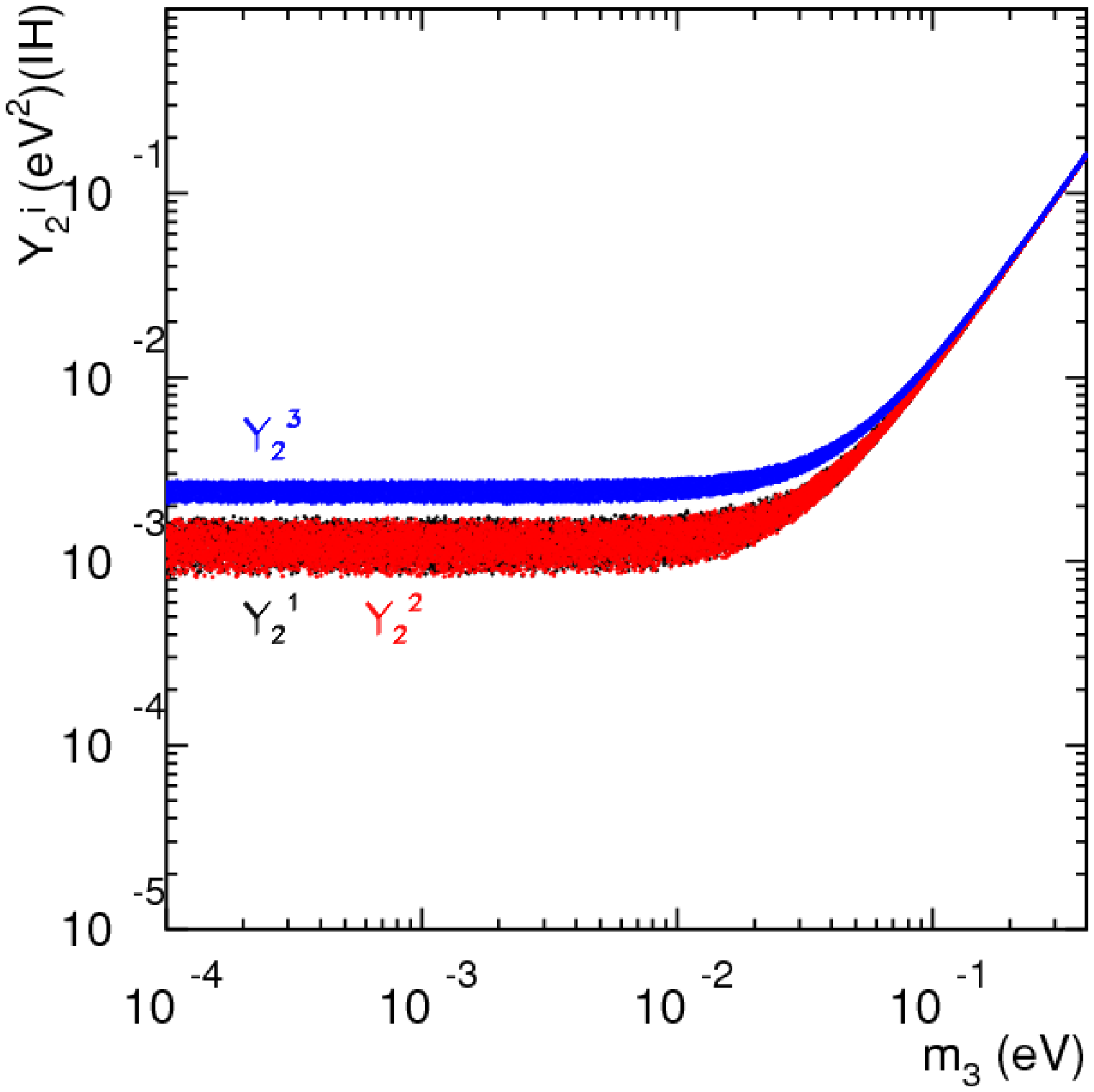}
\end{tabular}
\end{center}
\caption{Constraints on $Y_2^{i}=\sum^3_{j=1}|\Gamma_2^{ij}|^2\times v_\Delta^2$ versus
the lowest neutrino mass for NH (left) and IH (right)  when all the phases vanish.}
\label{y1i}
\end{figure}

As for the $\Psi_2$ decays, we sum
over the final state neutrinos since they are experimentally unobservable.
Thus the relevant couplings are written as
\begin{eqnarray}
&&Y_2^{i}=\sum^3_{j=1}|\Gamma_2^{ij}|^2\times v_\Delta^2 \ \ (i=1,2,3 \ {\rm for \ down \ quarks} \ d,s,b).
\end{eqnarray}
The allowed values are shown in Fig.~\ref{y1i}. In this case,
$Y_2^3\ll Y_2^1,Y_2^2$ in the NH and $Y_2^3>Y_2^1,Y_2^2$ in the IH when the lightest 
neutrino mass is smaller than $10^{-2}$ eV. The explicit expressions 
of $Y_1$ and $Y_2$ obtained via $\Gamma_1$ and $\Gamma_2$ are collected in Appendix A.
\subsection{Constraints from Rare Decays}
Here we discuss the constraints coming from meson decays, meson-antimeson mixing
and lepton flavor violating processes. The most important constraints for the leptoquark
parameters are from the $K_L^0$ decays to dileptons~\cite{PDG,constraint1,constraint2} 
and searches for $\mu-e$ conversion in nuclei~\cite{PDG}.
\subsubsection{$K_L$ decays}
The meson leptonic decay rate is given by
\begin{eqnarray}
\Gamma(M(\bar{q}^jq^n)\to \ell^r\bar{\ell}^i)&=&{k\over 4\pi}{1\over M_{\Psi_1}^4}
{|\Gamma_1^{nr}|^2\over 2}{|\Gamma_1^{ji}|^2\over 2}\left({f_M\over 2}\right)^2(E_iE_r-k^2),
\end{eqnarray}
where
\begin{eqnarray}
E_{i,r}=\sqrt{m_{i,r}^2+k^2}, \ \ \ k&=&{m_M\over 2}\lambda^{1/2}(1,{m_i^2\over m_M^2},{m_r^2\over m_M^2}).
\end{eqnarray}
For $K_L^0$ decay, we take $m_{K}=497.648~{\rm MeV}, f_K=160~{\rm MeV}$ and $\tau_{K_L}=5.116\times 10^{-8}~{\rm s}$~\cite{PDG}.
\begin{table}[tb]
\begin{center}
\begin{tabular}[t]{|c|c|c|}
\hline
Mesons & Final states & Branching fraction
\\
\hline
$K_L^0$ & $\mu^+\mu^-$ & $(6.84\pm 0.11)\times 10^{-9}$\\
        & $e^+e^-$ & $(0.087^{+0.057}_{-0.041})\times 10^{-10}$\\
        & $e^-\mu^+$ & $<4.7\times 10^{-12}$ \\
\hline
\end{tabular}
\caption{Experimental Constraints on $K_L^0$ pure leptonic decays~\cite{PDG}.}
\label{Kb}
\end{center}
\end{table}
We list the most precise results for the $K_L^0$ decay branching fractions $B(K_L\to X)$  in Table~\ref{Kb}.
In the case of the $\mu^+\mu^-$ and $e^+e^-$ final states, SM expectations involve the sum of short-distance (SD) and long-distance (LD) contributions~\cite{smee,GomezDumm:1998pj}. The latter are dominant, as the predicted short-distance contribution gives roughly one sixth of the measured branching ratio. Using chiral perturbation theory to perform a model-independent analysis of the long-distance contributions, the authors of Ref.~\cite{smee} find that they are nearly saturated by the lowest order (l.o.) term in the chiral expansion that is fixed from the measured $K_L\to\gamma\gamma$ branching fraction. The sub-leading long-distance contribution depends on an {\em a priori} unknown scale-dependent low-energy constant $\chi(\mu)$ whose value is obtained from analysis of $K_L\to\mu^+\mu^-$, $\eta\to\mu^+\mu^-$, and $\pi^0\to e^+e^-$ decays\footnote{The scale $\mu$ is typically chose to be the $\rho$-meson mass.}. Of these, $B(K_L\to\mu^+\mu^-)$ is known most precisely. The difference between its experimental value and the sum of the short-distance and leading order long-distance contributions
\begin{equation}
\Delta_K(\mu\mu)\equiv  B(K_L\to\mu^+\mu^-)^{\mathrm{exp}}-\left[B(K_L\to\mu^+\mu^-)^{\mathrm{SD}}+B(K_L\to\mu^+\mu^-)^{\mathrm{LD}}_{\mathrm{l.o.}}\right]
\end{equation}
determines the value of $\chi$ implied by this decay. 

Any non-SM contribution to $K_L\to\mu^+\mu^-$ would enter this difference and, thus, affect the extracted value of $\chi$. So long as the non-SM contribution is smaller than the uncertainty in $\Delta_K(\mu\mu)$, it could be absorbed by a shift in $\chi$ within its error bars without appreciably affecting the analysis of $K_L$, $\pi^0$, and $\eta$ decays into $\ell^+\ell^-$. The uncertainty $\delta\Delta_K$ is dominated by the theoretical uncertainty in $B(K_L\to\mu^+\mu^-)^{\mathrm{SD}}$:
\begin{equation}
\label{eq:kdecay}
\delta\Delta_K(\mu\mu) \simeq (\pm 0.6)\times 10^{-9} \ \ \ .
\end{equation}
To be conservative, we thus require that the LQ contribution to $B(K_L\to\mu^+\mu^-)$ be smaller than the magnitude of $\delta\Delta_K(\mu\mu)$ given in Eq.~(\ref{eq:kdecay}). 

Once the constant $\chi$ has been obtained from the other processes discussed above, it is possible to make a model-independent prediction for the $K_L\to e^+e^-$ branching fraction. In this case, however, the comparison of the experimental result with the Standard Model prediction,  $\delta\Delta_K(ee)$, is dominated by the experimental error in $B(K_L\to e^+e^-)$. Consequently, we require that the LQ contribution to this channel be less than the average of the upper and lower errors quoted in Table \ref{Kb}. 

\begin{figure}[tb]
\begin{center}
\begin{tabular}{cc}
\includegraphics[scale=1,width=8cm]{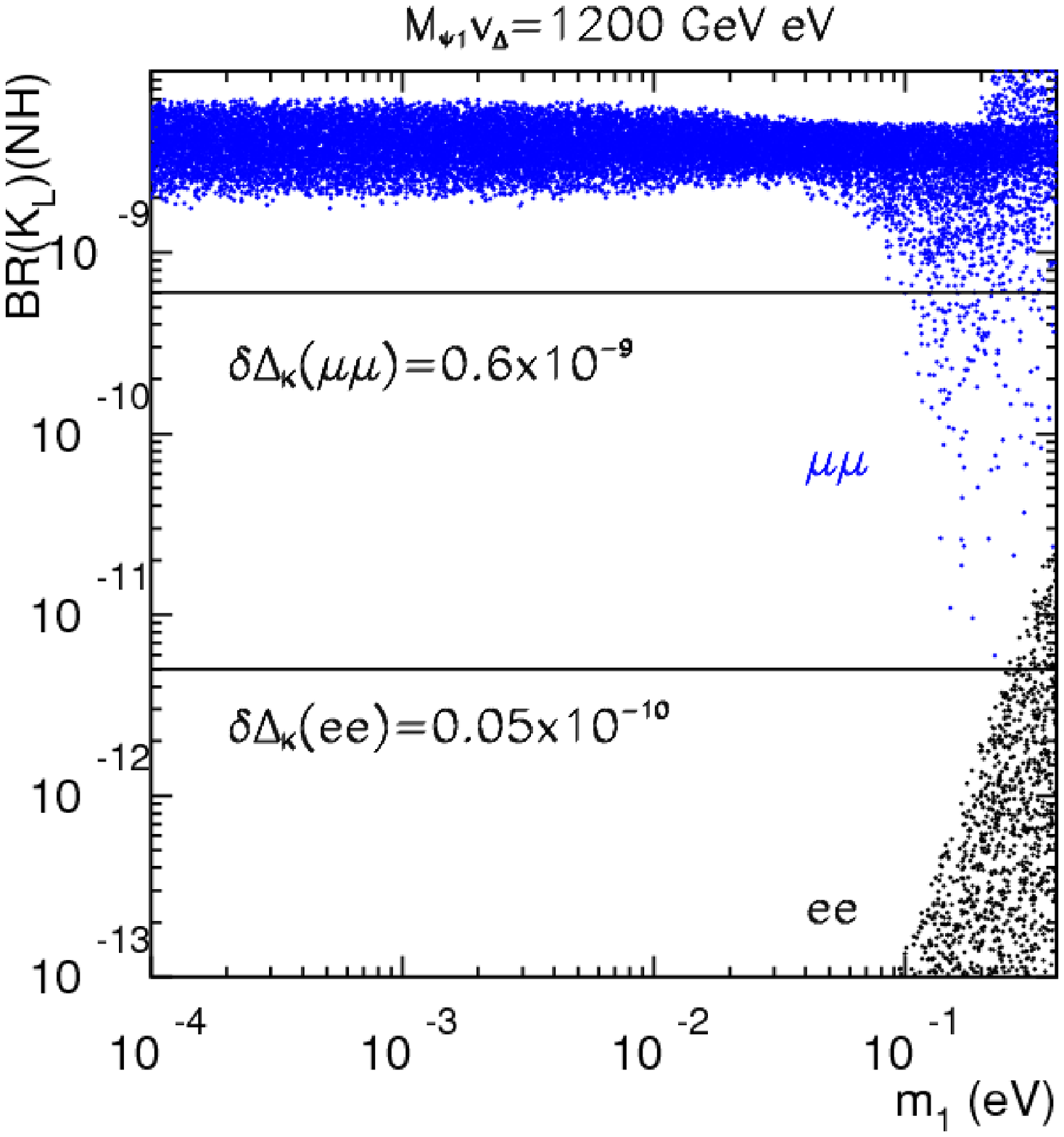}
\includegraphics[scale=1,width=8cm]{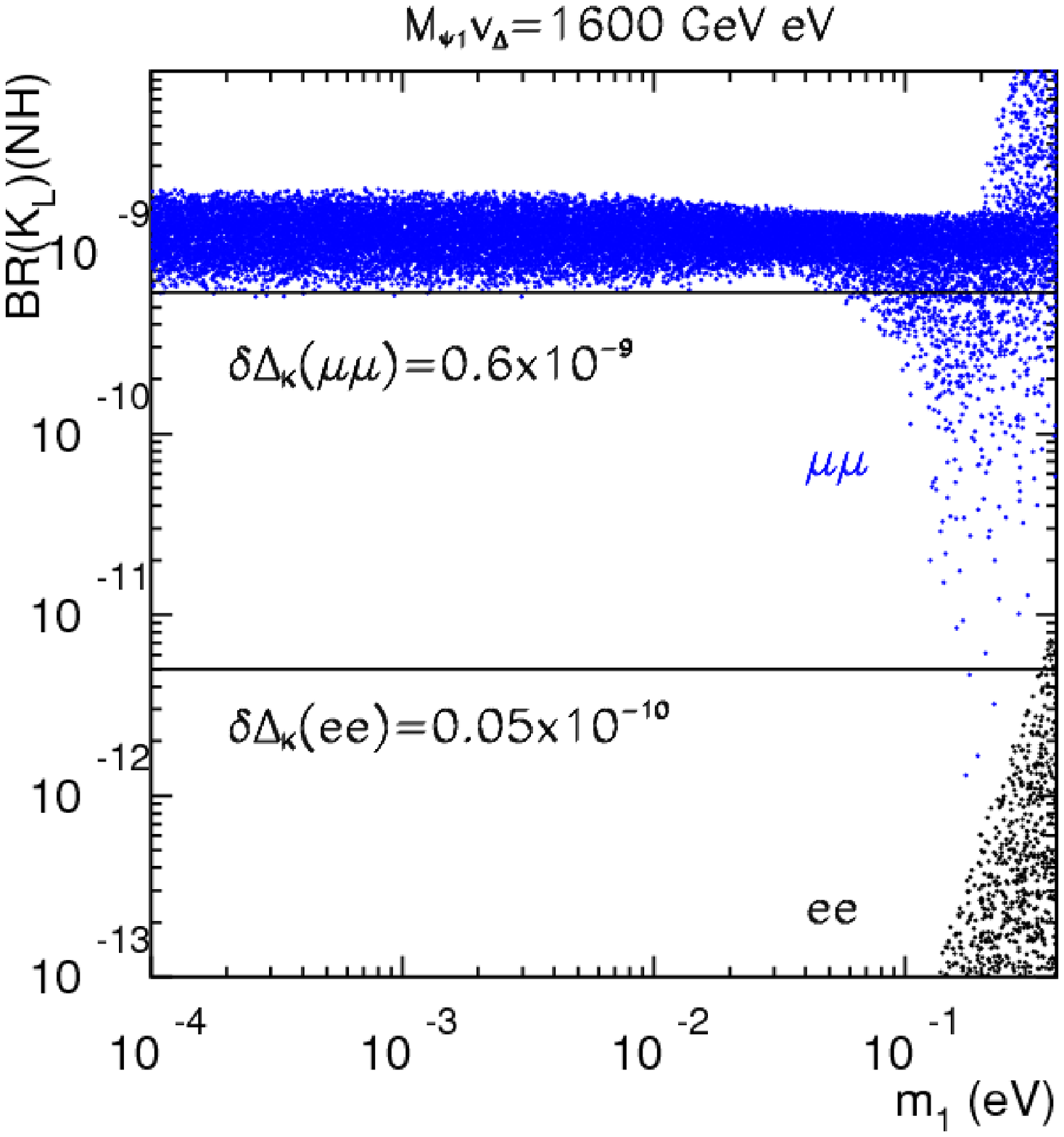}
\end{tabular}
\end{center}
\caption{Branching fractions for $K_L^0\to e^+e^-$(black) and $K_L^0\to \mu^- \mu^+$(blue) 
versus the lowest neutino mass for NH neglecting the phases,
$M_{\Psi_1}v_\Delta=1200~({\rm left}),~1600~({\rm right})~{\rm GeV}\cdot {\rm eV}$.
The horizontal lines represent the current $1~{\sigma}$ experimental bounds~\cite{PDG}.}
\label{nheemm}
\end{figure}

\begin{figure}[tb]
\begin{center}
\begin{tabular}{cc}
\includegraphics[scale=1,width=8cm]{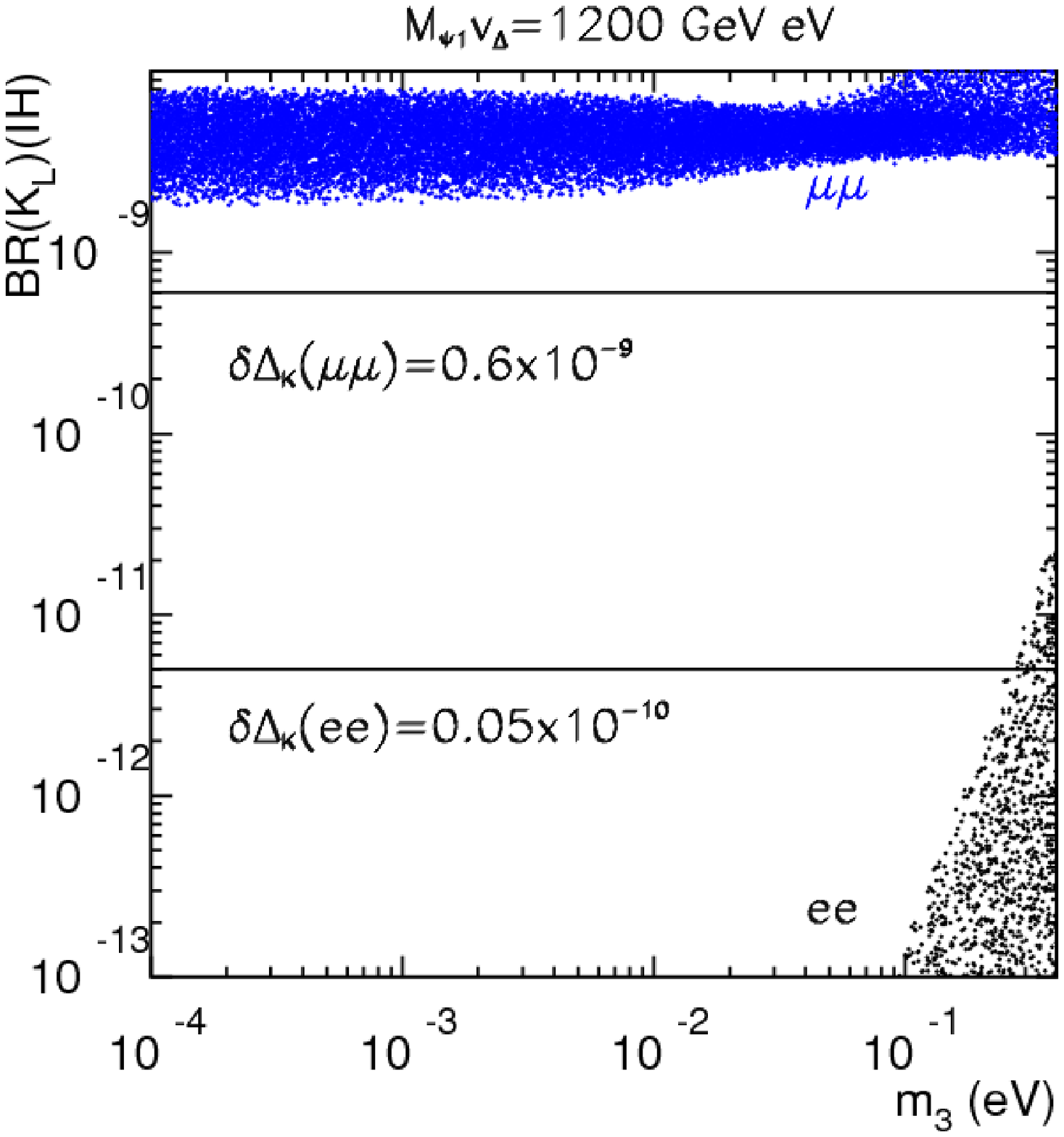}
\includegraphics[scale=1,width=8cm]{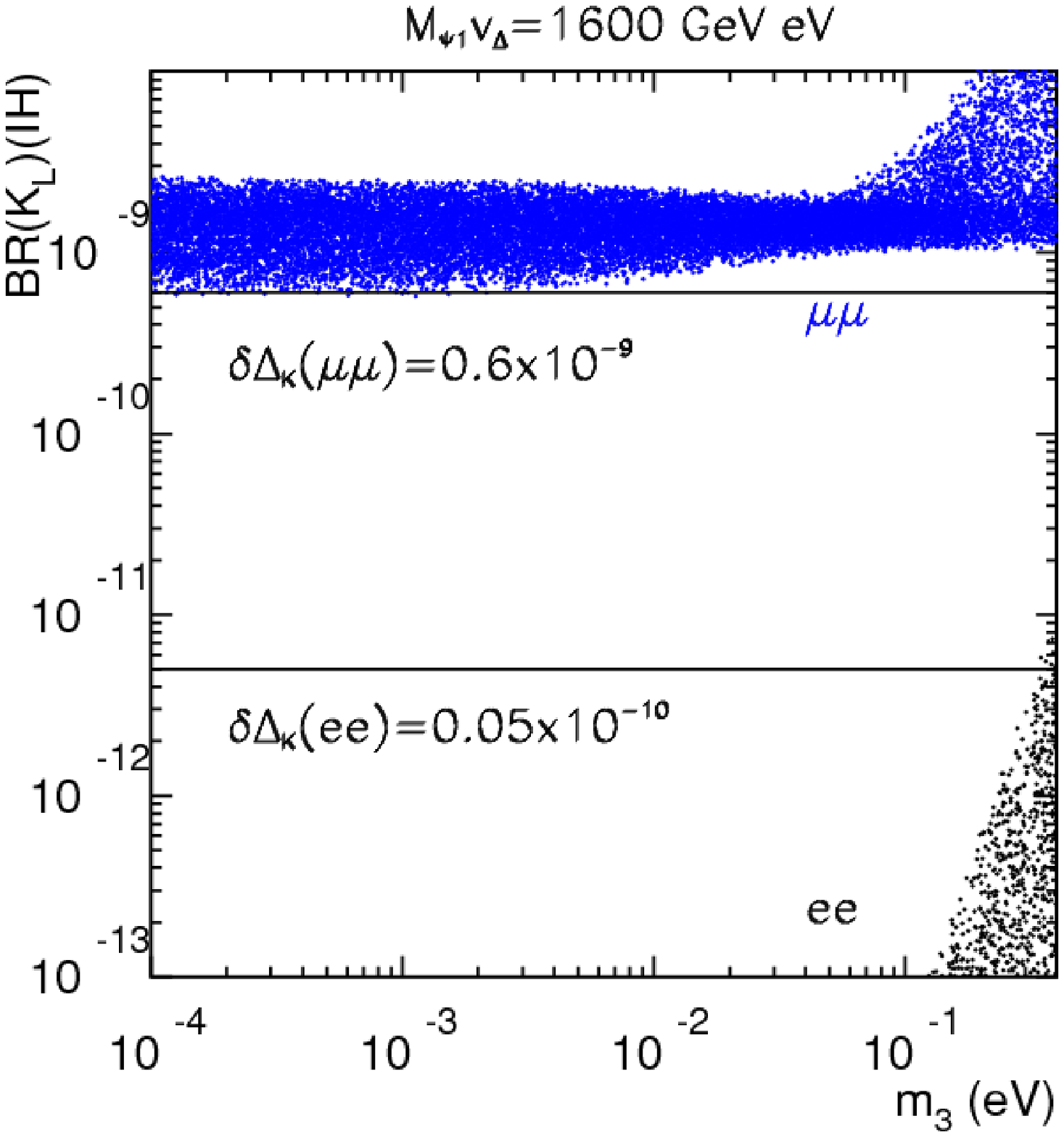}
\end{tabular}
\end{center}
\caption{Branching fractions for $K_L^0\to e^+e^-$(black) and $K_L^0\to \mu^- \mu^+$(blue) versus the lowest neutino mass for IH neglecting the phases, $M_{\Psi_1}v_\Delta=1200~({\rm left}),~1600~({\rm right})~{\rm GeV}\cdot {\rm eV}$. The horizontal lines represent the current $1~{\sigma}$ experimental bounds~\cite{PDG}. }
\label{iheemm}
\end{figure}

In Figs.~\ref{nheemm} and \ref{iheemm},
we show the branching fractions for $K_L^0\to e^+e^-,\ \mu^+\mu^-$ for the NH and IH, respectively, 
taking into account the leptoquark $\Psi_1$ contribution versus the lightest neutrino mass
for $M_{\Psi_1} v_\Delta > $1200 GeV$\cdot$ eV 
(left panel) and 1600 GeV$\cdot$ eV (right panel).
Note that the decay rate is proportional to $(M_{\Psi_1} v_\Delta)^{-4}$ since each $\Gamma^{k\ell}\sim m_\nu/v_\Delta$.
We see that there exist parameter space points for which the
leptoquark contribution is smaller than uncertainty in $\Delta_K(\ell\ell)$ as indicated by the horizontal lines 
for $M_{\Psi_1} v_\Delta > $1600 GeV$\cdot$ eV.

\begin{figure}[tb]
\begin{center}
\begin{tabular}{cc}
\includegraphics[scale=1,width=8cm]{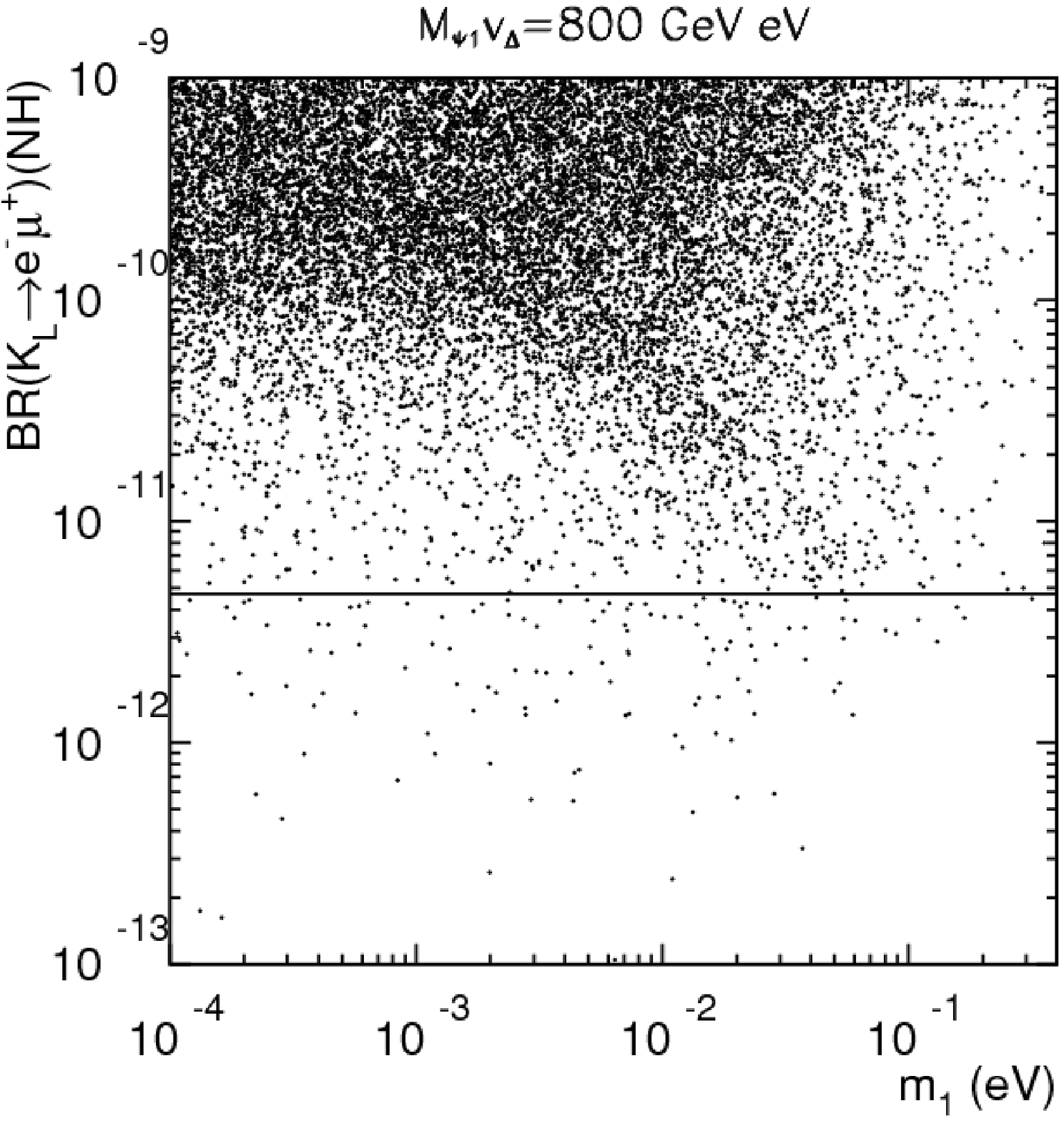}
\includegraphics[scale=1,width=8cm]{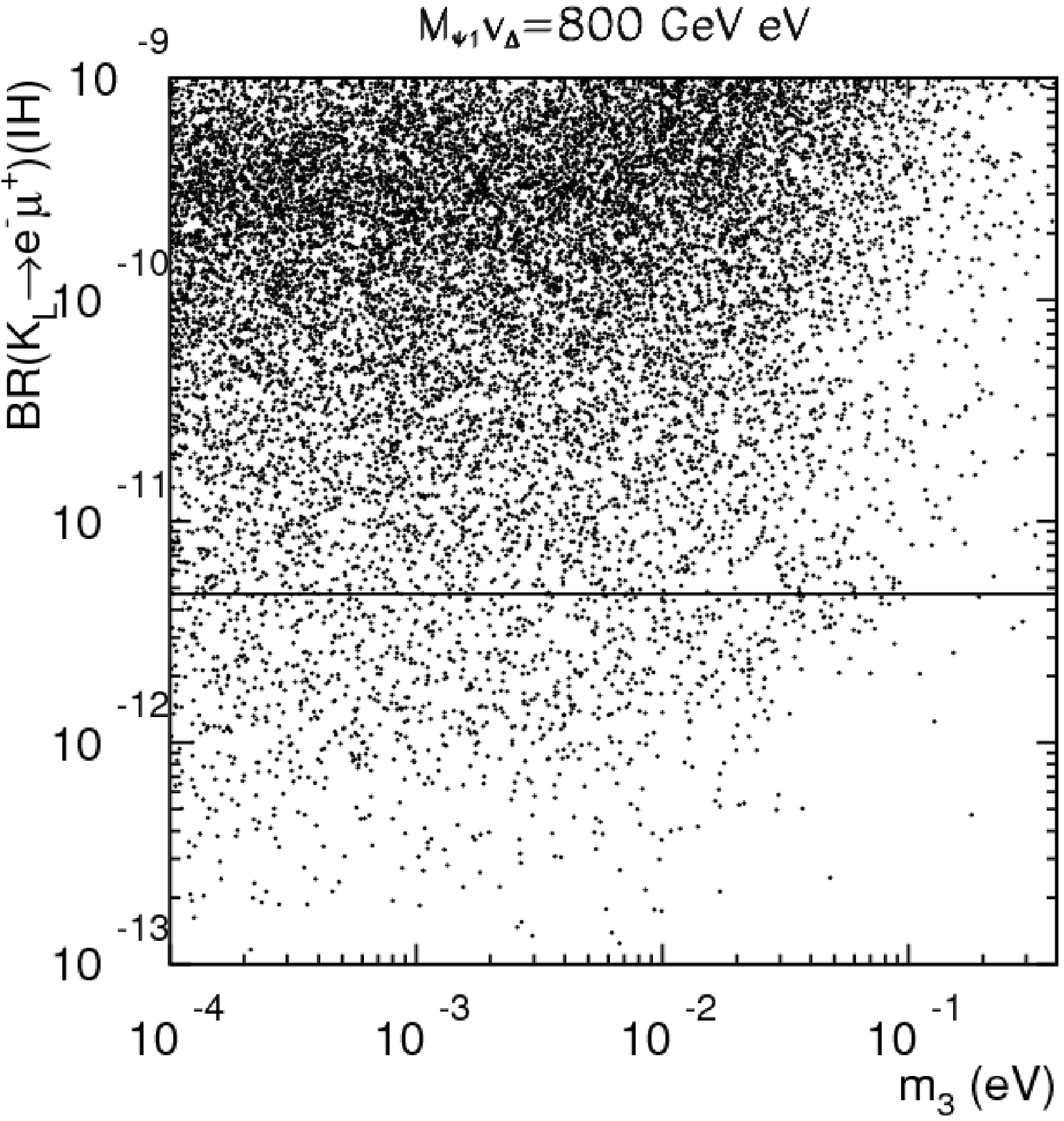}
\end{tabular}
\end{center}
\caption{Branching fractions for $K_L^0\to e^- \mu^+$ versus the lowest neutrino mass for 
NH (left) and IH (right) 
neglecting the phases, $M_{\Psi_1}v_\Delta=800~{\rm GeV}\cdot {\rm eV}$. 
The horizontal line represents the current $1~{\sigma}$ experimental bound~\cite{PDG}.}
\label{nhem}
\end{figure}

In Fig.~\ref{nhem} we show the predictions for LQ contributions to the decay channel $e^-\mu^+$ with 
$M_{\Psi_1} v_\Delta > $800 GeV$\cdot$ eV. In this case, the SM prediction is highly suppressed since flavor violation arises solely through the neutrino sector, so any observable effect would have to arise from new physics. The experimental BR limits, indicated by the horizontal lines in the left and right panels of Fig. \ref{nhem} thus translate into direct constraints on the LQ parameters.  We find the BR results for this channel more spread out and, thus, less constraining than are those for the lepton flavor conserving decays.
This observation leads to an important point: Although the experimental bounds on
the $e\mu$ and $ee$ channels are stronger than that of $\mu\mu$, 
they actually provide less stringent 
constraints on the model parameters, due to the suppressed couplings governed
by the neutrino oscillation data. 

Additional constraints follow from other meson properties and low-energy semileptonic interactions, such as
the mass difference between
$K_L$ and $K_S$: $\Delta m_K=3.48\times 10^{-15}$ GeV~\cite{PDG}.
The box diagrams for a leptoquark with couplings to leptons give a contribution
to $\Delta m_K$~\cite{constraint1,constraint2},  that yields a constraint
\begin{eqnarray}
\left( (\Gamma_1 \Gamma_1^\dagger)^{21} \right)^2 \lesssim {192\pi^2 M_{\Psi_1}^2 \Delta m_K\over f_K^2 m_K}
\approx 5.2 \times 10^{-10} M_{\Psi_1}^2.
\end{eqnarray}
As we have shown in the previous section when the lightest neutrino mass is smaller than $10^{-2}$ eV,
$(Y_1 Y_1^\dagger)^{21}$ is smaller than $25 (24) \times 10^{-4}$ \ \text{eV}$^2$ in the NH (IH)
spectrum. Using these results one finds
\begin{equation}
\frac{v_\Delta}{1 \text{eV}} \gtrsim 1.05 \times \left(\frac{10^2 \ \text{GeV}}{M_{\Psi_1}} \right)^{1/2}.
\end{equation}
This leads to a weaker bound than that obtained from the rare decays. 
The same constraints are valid for the case of $\Psi_2$. Notice that one can satisfy these
bounds easily. 

We emphasize that the most stringent constraints, which follow from $K_L^0\to \mu^- \mu^+$, imply that for 
$M_{\Psi_1} v_\Delta > $1600 GeV$\cdot$ eV it is possible to find values for the Yukawa couplings that are consistent with both neutrino oscillation data and the quantity $\Delta_K(\mu\mu)$. As we discuss below, one could expect to observe a significant number of LQ events at the LHC for LQ masses in the approximate range 
$400\lesssim M_{\Psi_1}\lesssim 1000$ GeV, or $v_\Delta$ lying between about two and four eV. 
Since we have no independent handle on the value of $v_\Delta$ nor any reason to preclude 
values lying in this range, we conclude that consistency of an LHC LQ discovery and the 
flavor changing decay $K_L^0\to \mu^- \mu^+$ would imply a relatively large triplet vev in this model. 

\subsubsection{$\mu-e$ Conversion Constraints}
\begin{figure}[tb]
\begin{center}
\begin{tabular}{ccc}
\includegraphics[scale=1,width=8cm]{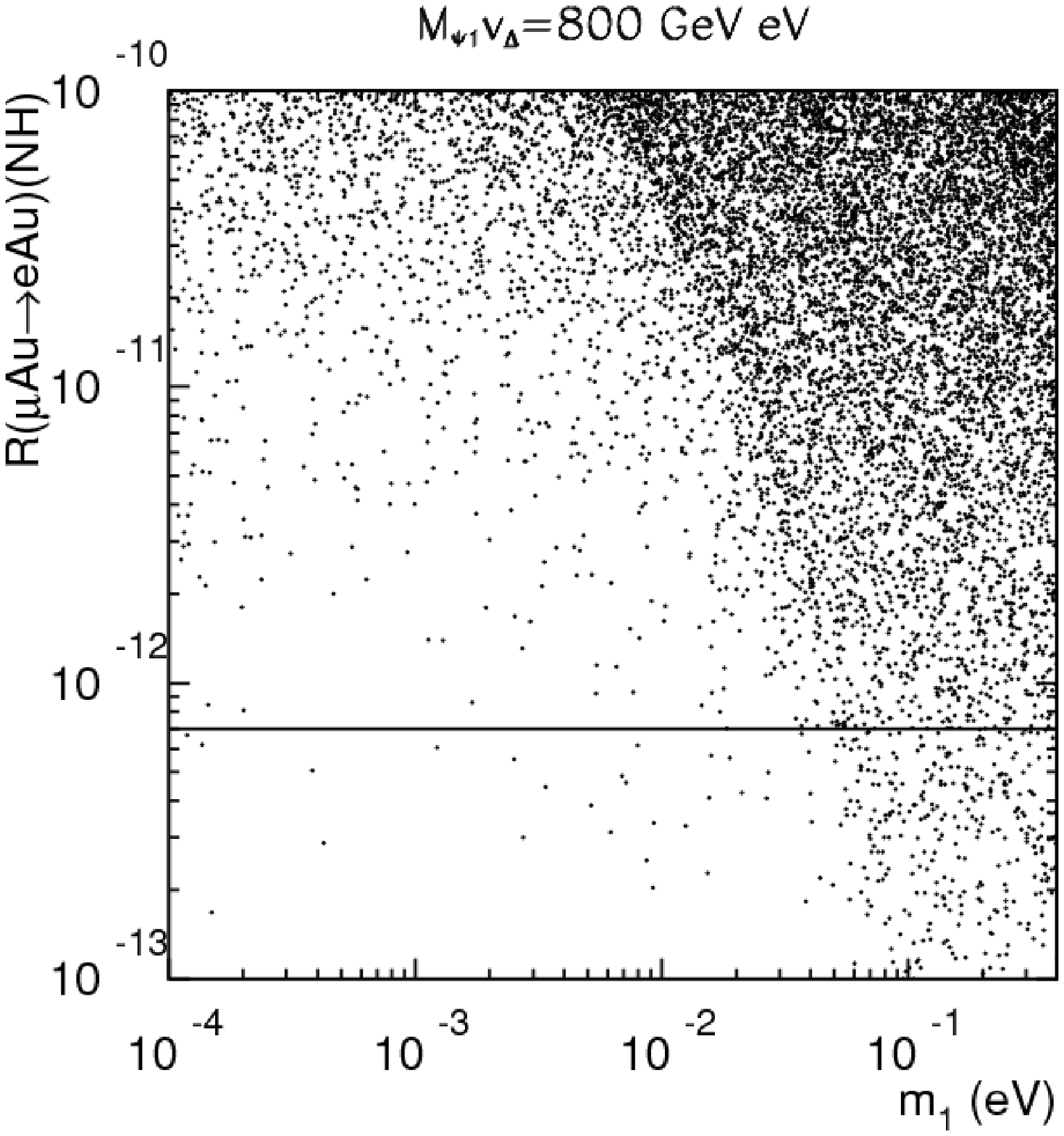}
\includegraphics[scale=1,width=8cm]{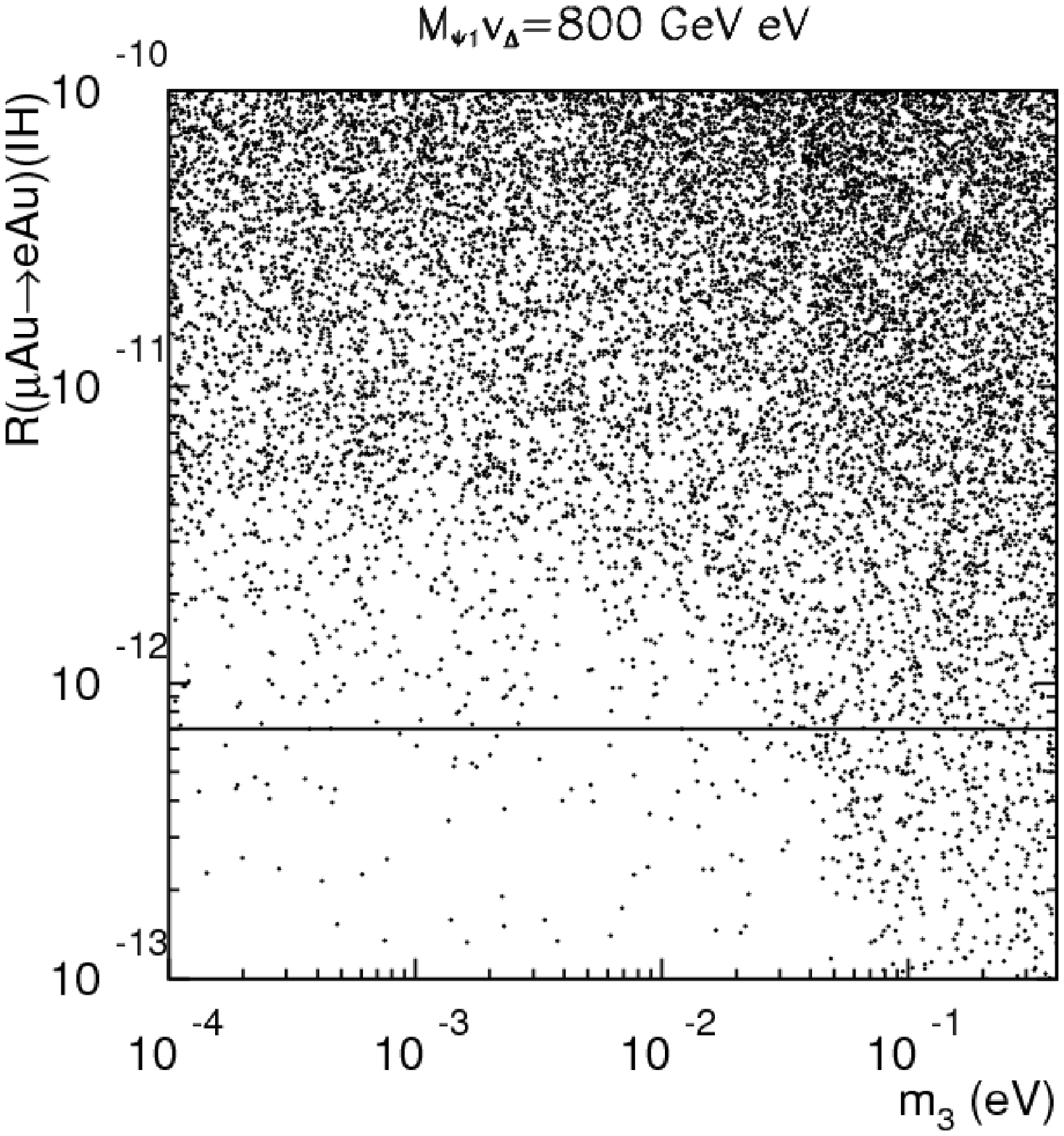}
\end{tabular}
\end{center}
\caption{The rate of $\mu-e$~conversion in $\rm Au$ nuclei versus the lowest neutrino mass for NH(left) and IH(right) without any phases $M_{\Psi_1}v_\Delta=800~{\rm GeV\cdot eV}$. The horizontal line represents the current 90\% C.L. experimental bound \cite{PDG}.}
\label{nhAu}
\end{figure}

Searches for charged lepton flavor violation (CLFV) in processes that conserve quark flavor can impose severe constraints on new physics. In the present case, searches for $\mu$-$e$ conversion in nuclei yield the most stringent constraints on the LQ couplings and masses. The tightest bound on the rate for this process has been obtained by the SINDRUM Collaboration\cite{Bertl:2006up} using the gold nucleus. The result for the conversion to capture ratio $R^{\rm Au}_{\mu\to e}$ is
\begin{equation}
R^{\rm Au}_{\mu\to e} = {\Gamma[\mu^-+A(Z,N) \to e^- + A(Z,N)]\over \Gamma[\mu^-+A(Z,N) \to \nu + A(Z-1,N+1)]} 
 < 7.0\times 10^{-13} \quad (90\% {\mathrm{C.L.}})\ \ \ .
 \end{equation}
 This ratio can be expressed in terms of the reduced conversion and  capture rates $R^{\rm Au}_{\mu\to e}=\omega_{\rm conv}/\omega_{\rm capt}$ where \cite{mecon}
 \begin{equation}
 \omega_{\rm conv}={|\Gamma_1^{11}|^2|\Gamma_1^{12}|^2 m_\mu^5\over 4M_{\Psi_1}^4}(V^{(p)}+2V^{(n)})^2\ \ \ ,
 \end{equation}
 where $V^{(p)}$ and $V^{(n)}$ are overlap integrals involving the upper and lower components of the muon and electron wavefunctions and the proton (p) and neutron (n) number densities. Using the values for these integrals computed in Ref.~\cite{mecon}, the experimental value for $\omega_{\rm capt}$ and experimental limit on $R^{\rm Au}_{\mu\to e}$ shown in  Table.~\ref{nucl}, we obtain 90\% C.L. constraints on the LQ parameters as shown in Fig.~\ref{nhAu} . There we give 
$R^{\rm Au}_{\mu\to e}$ corresponding
to lightest neutrino mass in NH and IH when $M_{\Psi_1}v_\Delta=800~{\rm GeV\cdot eV}$. We observe that in either case, there exist values of the LQ parameters for which the predicted ratio lies below the 90\% C.L. limit (indicated by the horizontal line). Although the present limits exclude large portions of parameter space, they do not preclude the model entirely. Future searches for $\mu\to e$ conversion, such as the Mu2e experiment proposed for Fermilab or the PRIME experiment at JPARC, could improve the sensitivity be several orders of magnitude. A null result from these future experiments could rule out the LQ model under consideration here unless the product $M_{\Psi_1}v_\Delta$ is quite heavy. We will explore this possibility in a forthcoming study. 

Finally, we note that the leptoquark contribution to other lepton flavor violating processes, such as
$\mu\to e\gamma$ and $\mu\to 3e$, arise at one-loop level, and thus  provide
less stringent constraints.

\begin{table}[tb]
\begin{center}
\begin{tabular}{|c|c|c|c|c|}
\hline
Nucleus & $V^{(p)}$ & $V^{(n)}$ & $\omega_{\rm capt}(10^6s^{-1})$ & $\omega_{\rm conv}/\omega_{\rm capt}$
\\
\hline
$^{197}_{79}{\rm Au}$ & 0.0974 & 0.146 & 13.07 & $<7\times 10^{-13}$\\
\hline
\end{tabular}
\end{center}
\caption{Values of the relevant parameters for $\mu-e$ conversion in Au~\cite{PDG,mecon}.}
\label{nucl}
\end{table}
\subsection{Other Constraints}

\subsubsection{Oblique Parameters}
In this section we study the constraints implied by leptoquark loop contributions to electroweak precision observables (EWPO). Since the LQ-matter field couplings are tiny due to their dependence on $Y_\nu$, we may safely neglect vertex, fermion propagator, and box graph loop corrections and concentrate on corrections to the gauge boson propagators. For sufficiently heavy LQs, one may characterize the leading effects of these corrections in terms of the oblique parameters,  $S$, $T$, and $U$ \cite{Peskin}. The $U$ parameter is typically quite small and does not add a significant constraint in this case. We have computed the full set of LQ contributions to the gauge boson self-energy functions, and the resulting contributions to $T$ and $S$ are:
\begin{eqnarray}
\label{eq:Tparam}
\hat{\alpha}(M_Z)T&\equiv& \frac{1}{M_W^2}
\Biggl\{ {\hat \Pi}_{WW}(0)-
{\hat c}^2\left( {\hat \Pi}_{ZZ}(0)+\frac{2{\hat s}}{\hat c}
{\hat \Pi}_{Z\gamma}(0) \right) \Biggr\}^{\rm
New}
\approx {N_C\over 8\pi^2 v^2}\left(-{2\Delta M^2\over 3}\right),\\
\nonumber
{\hat{\alpha}(M_Z)\over 4\hat{s}_Z^2\hat{c}_Z^2}S&\equiv& \frac{1}{M_Z^2}{\rm Re}\Biggl\{
{\hat \Pi}_{ZZ}(M_Z^2)-{\hat \Pi}_{ZZ}(0)+\frac{{\hat c}^2-{\hat
s}^2}{{\hat c}{\hat s}} \left[{\hat \Pi}_{Z\gamma}(0)-{\hat
\Pi}_{Z\gamma}(M_Z^2)\right] -{\hat \Pi}_{\gamma\gamma}(M_Z^2)
\Biggr\}^{\rm New} \\
\label{eq:Sparam}
&\approx& -{N_C\over 24\pi^2 v^2}\left({M_Z^2\Delta M\over 3M_{\Psi_1}}\right)
\end{eqnarray}
where $\Delta M=M_{\Psi_2}-M_{\Psi_1}$, ${\hat s}$ (${\hat c}$) is the sine (cosine) of the weak mixing angle in the $\overline{\mathrm{MS}}$ scheme, and \lq\lq New" indicates the contribution from new physics ($\Psi_{1,2}$ ). We have carried out renormalization at a scale $\mu=M_Z$. The complete expressions for LQ contributions are given in  Appendix \ref{app:oblique}, while in Eqs.~(\ref{eq:Tparam},\ref{eq:Sparam}) we give approximate expressions in the limit that the mass splitting $|\Delta M|$ and $Z$-boson mass are small compared to the LQ mass $M_{\Psi_1}$. 
The latest global fit to EWPO yields for these parameters
are $S=-0.10\pm 0.10(-0.08)$, $T=-0.08\pm 0.11(+0.09)$, and $U=0.15\pm 0.11 (+0.01)$, assuming a value for the SM Higgs boson mass $M_H=117$ GeV (300 GeV)~\cite{PDG}.
The most important constraint is from the $T$ parameter which gives us a 
$1\sigma$ bound of leptoquark mass splitting $|\Delta M| \lesssim 60~{\rm GeV}$.
\subsubsection{Collider Constraints and $\rho$-parameter}
The current constraint on the LQ mass, $M_{\Psi_1}$ and $M_{\Psi_2}$, comes from the
direct searches at the Tevatron~\cite{PDG}
\begin{eqnarray}
M_{\Psi_{1,2}}\gtrsim 250~{\rm GeV}.
\end{eqnarray}
while for the values of the triplet vev one has,
\begin{eqnarray}
1~{\rm eV}\lesssim v_\Delta\lesssim 1~{\rm GeV},
\end{eqnarray}
where the lower bound follows from the assumption that $m_\nu\sim 1$ eV and that the Yukawa interactions are perturbative, while  the upper bound is from the constraint on the electroweak $\rho-$parameter~\cite{rho}.
\section{Leptoquark Decays and Neutrino Masses}
In this section we study the main features of the LQ decays taking into 
account the constraints on the neutrino masses and mixing.
From Eq.~(\ref{G12}) one can conclude that the leptoquark (LQ)
decays could be different in each spectrum for neutrinos 
and we will explore this interesting connection.
\subsection{Main Features of the Leptoquarks Decays}
The decays of the leptoquarks into fermions
$\Psi_1 \ \to \ d_i \ e^+_j$  and $\Psi_2 \to d_i \ \bar{\nu}$, where $d_i=d,s,b$ and $e_j=e,\mu,\tau$,
are of most interest. The partial widths are given by
\begin{eqnarray}
\Gamma(\Psi_1\to d_i e^+_j)={|\Gamma_1^{ij}|^2\over 16\pi}M_{\Psi_1},
\qquad
\Gamma(\Psi_2\to d_i \bar{\nu}_j)={|\Gamma_2^{ij}|^2\over 16\pi}M_{\Psi_2}.
\end{eqnarray}
The widths are proportional to the Yukawa coupling squared, governed by
$m^2_\nu/ v_\Delta^2$.

\begin{figure}[tb]
\begin{center}
\begin{tabular}{cc}
\includegraphics[scale=1,width=8cm]{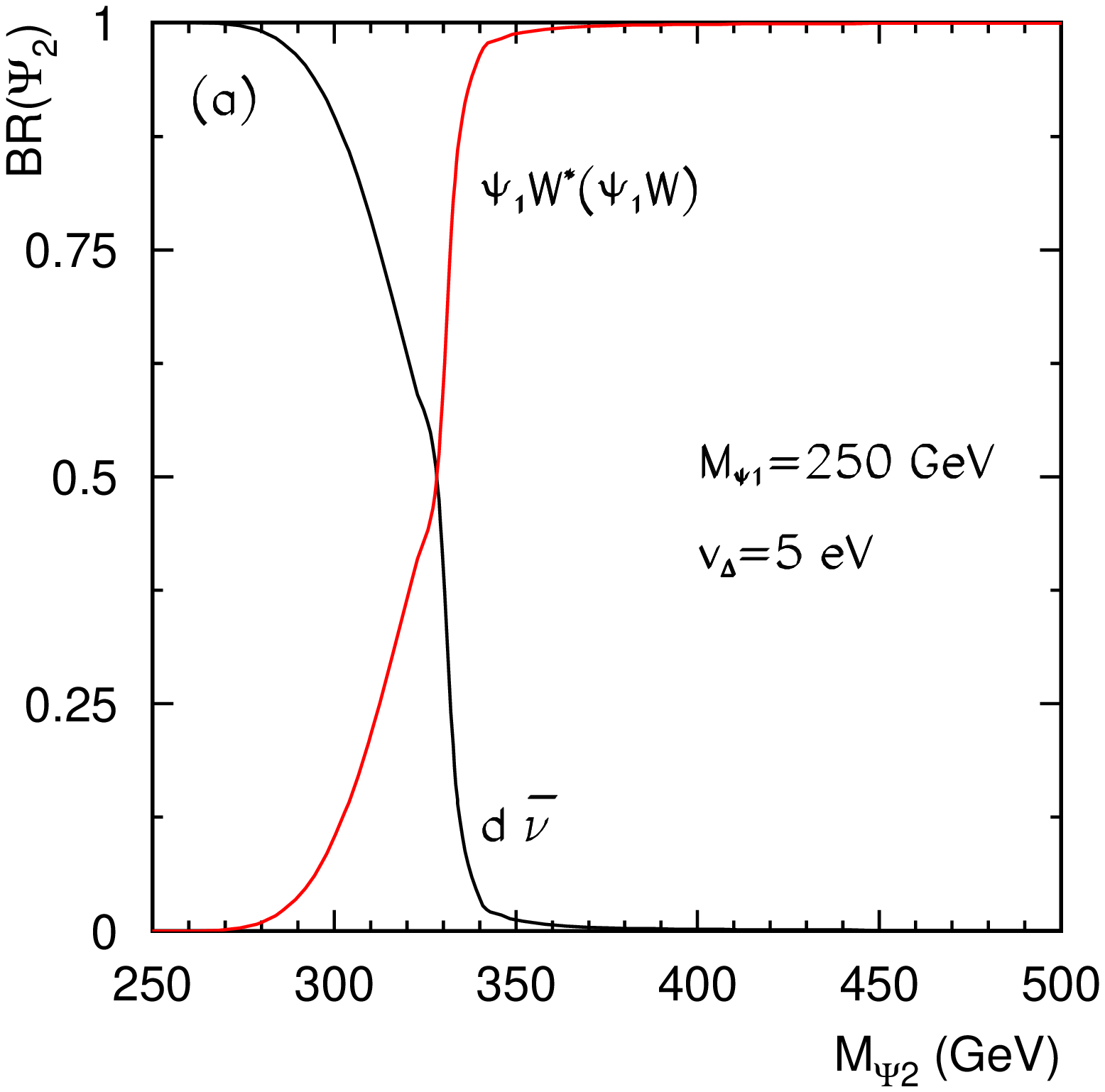}
\includegraphics[scale=1,width=8cm]{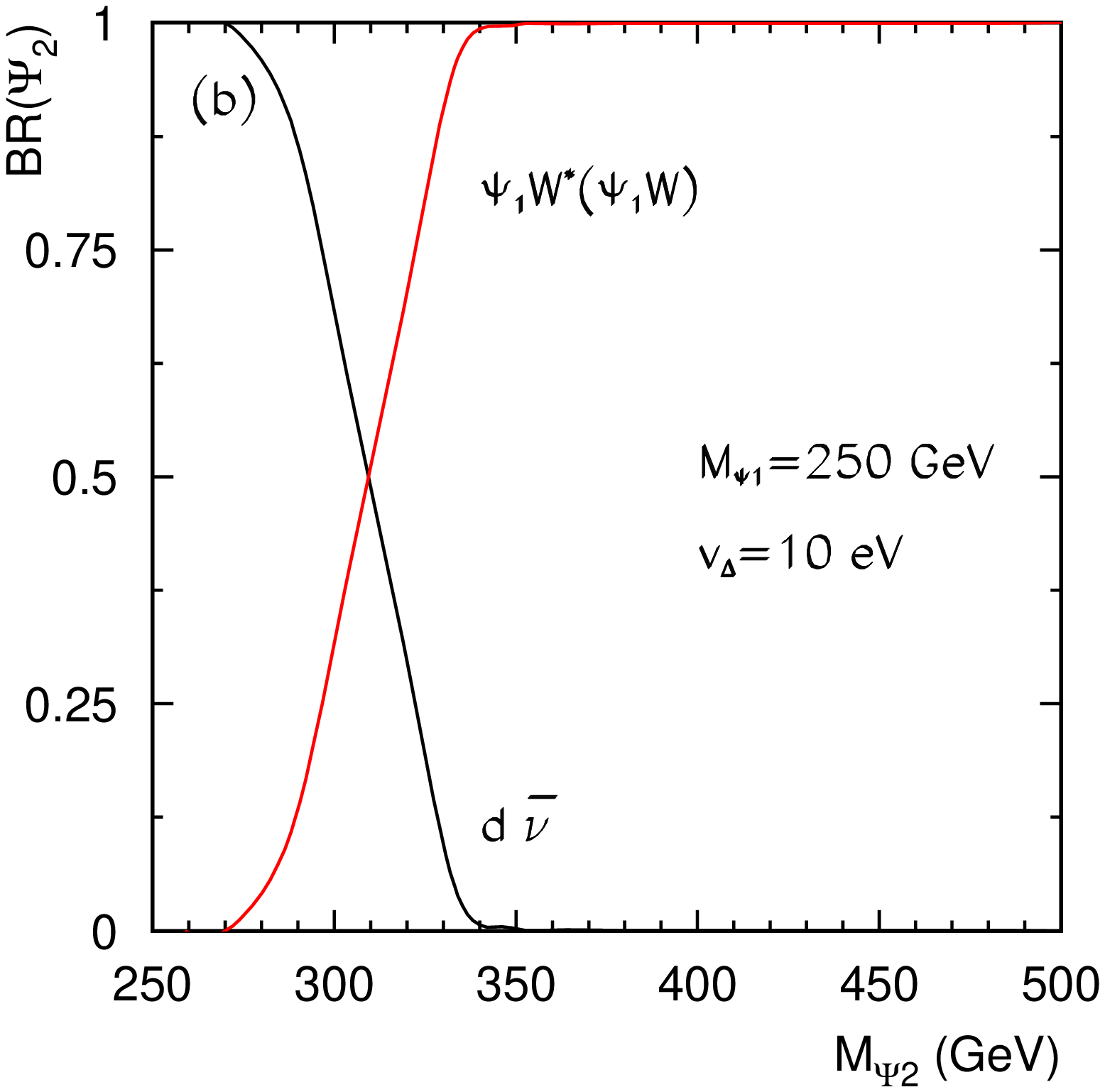}
\end{tabular}
\end{center}
\caption{Branching fractions of $\Psi_2$ decay for (a) $v_\Delta=5~{\rm eV}$ and 
(b) $10~{\rm eV}$, respectively, with $M_{\Psi_1}=250~{\rm GeV}$.}
\label{brm}
\end{figure}

\begin{figure}[tb]
\begin{center}
\begin{tabular}{cc}
\includegraphics[scale=1,width=8cm]{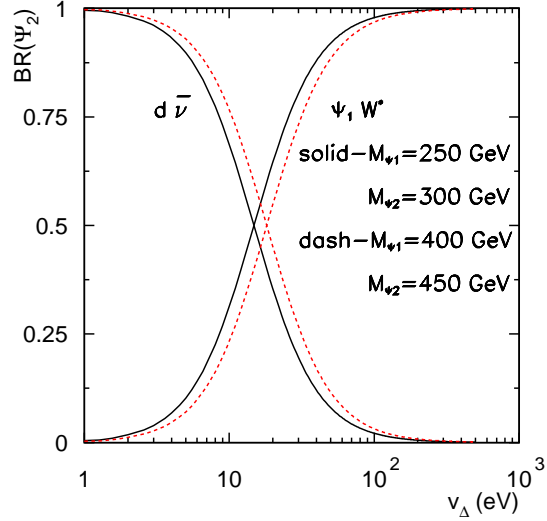}
\end{tabular}
\end{center}
\caption{Branching fractions of $\Psi_2$ decay for $M_{\Psi_1}=250~{\rm GeV}$, $M_{\Psi_2}=300~{\rm GeV}$ and $M_{\Psi_1}=400~{\rm GeV}$, $M_{\Psi_2}=450~{\rm GeV}$.}
\label{brv}
\end{figure}

\begin{figure}[tb]
\begin{center}
\begin{tabular}{cc}
\includegraphics[scale=1,width=8cm]{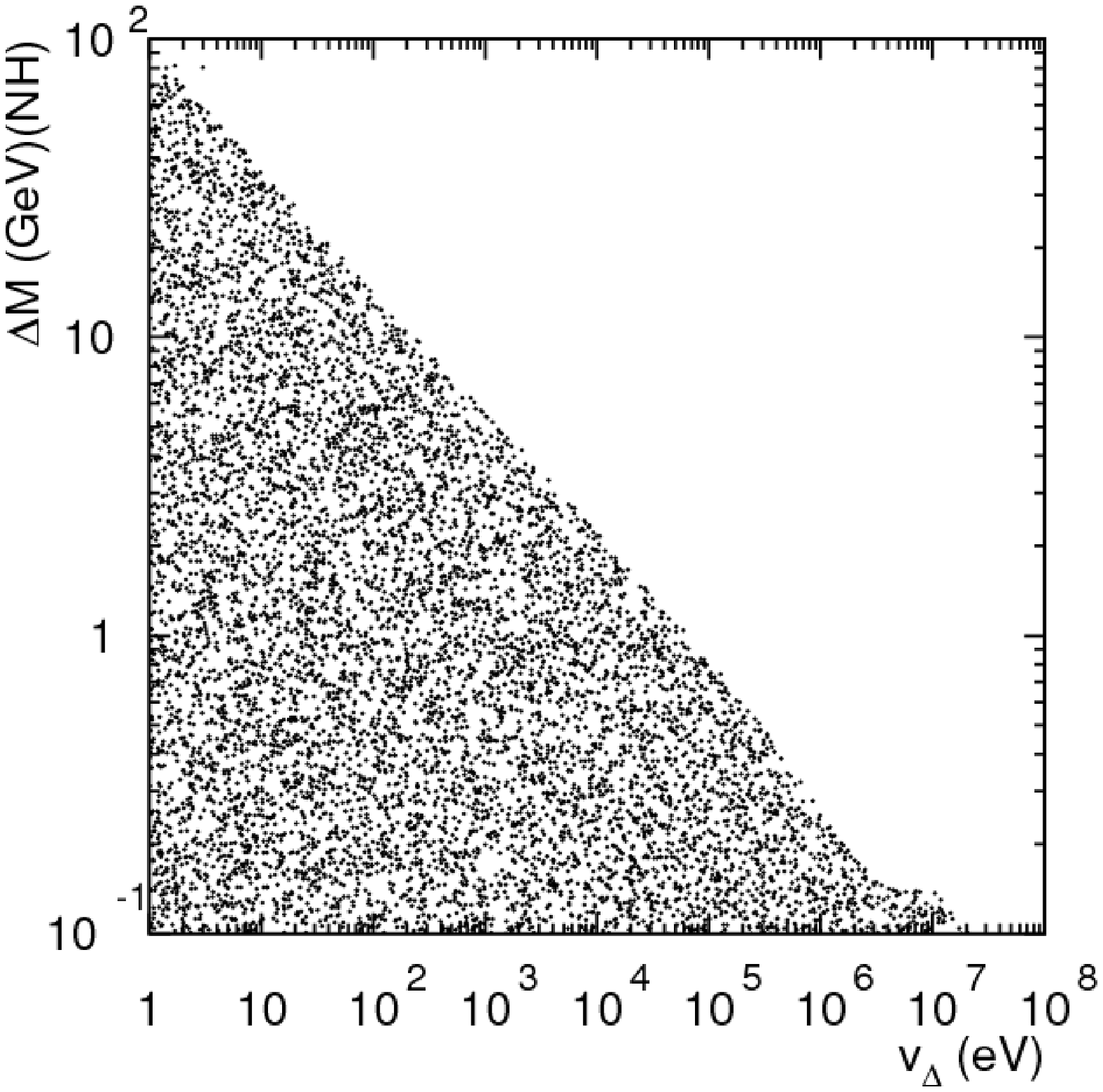}
\includegraphics[scale=1,width=8cm]{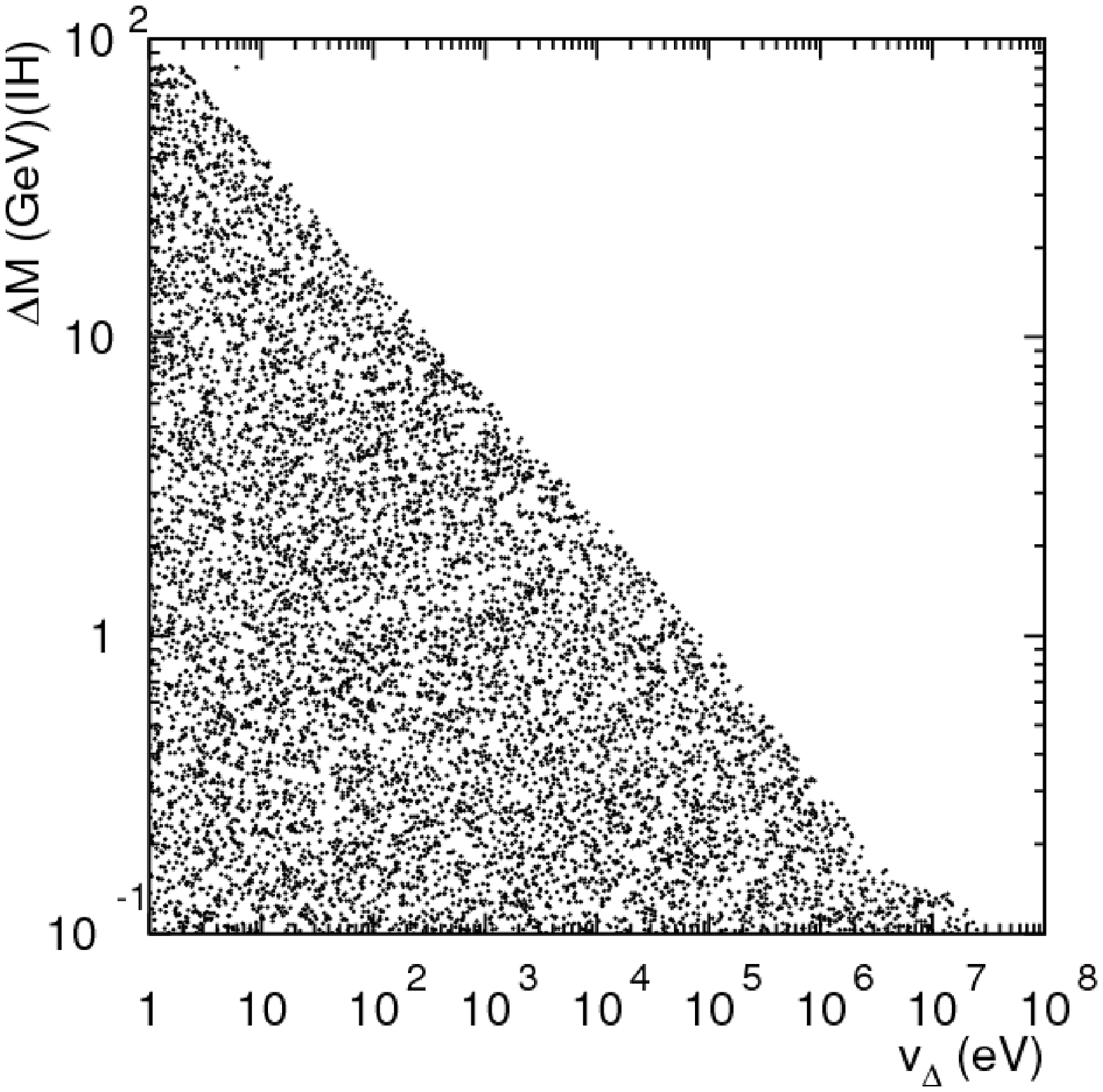}
\end{tabular}
\end{center}
\caption{Allowed region of the parameter space when the
leptonic decays of $\Psi_2$ are dominant for NH (left) and IH (right). $\Delta M$ versus $v_\Delta$
assuming $M_{\Psi_1}=250~{\rm GeV}$ when $10^{-4}~{\rm eV}\leq m_0\leq 10^{-2}~{\rm eV}$, $m_0$ is the lightest neutrino mass.}
\label{mdv}
\end{figure}

The other competing decay modes come from the charged current gauge interaction.
Depending on the mass splitting between the leptoquarks, one has the decays of
the heavy leptoquark into the lighter one plus a real ($W$) or virtual ($W^*$)
gauge boson. For example, if $\Delta M=M_{\Psi_2}-M_{\Psi_1}>0$,
the $\Psi_2$ decay rates for these processes are given by
\begin{eqnarray}
\Gamma(\Psi_2\to \Psi_1 W^-)&=&{M_{\Psi_2}g_2^2\over 32\pi r_W^2}\lambda^{3\over 2}(1,r_W^2,r_{\Psi_1}^2),
\\
\Gamma(\Psi_2\to \Psi_1 W^{-\ast}\to \Psi_1\pi^-)&=&{g_2^4V_{ud}^2\Delta M^3 f_\pi^2\over 32\pi M_W^4}\sqrt{1-{m_\pi^2\over \Delta M^2}},
\\
\Gamma(\Psi_2\to \Psi_1 W^{-\ast}\to \Psi_1e^-(\mu^-)\bar{\nu}_e(\bar{\nu}_\mu))&=&{g_2^4\Delta M^5\over 480\pi^3M_W^4}, \\
\label{eq:qqbar}
\Gamma(\Psi_2\to \Psi_1 W^{-\ast}\to \Psi_1q\bar{q}')&=&3\Gamma(\Psi_2\to \Psi_1e^-(\mu^-)\bar{\nu}_e(\bar{\nu}_\mu)).
\end{eqnarray}
where we have omitted the CKM factor associated with the $W q{\bar q}^\prime$ coupling in Eq.~(\ref{eq:qqbar}) 
since we are considering the inclusive channels of $q\bar q'$, and 
where $r_i=M_i/M_{\Psi_2}$.
In Fig.~\ref{brm} we show the branching fractions for the decays of $\Psi_2$
versus its mass with $M_{\Psi_1}=250~{\rm GeV}$ and $v_\Delta=5$ eV and $10$ eV
in (a) and (b), respectively, assuming that the Yukawa coupling is diagonal for simplicity.
We see that once the mass difference is large and a real gauge boson channel is open,
then it takes over the fermionic channels. In Fig.~\ref{brv} we plot the branching
fractions versus $v_\Delta$ with $M_{\Psi_1}=250~{\rm GeV}$ and $M_{\Psi_2}=300~{\rm GeV}$. 
Also in the second case: $M_{\Psi_1}=400$ GeV, and $M_{\Psi_2}=450$ GeV.
We see that for $v_\Delta<10~{\rm eV}$, the Yukawa coupling is sufficiently large so that
the fermionic channels become dominant. The features are the same for $\Psi_1$ if it is heavier than $\Psi_2$.
The presence of charged leptons  in the final state (rather than missing energy associated with neutrinos) facilitates the use of collider observables to identify  the connection between the leptoquark properties and the neutrino mass spectrum. Consequently,
we will focus on the parameter region in which $M_{\Psi_2}> M_{\Psi_1}$.
To further quantify this situation, we present  the scatter plots in Fig.~\ref{mdv}
in a plane of  $\Delta M-v_\Delta^{}$ under the condition
$\Gamma(\Psi_2\to d_i \bar{\nu}_j)>\Gamma(\Psi_2\to \Psi_1 W^*)$.
We see that if the mass difference is small enough, there is a broad range for the
vev $v_\Delta^{}\sim 1-10^6~{\rm eV}$, although an $\rm eV$ value may be more natural for the
light neutrino mass generation.
\subsection{Leptoquark Decays and Neutrino Spectra}
\begin{figure}[tb]
\begin{center}
\begin{tabular}{cc}
\includegraphics[scale=1,width=8cm]{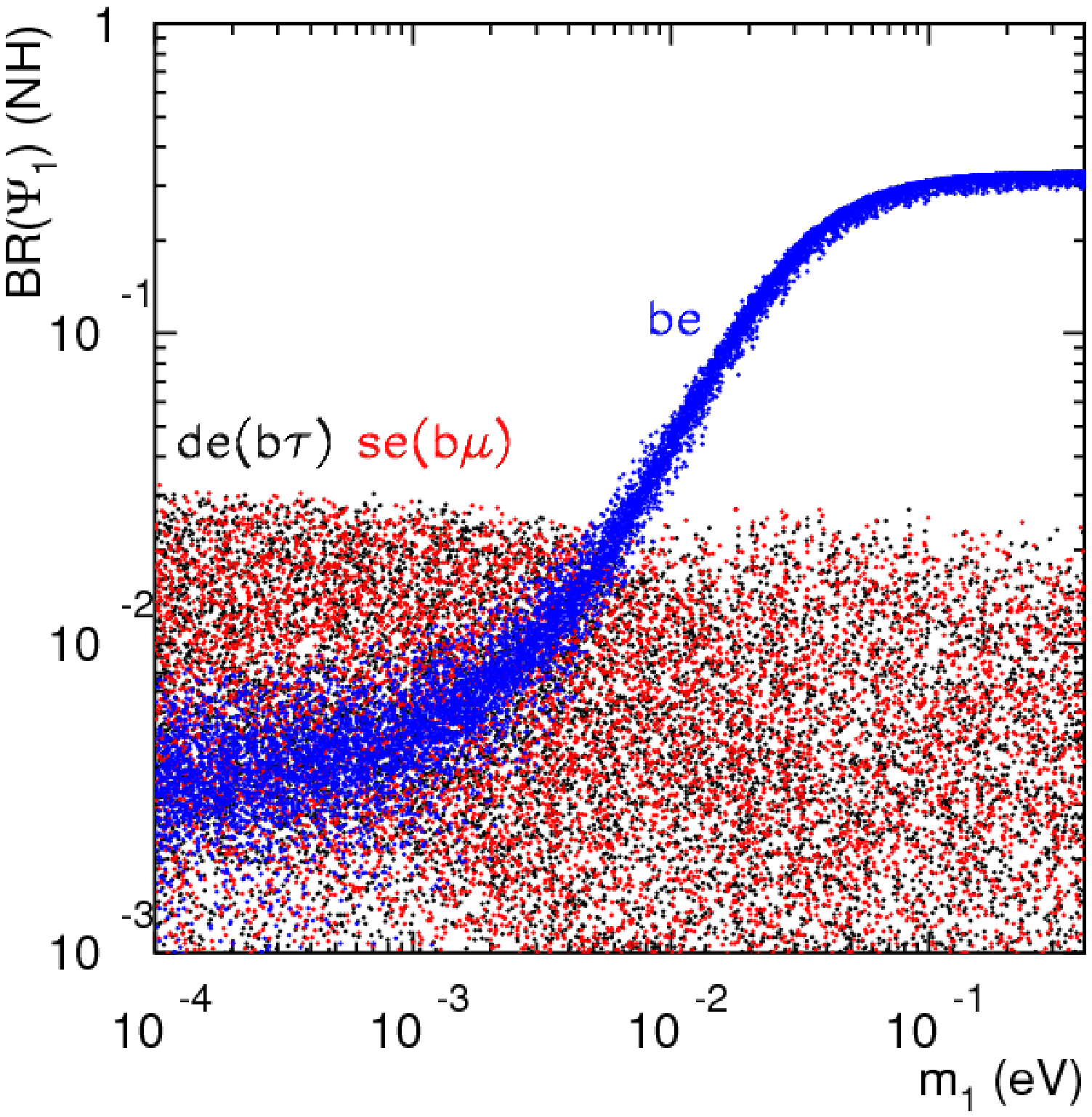}
\includegraphics[scale=1,width=8cm]{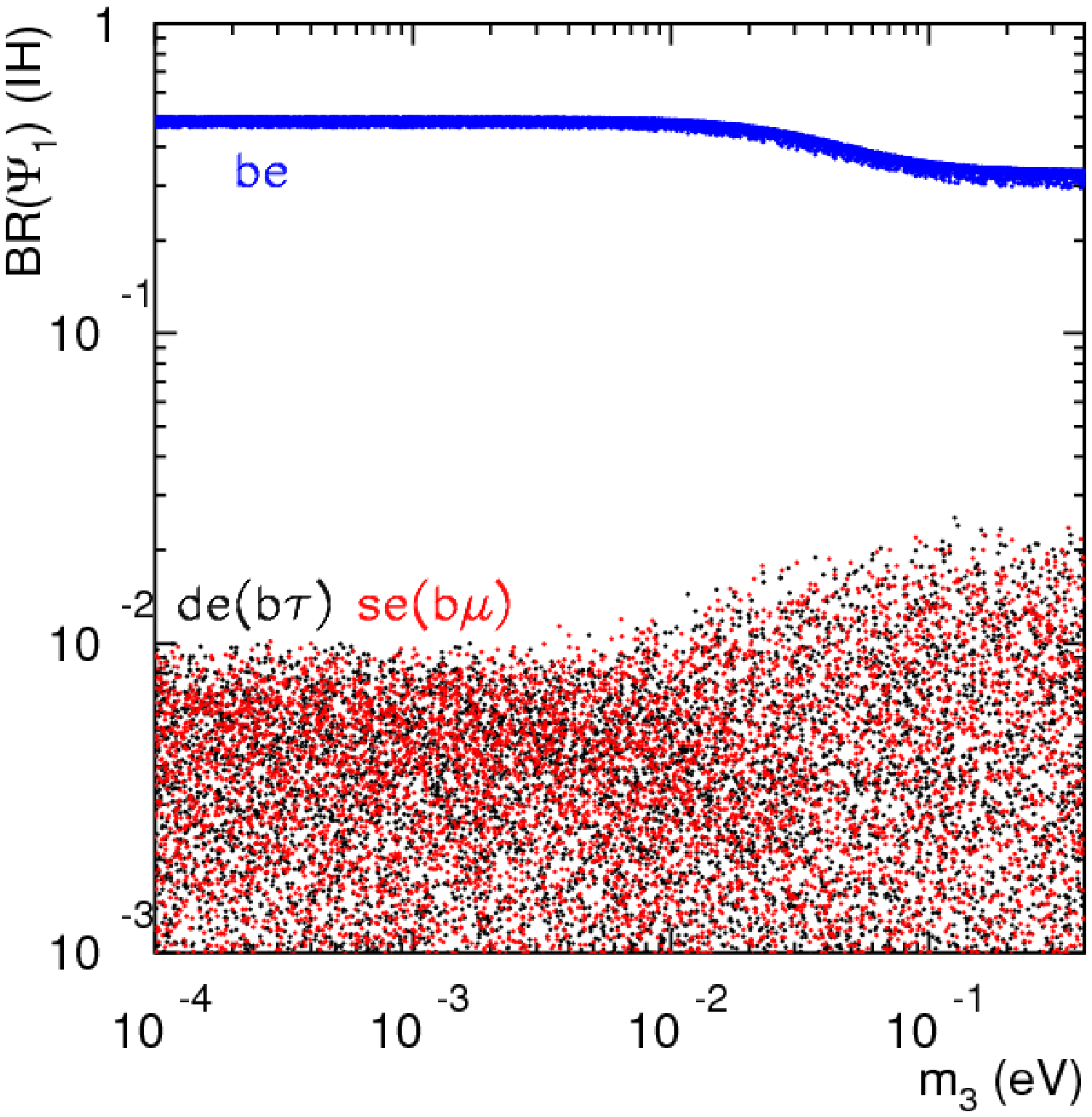}
\end{tabular}
\end{center}
\caption{The branching fractions of $\Psi_1\to d_ie \ (d_i=d,s,b)$ 
versus the lowest neutrino mass for NH (left) and IH (right) 
 when all the phases vanish.
Due to the symmetry as in Eq.~(\ref{nmy}), the equal channels are also indicated in the
parentheses. }
\label{p1bre}
\end{figure}

\begin{figure}[tb]
\begin{center}
\begin{tabular}{cc}
\includegraphics[scale=1,width=8cm]{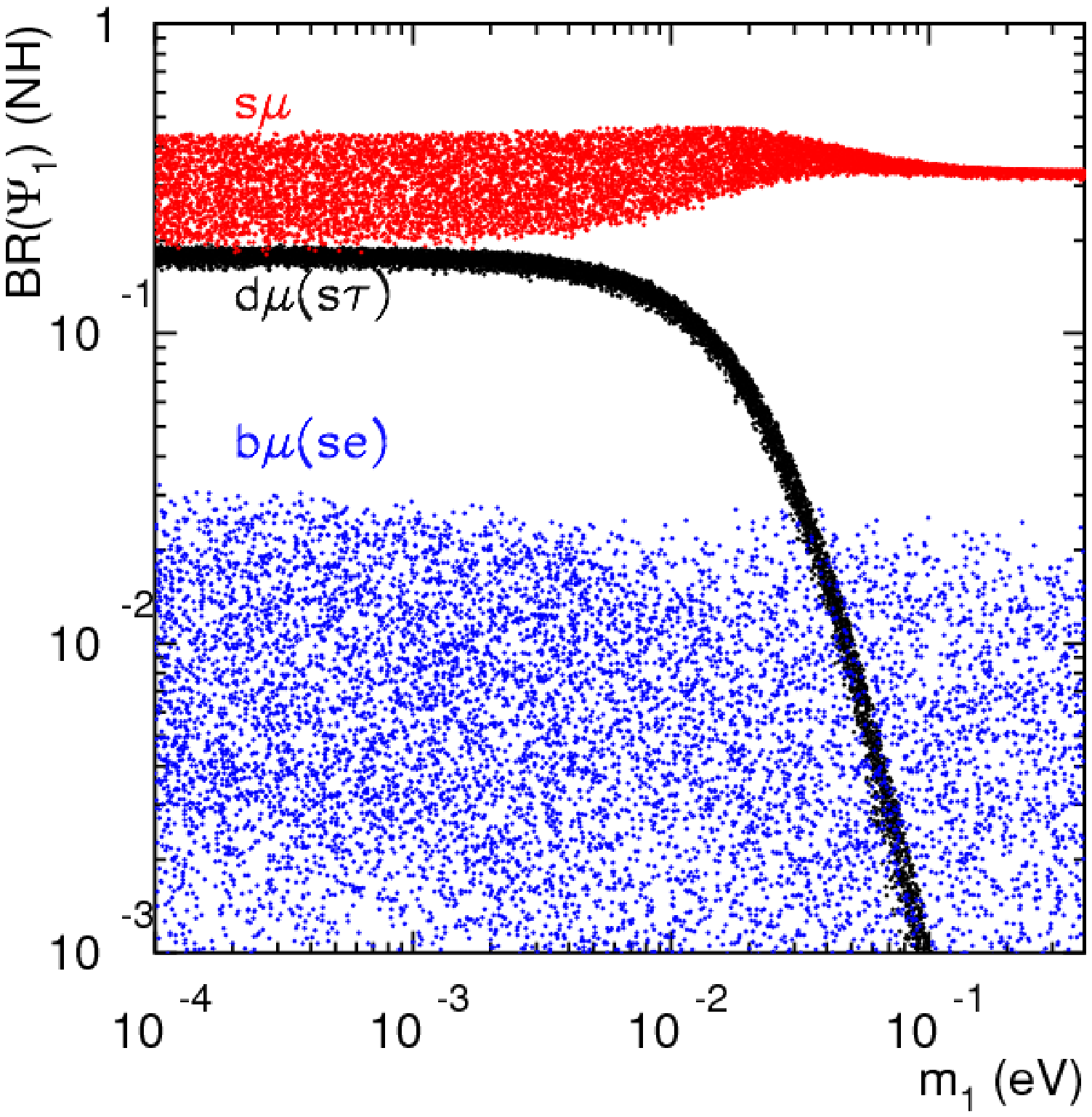}
\includegraphics[scale=1,width=8cm]{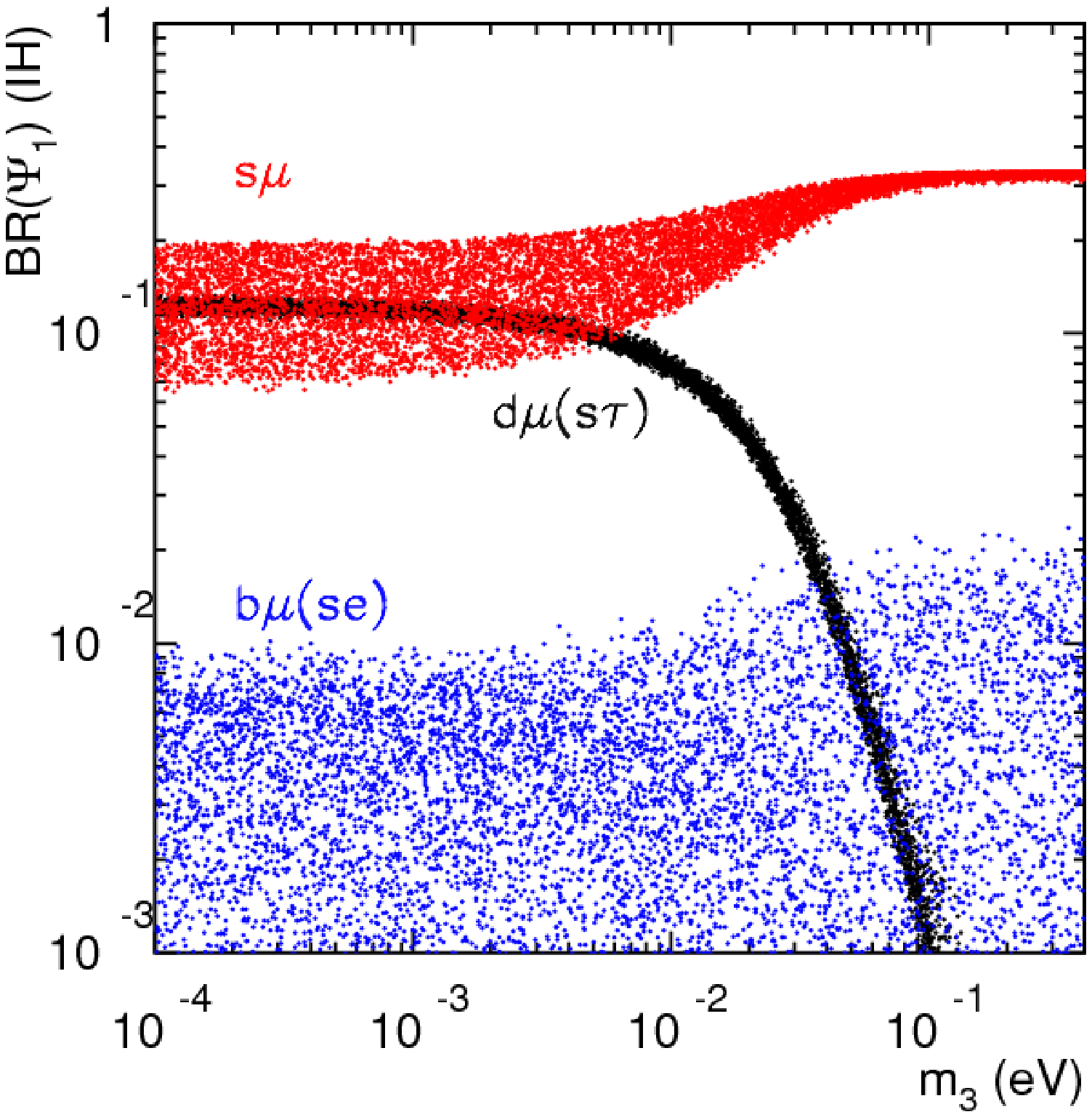}
\end{tabular}
\end{center}
\caption{The branching fractions of $\Psi_1\to d_i\mu \ (d_i=d,s,b)$ 
versus the lowest neutrino mass for NH (left) and IH (right) 
 when all the phases vanish.
Due to the symmetry as in Eq.~(\ref{nmy}), the equal channels are also indicated in the
parentheses. }
\label{p1brm}
\end{figure}

\begin{figure}[tb]
\begin{center}
\begin{tabular}{cc}
\includegraphics[scale=1,width=8cm]{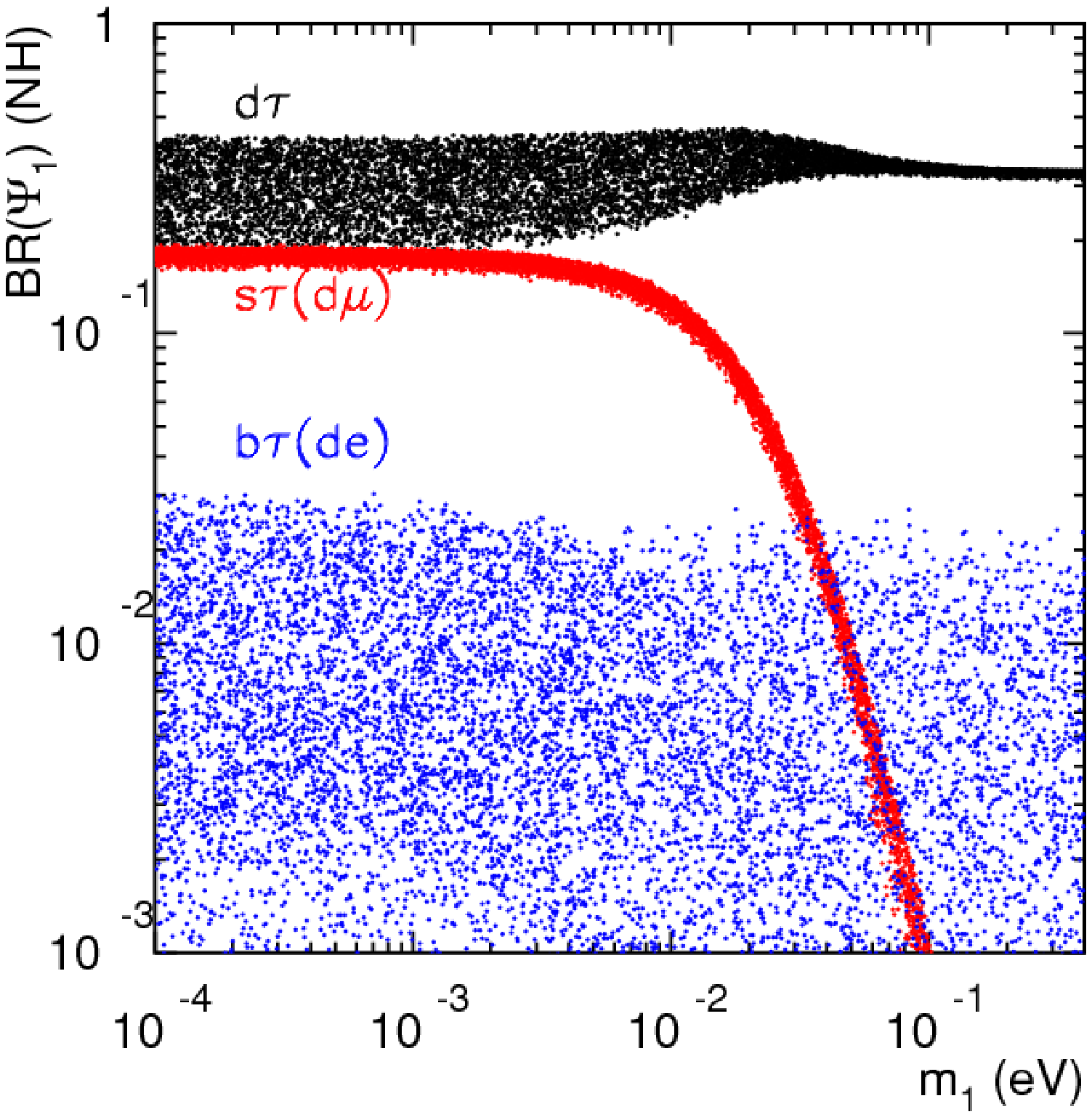}
\includegraphics[scale=1,width=8cm]{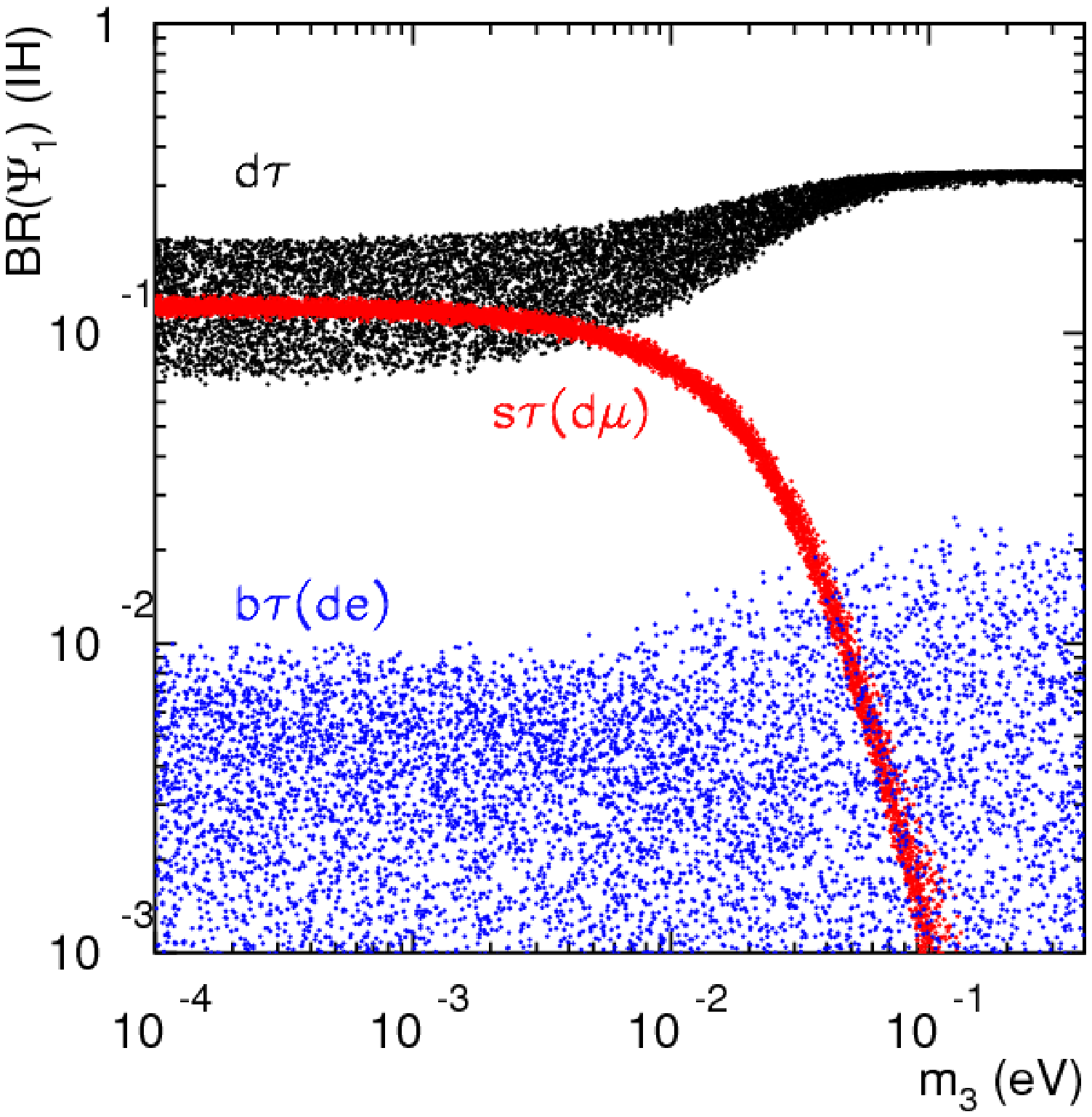}
\end{tabular}
\end{center}
\caption{The branching fractions of $\Psi_1\to d_i\tau \ (d_i=d,s,b)$ 
versus the lowest neutrino mass for NH (left) and IH (right) 
 when all the phases vanish.
Due to the symmetry as in Eq.~(\ref{nmy}), the equal channels are also indicated in the
parentheses. }
\label{p1brt}
\end{figure}

As discussed earlier, the lepton-flavor contents of leptoquark decays will be
different for each neutrino spectrum. Here, we study this
issue in great detail. In Fig.~\ref{p1bre},~\ref{p1brm} and~\ref{p1brt} we show the
impact of the neutrino masses and mixing angles on the branching fractions of $\Psi_1$ 
decaying into $e,\mu,\tau$ lepton respectively, with the left panels for the
Normal Hierarchy (NH) and the right panels for the Inverted Hierarchy (IH).
We first note that  the absence of the decay channel $be$ would indicate a NH
with  $m_{1} < 10^{-2}$ eV,  as seen in Fig.~\ref{p1bre}. In contrast, the $be$ channel is
the leading one in the IH. 
The $s\mu,\ d\mu$ and $d\tau,\ s\tau$ channels have the 
leading branching fractions in the NH,  but those channels alone 
do not seem to provide sufficient information to  discriminate 
among the various possible neutrino mass spectra.
If the lightest neutrino mass is above 0.1 eV, then this approaches 
the ``quasi-degenerate" (QD) scenario of the neutrino mass pattern. 
The channels ($be,\ s\mu,\ d\tau$) governed by the diagonal neutrino
matrix elements reach to about $30\%$ in this case. 
The interesting feature here is that $d\mu$ and the 
$s\tau$ channels vanish due to the unitarity cancelation 
of the mixing matrix.

\begin{figure}[tb]
\begin{center}
\begin{tabular}{cc}
\includegraphics[scale=1,width=8cm]{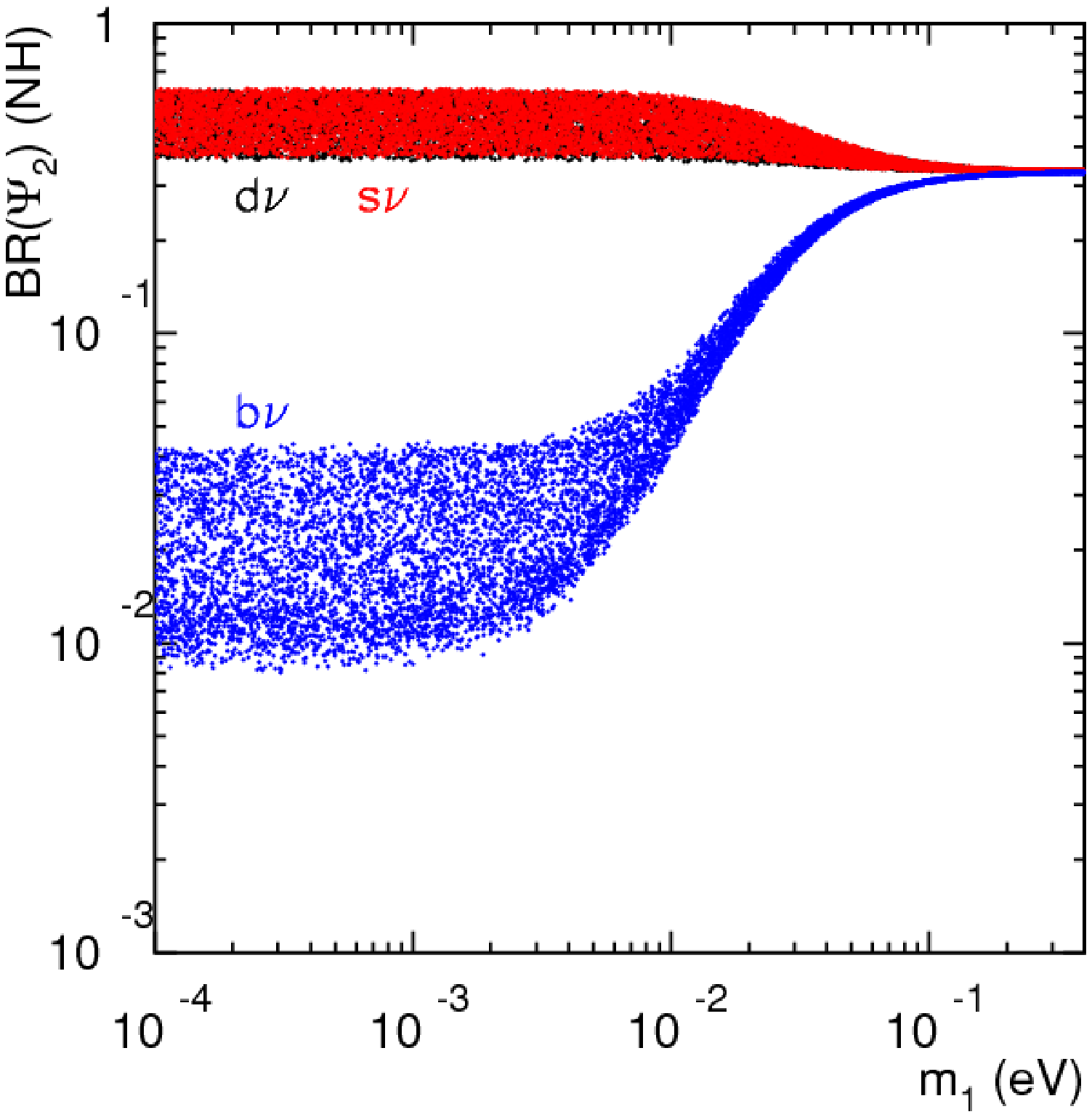}
\includegraphics[scale=1,width=8cm]{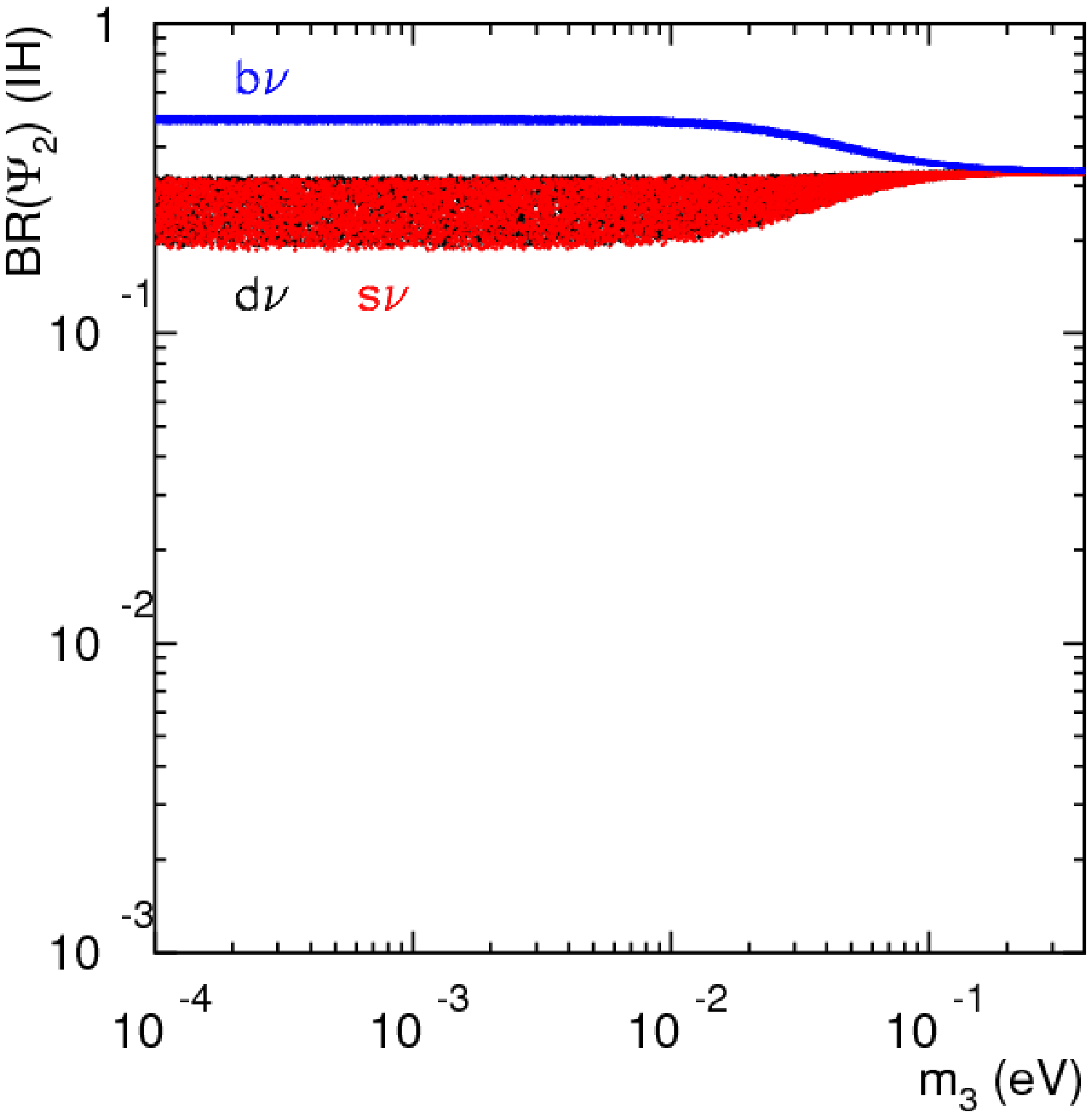}
\end{tabular}
\end{center}
\caption{$\Psi_2$ branching fractions versus the lowest neutrino mass
for NH (left) and IH (right), independent of the phases.} 
\label{p2i}
\end{figure}

\begin{figure}[tb]
\begin{center}
\begin{tabular}{cc}
\includegraphics[scale=1,width=8cm]{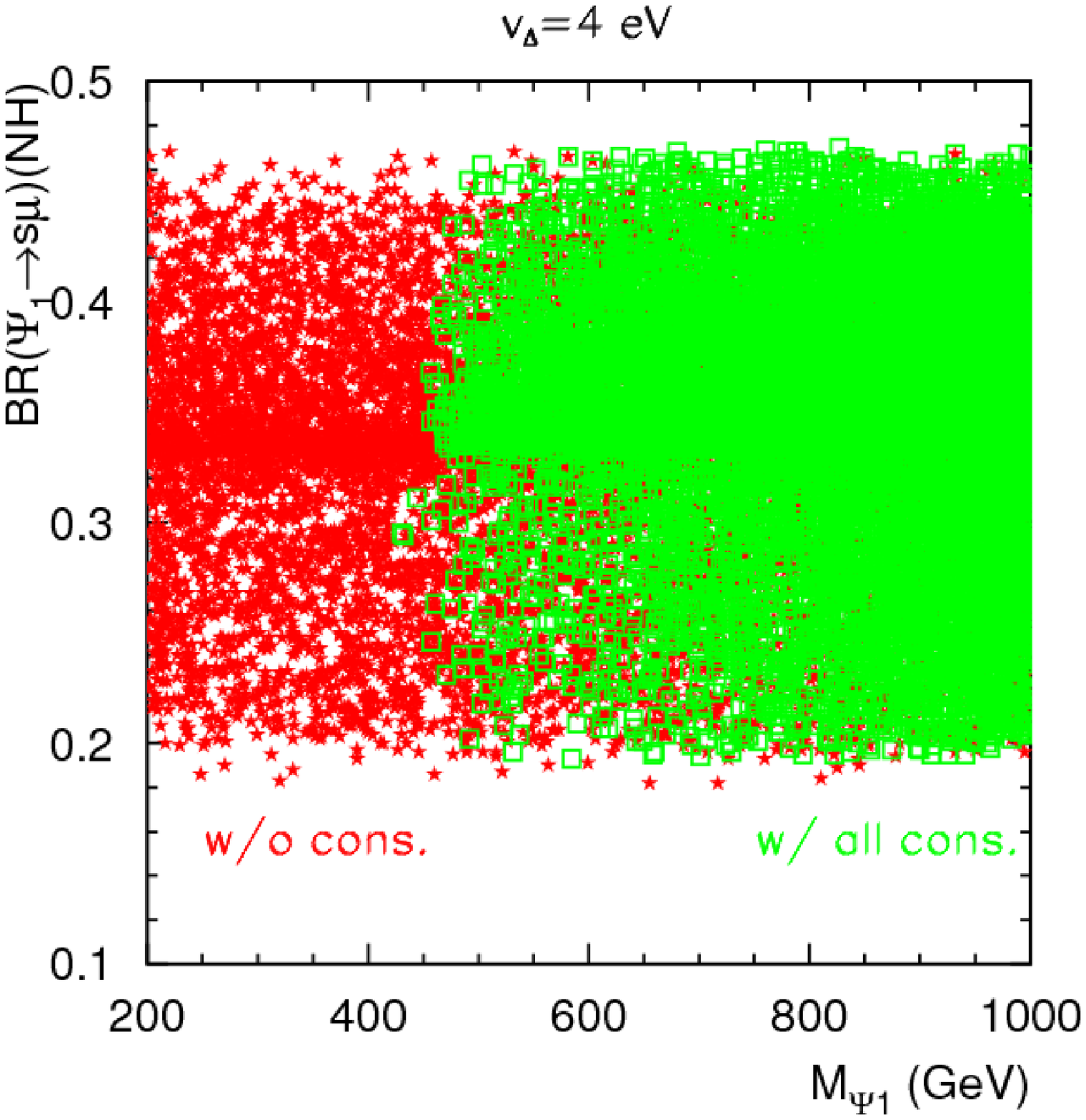}
\includegraphics[scale=1,width=8cm]{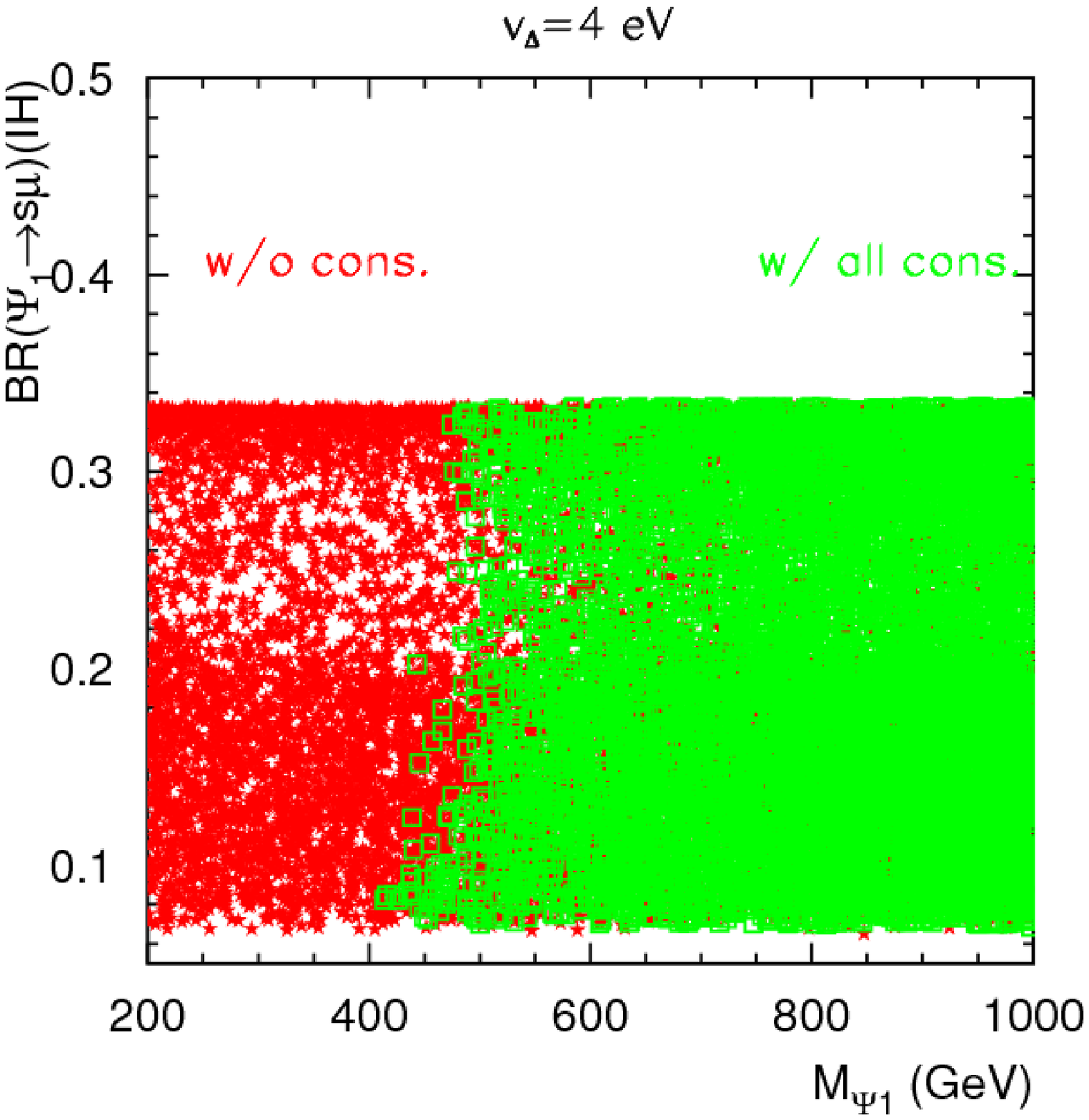}
\end{tabular}
\end{center}
\caption{The branching fraction of leading channel $\Psi_1\to s\mu$ versus leptoquark mass for NH (left) and IH (right) without any constraints (red solid star), and with both $K_L$ decay and $\mu-e$ conversion constraints (green empty square), $v_\Delta=4~{\rm eV}$.}
\label{p1brwii}
\end{figure}

\begin{figure}[tb]
\begin{center}
\begin{tabular}{cc}
\includegraphics[scale=1,width=8cm]{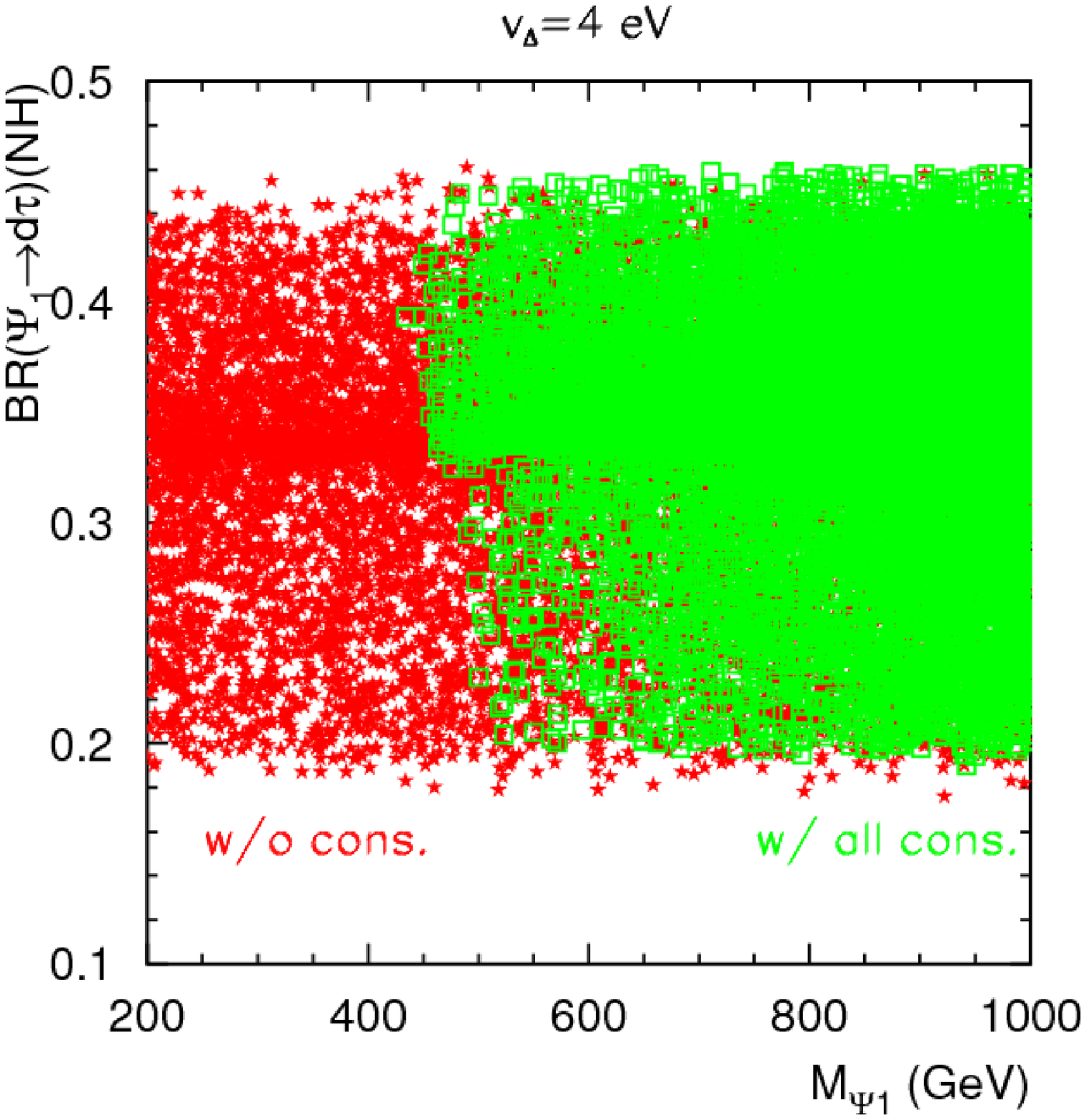}
\includegraphics[scale=1,width=8cm]{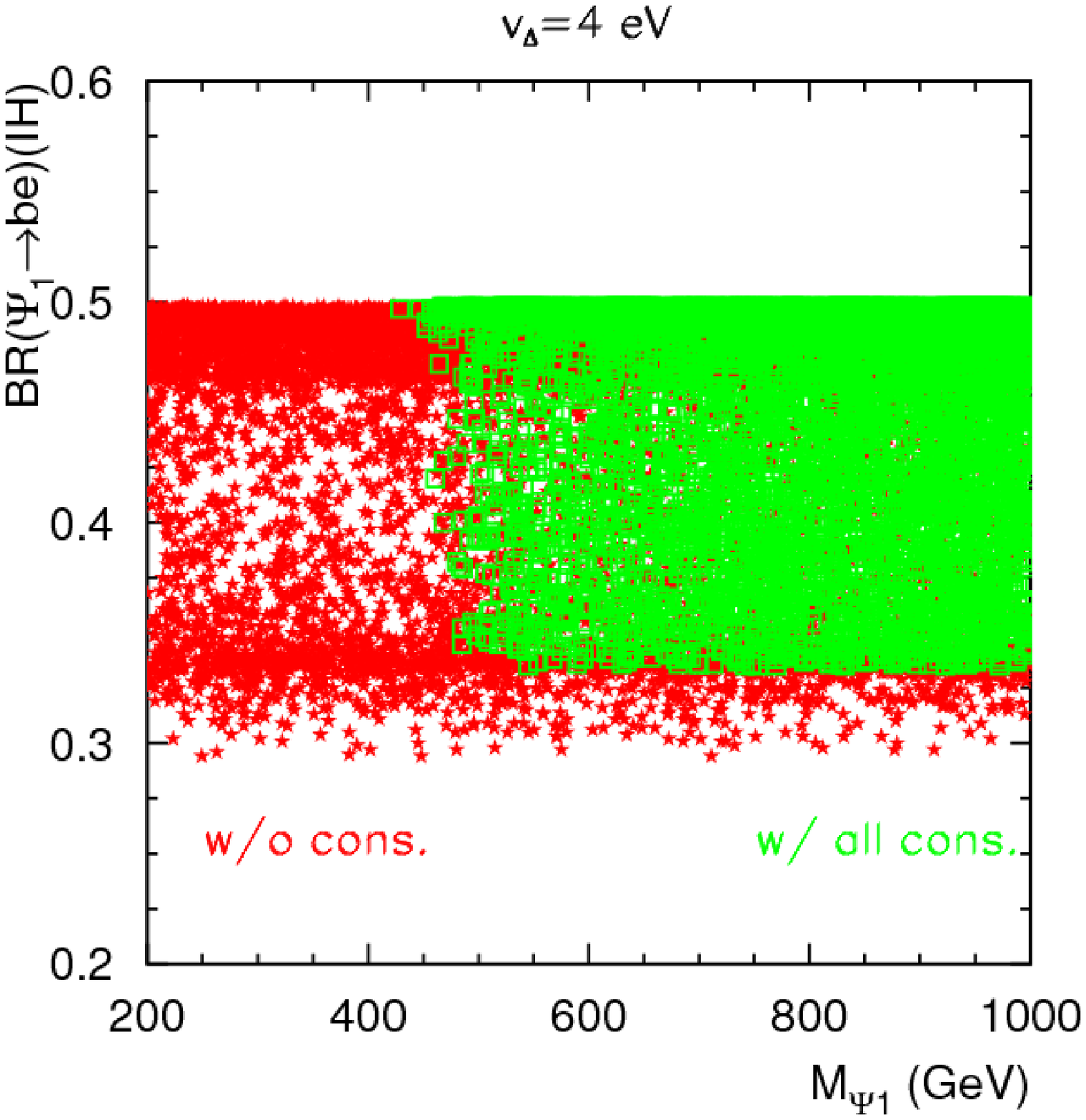}
\end{tabular}
\end{center}
\caption{The branching fraction of leading channel $\Psi_1\to d\tau$ for NH (left) and $\Psi_1\to be$ for IH (right) versus leptoquark mass without any constraints (red solid star), and with both $K_L$ decay and $\mu-e$ conversion constraints 
(green empty square), $v_\Delta=4~{\rm eV}$.}
\label{p1brwij}
\end{figure}

The predictions for the decays of $\Psi_2$ taking into account the
experimental constraints on neutrino mass and mixing angles are shown in Fig.~\ref{p2i}.
Because of the existence of the missing neutrino in the final state, we must sum over the
contributing neutrinos incoherently. As one can see that, in the NH case the
decay $\Psi_2\to d\bar{\nu},s\bar{\nu}$ is the dominant channel, and in the
IH case $\Psi_2 \to b \bar{\nu}$ is the leading one.
In Figs.~\ref{p1brwii} and \ref{p1brwij} we plot the branching fractions of $\Psi_1$ the leading 
decay channels for NH and IH versus $M_{\Psi_1}$ without any constraints (red stars), and
with $K_L$ decay constraints and $\mu-e$ conversion constraints (green squares) 
when $v_\Delta=4~{\rm eV}$. As noted earlier, present experimental results for these studies strongly constrain the LQ parameters, but do not preclude the possibility of scenarios in the LQ is sufficiently light to be discovered at the LHC. In the event of such a discovery, more precise measurements of $B(K_L\to\mu^+\mu^-)$ and more sensitive searches for $\mu-e$ conversion would provide a consistency test for this model. 

\subsection{Impact of Majorana Phases in Leptoquark Decays}
Although the decays of the leptoquark $\Psi_1$ are independent
of the phases in the matrices $K_3$ and $B$, the unknown Majorana phases
could modify the predictions for their decays. In this section we study the
predictions for the LQ decays including the impact of the
Majorana phases. It is important to note that the  $\Psi_2$ decay branching ratios are
independent of all unknown phases, including the phases 
in the $K_3$ and $B$ matrices 
and the Majorana phases, after summing over all final state neutrino 
flavors incoherently.  This feature could allow us
to probe the neutrino mass spectrum as well as the Majorana phases when
combining the information from the $\Psi_1$ decays. To illustrate this point, we consider the limiting cases of $m_1\approx 0$ ($m_3\approx 0$) for the NH (IH), an approximation that we expect to hold when the mass of the lightest neutrino in either case is smaller than $\sim 10^{-2}$ eV. 
For a heavier mass of  $m_{1,3} > 10^{-1}$ eV, 
the situation approaches  the quasi-degenerate, as we discuss below. 
\subsubsection{Normal Hierarchy with one massless neutrino $(m_1\approx 0)$}
\begin{figure}[tb]
\begin{center}
\begin{tabular}{cc}
\includegraphics[scale=1,width=8cm]{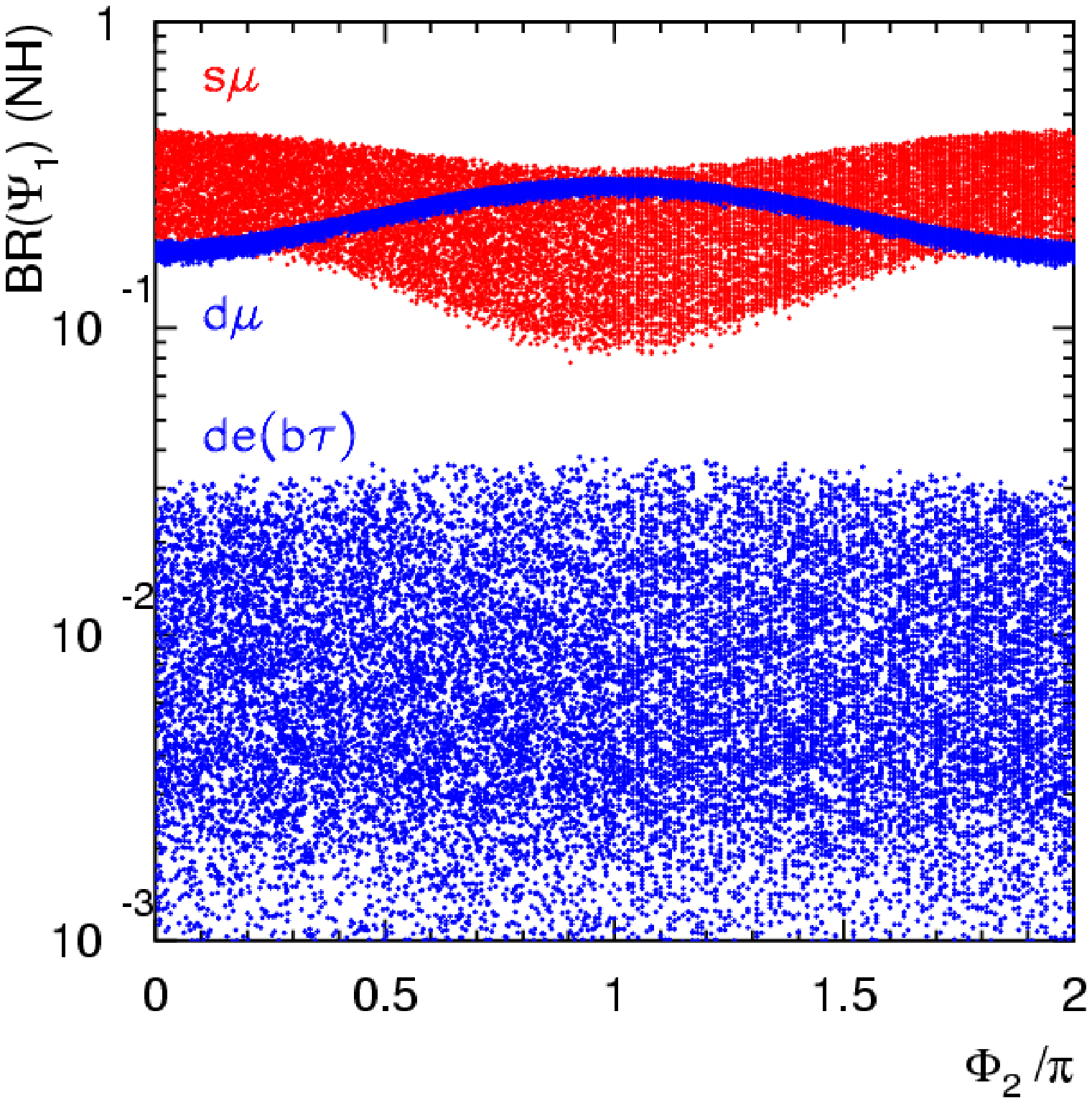}
\includegraphics[scale=1,width=8cm]{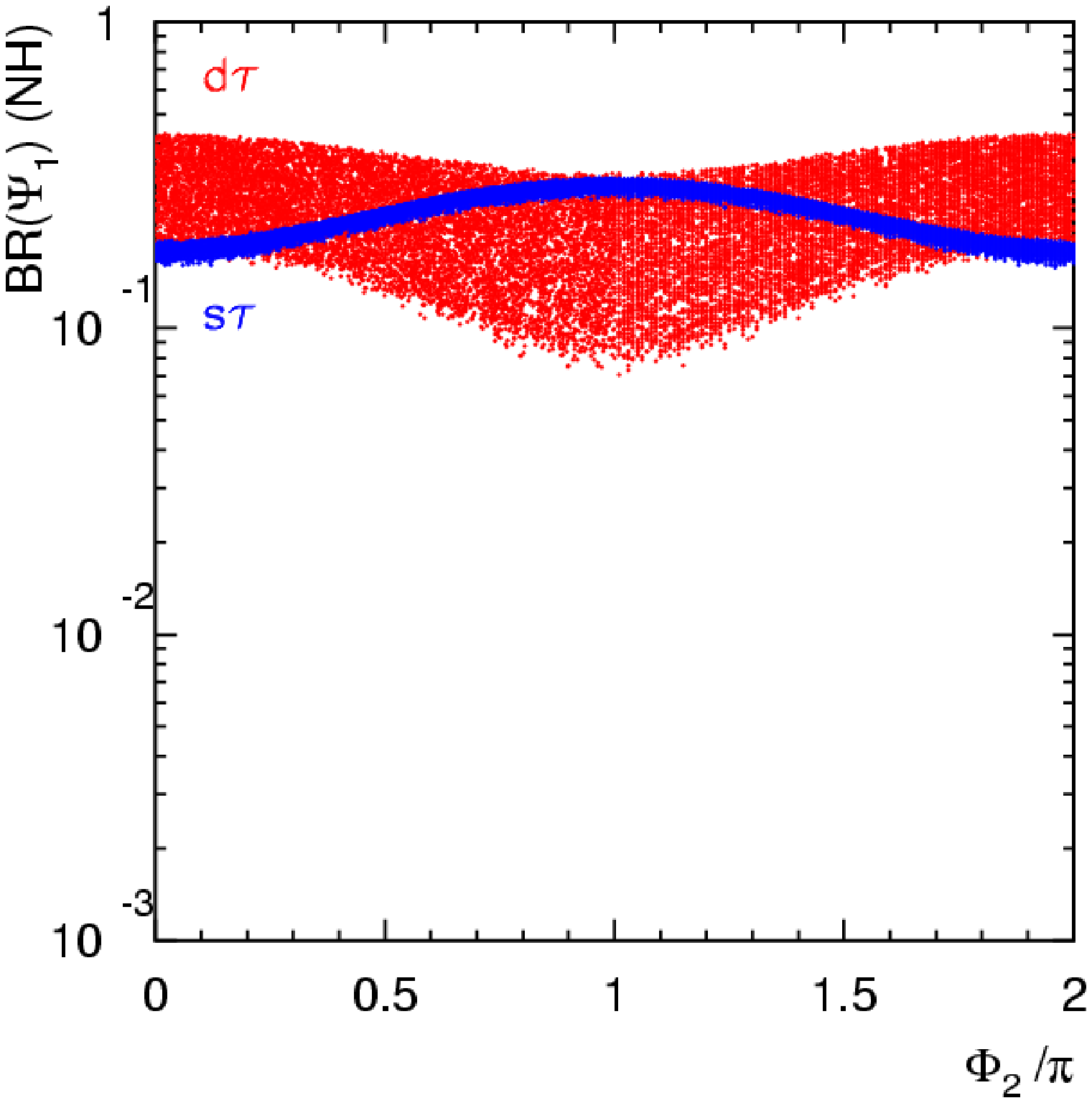}
\end{tabular}
\end{center}
\caption{$\Psi_1$ branching fractions versus the Majorana
phase $\Phi_2$ for the $m_1=0$ scenario.}
\label{phinh}
\end{figure}

The leptoquark decay rates depend on only one Majorana phase $\Phi_2$
when $m_1\approx 0$ in the NH case. Taking $s_{13}=0$ for simplicity,
one finds the expressions for the physical couplings:
\begin{eqnarray}
e^{-i\beta_3}\Gamma_1^{31}&=&{1\over v_\Delta}e^{-2i\alpha_1}\left(\sqrt{\Delta m_{21}^2}s_{12}^2\right),
\\
e^{-i\beta_2}\Gamma_1^{21}&=&e^{-i\beta_3}\Gamma_1^{32}={1\over v_\Delta}e^{-i(\alpha_1+\alpha_2)}\left(\sqrt{\Delta m_{21}^2}s_{12}c_{12}c_{23}\right),
\\
e^{-i\beta_1}\Gamma_1^{11}&=&e^{-i\beta_3}\Gamma_1^{33}={1\over v_\Delta}e^{-i(\alpha_1+\alpha_3)}\left(-\sqrt{\Delta m_{21}^2}s_{12}c_{12}s_{23}\right),
\\
e^{-i\beta_2}\Gamma_1^{22}&=&{1\over v_\Delta}e^{-2i\alpha_2}\left(\sqrt{\Delta m_{21}^2}c_{12}^2c_{23}^2+\sqrt{\Delta m_{31}^2}e^{-i\Phi_2}s_{23}^2\right)\approx \sqrt{{\Delta m_{31}^2\over v_\Delta^2}}e^{-i(2\alpha_2+\Phi_2)}s_{23}^2,
\\
e^{-i\beta_1}\Gamma_1^{12}&=&e^{-i\beta_2}\Gamma_1^{23}={1\over v_\Delta}e^{-i(\alpha_2+\alpha_3)}\left(-\sqrt{\Delta m_{21}^2}c_{12}^2+\sqrt{\Delta m_{31}^2}e^{-i\Phi_2}\right)s_{23}c_{23}\nonumber \\
&\approx& \sqrt{{\Delta m_{31}^2\over v_\Delta^2}}e^{-i(\alpha_2+\alpha_3+\Phi_2)}s_{23}c_{23},
\end{eqnarray}
and
\begin{eqnarray}
&&e^{-i\beta_1}\Gamma_1^{13}={1\over v_\Delta}e^{-2i\alpha_3}\left(\sqrt{\Delta m_{21}^2}c_{12}^2s_{23}^2+\sqrt{\Delta m_{31}^2}e^{-i\Phi_2}c_{23}^2\right)\approx \sqrt{{\Delta m_{31}^2\over v_\Delta^2}}e^{-i(2\alpha_3+\Phi_2)}c_{23}^2.
\end{eqnarray}
The behavior of the branching fractions for the dominant channels is
shown in Fig.~\ref{phinh}. We can see the rather weak dependence
of the decay branching fractions on the phase $\Phi_2$ that can be also
understood from the above analytical expressions.  When the
phase $\Phi_2=\pi$, one obtains the maximal suppression (enhancement)
for the channels $\Psi_1\to s\mu^+$ and $\Psi_1\to d\tau^+$ ($\Psi_1\to d\mu^+, s\tau^+$),
by a factor two at most. 
It is important to note that the braching ratio of the sum over all the 
quark contributions for each lepton flavor, 
$s\mu^+ + d\mu^+$ or $d\tau^+ + s\tau^+$, remains the same.
Thus, from the observational point of view in collider experiments, the signals of
$\mu+$jet and $\tau+$jet will be unchanged, insensitive to the Majorana phase. 

\subsubsection{Inverted Hierarchy with one massless neutrino $(m_3 \approx 0)$}
\begin{figure}[tb]
\begin{center}
\begin{tabular}{cc}
\includegraphics[scale=1,width=8cm]{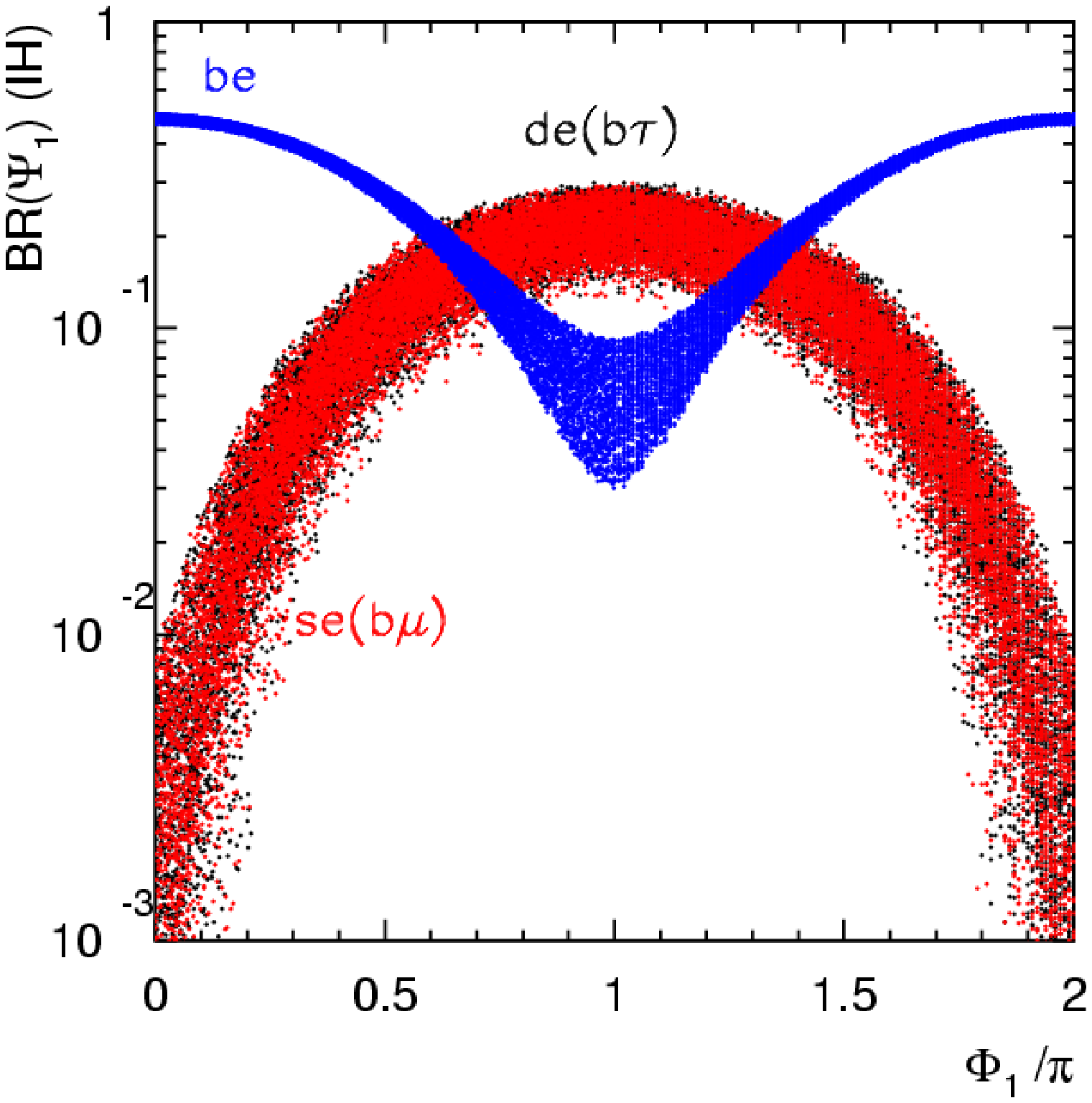}\\
\includegraphics[scale=1,width=8cm]{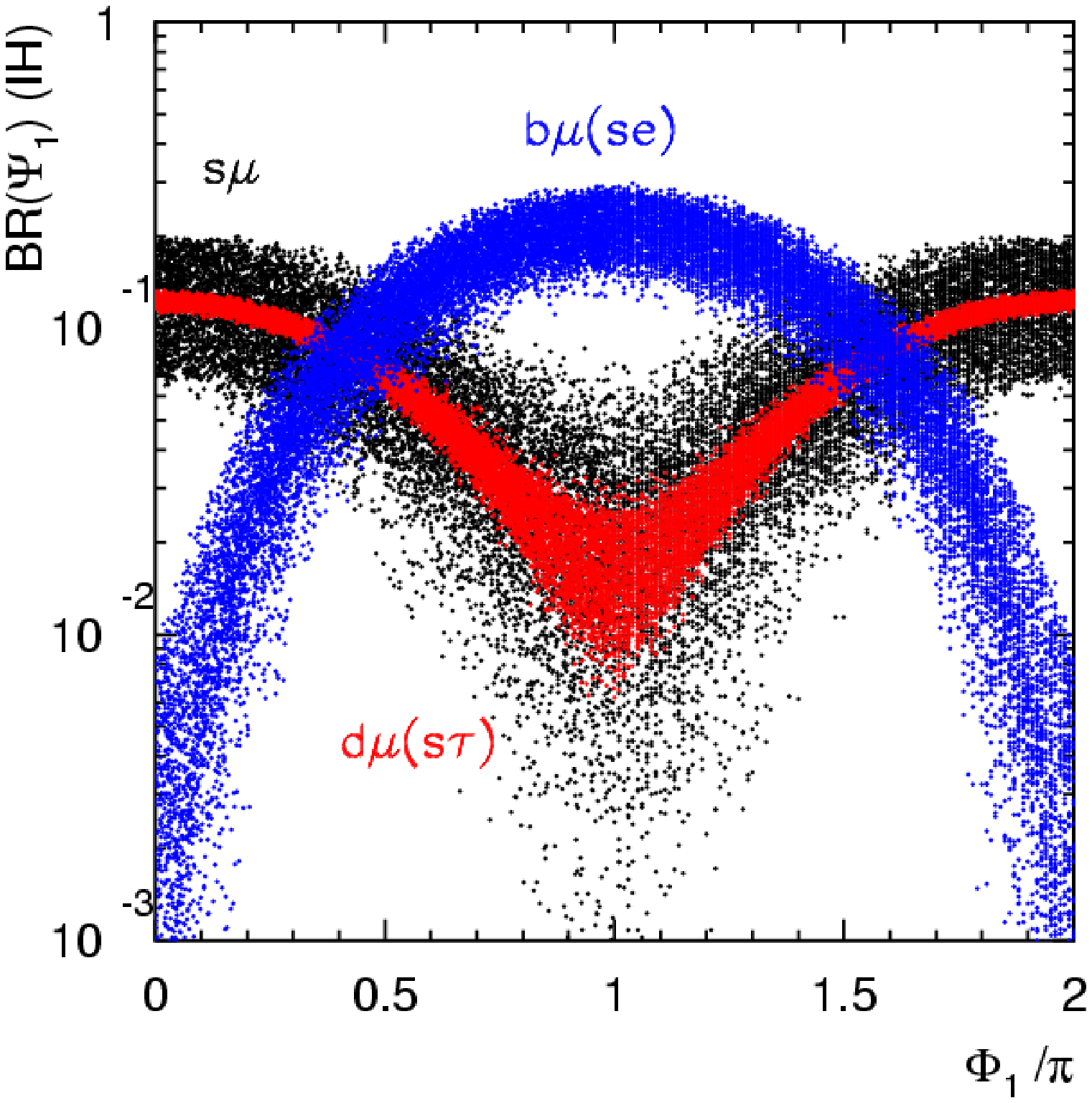}
\includegraphics[scale=1,width=8cm]{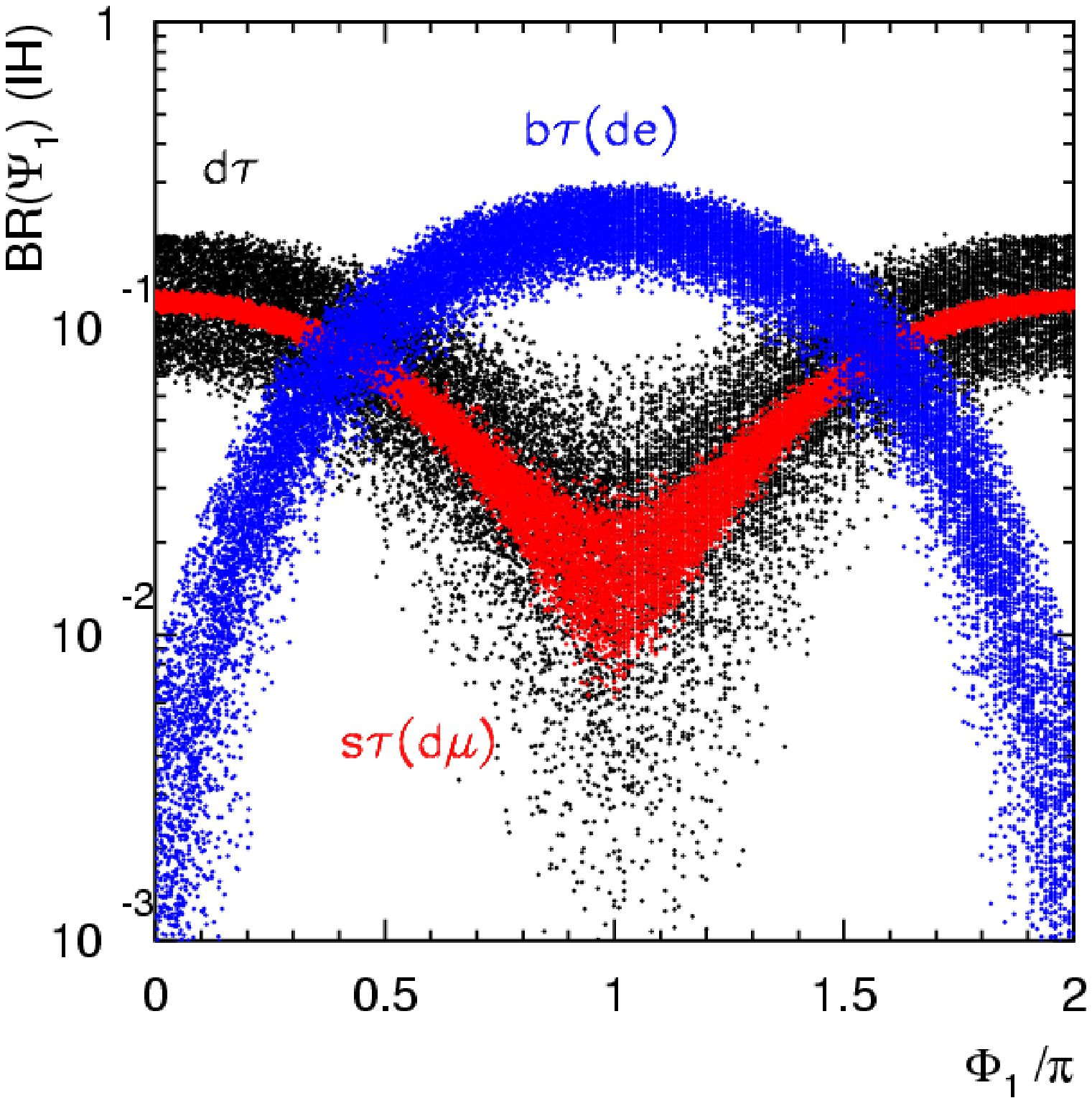}
\end{tabular}
\end{center}
\caption{$\Psi_1$ branching fractions versus the Majorana
phase $\Phi_1$ for the $m_3=0$ scenario and $\Phi_1\in (0,2\pi)$.}
\label{phiih}
\end{figure}

In the case of the Inverted Hierarchy with one massless neutrino
one finds that all relevant decays depend on only one
phase $\Phi_1$. 
The relevant couplings for $s_{13}=0$ are given by
\begin{eqnarray}
e^{-i\beta_3}\Gamma_1^{31}&=&{1\over v_\Delta}e^{-2i\alpha_1}\left(\sqrt{\Delta m_{21}^2+|\Delta m_{31}^2|}s_{12}^2+\sqrt{|\Delta m_{31}^2|}e^{-i\Phi_1}c_{12}^2\right)\nonumber \\
&\approx& \sqrt{{|\Delta m_{31}^2|\over v_\Delta^2}}e^{-2i\alpha_1}(s_{12}^2+e^{-i\Phi_1}c_{12}^2),
\end{eqnarray}
\begin{eqnarray}
e^{-i\beta_2}\Gamma_1^{21}&=&e^{-i\beta_3}\Gamma_1^{32}={1\over v_\Delta}e^{-i(\alpha_1+\alpha_2)}\left(\sqrt{\Delta m_{21}^2+|\Delta m_{31}^2|}-\sqrt{|\Delta m_{31}^2|}e^{-i\Phi_1}\right)c_{12}c_{23}s_{12}\nonumber \\
&\approx& \sqrt{{|\Delta m_{31}^2|\over v_\Delta^2}}e^{-i(\alpha_1+\alpha_2)}(1-e^{-i\Phi_1})c_{12}c_{23}s_{12},
\\
e^{-i\beta_1}\Gamma_1^{11}&=&e^{-i\beta_3}\Gamma_1^{33}={1\over v_\Delta}e^{-i(\alpha_1+\alpha_3)}\left(-\sqrt{\Delta m_{21}^2+|\Delta m_{31}^2|}+\sqrt{|\Delta m_{31}^2|}e^{-i\Phi_1}\right)c_{12}s_{23}s_{12}\nonumber \\
&\approx& \sqrt{{|\Delta m_{31}^2|\over v_\Delta^2}}e^{-i(\alpha_1+\alpha_3)}(-1+e^{-i\Phi_1})c_{12}s_{23}s_{12},
\\
e^{-i\beta_2}\Gamma_1^{22}&=&{1\over v_\Delta}e^{-2i\alpha_2}\left(\sqrt{\Delta m_{21}^2+|\Delta m_{31}^2|}c_{12}^2+\sqrt{|\Delta m_{31}^2|}e^{-i\Phi_1}s_{12}^2\right)c_{23}^2\nonumber \\
&\approx& \sqrt{{|\Delta m_{31}^2|\over v_\Delta^2}}e^{-2i\alpha_2}(c_{12}^2+e^{-i\Phi_1}s_{12}^2)c_{23}^2,
 \\
e^{-i\beta_1}\Gamma_1^{12}&=&e^{-i\beta_2}\Gamma_1^{23}={1\over v_\Delta}e^{-i(\alpha_2+\alpha_3)}\left(-\sqrt{\Delta m_{21}^2+|\Delta m_{31}^2|}c_{12}^2-\sqrt{|\Delta m_{31}^2|}e^{-i\Phi_1}s_{12}^2\right)s_{23}c_{23}\nonumber \\
&\approx& \sqrt{{|\Delta m_{31}^2|\over v_\Delta^2}}e^{-i(\alpha_2+\alpha_3)}(-c_{12}^2-e^{-i\Phi_1}s_{12}^2)s_{23}c_{23},
\end{eqnarray}
and
\begin{eqnarray}
e^{-i\beta_1}\Gamma_1^{13}&=&{1\over v_\Delta}e^{-2i\alpha_3}\left(\sqrt{\Delta m_{21}^2+|\Delta m_{31}^2|}c_{12}^2+\sqrt{|\Delta m_{31}^2|}e^{-i\Phi_1}s_{12}^2\right)s_{23}^2\nonumber \\
&\approx& \sqrt{{|\Delta m_{31}^2|\over v_\Delta^2}}e^{-2i\alpha_3}(c_{12}^2+e^{-i\Phi_1}s_{12}^2)s_{23}^2.
\end{eqnarray}
In Fig.~\ref{phiih} we show the dependence of the branching fractions
on this Majorana phase. The maximal suppression or enhancement takes
place also when $\Phi_1=\pi$. 
In this scenario the dominant channels also 
swap from one to the other. Unlike in the case of the NH where 
$s\mu \leftrightarrow d\mu$ and $s\tau \leftrightarrow d\tau$ at $\Phi_2=\pi$, 
the interchanges of the channels occur for 
$s\mu,d\mu \leftrightarrow b\mu$, and $d\tau,s\tau \leftrightarrow b\tau$, 
as well as  $be \leftrightarrow de, se$,  when $\Phi_1$ varies from $0$ to $\pi$.
Once again, we notice that the branching ratio for the sum over all the quark contributions for each lepton flavor, 
$be^+ + de^+ + se^+$ or $s\mu^+ + d\mu^++b\mu^+$ or $d\tau^+ + s\tau^+ + b\tau^+$
remains the same.
This fact makes the collider search for a leptonic channel rather phase-independent.
However, if we require a $b$-flavor tagging in the event selection along with the
lepton flavor identification, this could provide crucial information on  the
value of the Majorana phase $\Phi_1$. 
Specifically, we see the qualitative features in the IH
\begin{eqnarray}
{\rm BR}(be) \left\{
\begin{array}{lll}
\gg {\rm BR}(b\mu),\ {\rm BR}(b\tau)  &  & {\rm for}\ \ \Phi_1 \approx 0; \\
\ll {\rm BR}(b\mu),\ {\rm BR}(b\tau)  &  & {\rm for}\ \ \Phi_1 \approx \pi .
\end{array}
\right.
\end{eqnarray} 

\subsubsection{Quasi-Degenerate Hierarchy}
For a QD neutrino spectrum, the approximation of a nearly massless lightest neutrino 
no longer applies, and one must use the full expressions for the $\Gamma^{ij}$ that 
are given in Appendix \ref{app:gammas}. As one can readily appreciate from these 
expressions by setting $m_1 \approx  m_2 \approx  m_3$, 
the branching ratios for the $\Psi_1$ will in general depend on 
both $\Phi_1$ and $\Phi_2$. From a numerical scan over the neutrino masses and mixing angles, we find 
that the mild $\Phi_2$-dependences 
of the branching ratios for $\Phi_1=0$ are quite similar to those of the NH, 
and that the strong $\Phi_1$-dependences of the branching ratios 
for $\Phi_2=0$ are quite similar to those of the IH. 
Again, if $b$-flavor tagging is effective along with the lepton flavor identification, 
one could hope to probe crucial information on  the value of the Majorana phase 
$\Phi_1$. The only difference from the IH case is the numerical values for the 
branching ratio of the leading channels. 
\subsection{Total Decay Width of Leptoquarks}
\begin{figure}[tb]
\begin{center}
\begin{tabular}{cc}
\includegraphics[scale=1,angle=90,width=8.5cm]{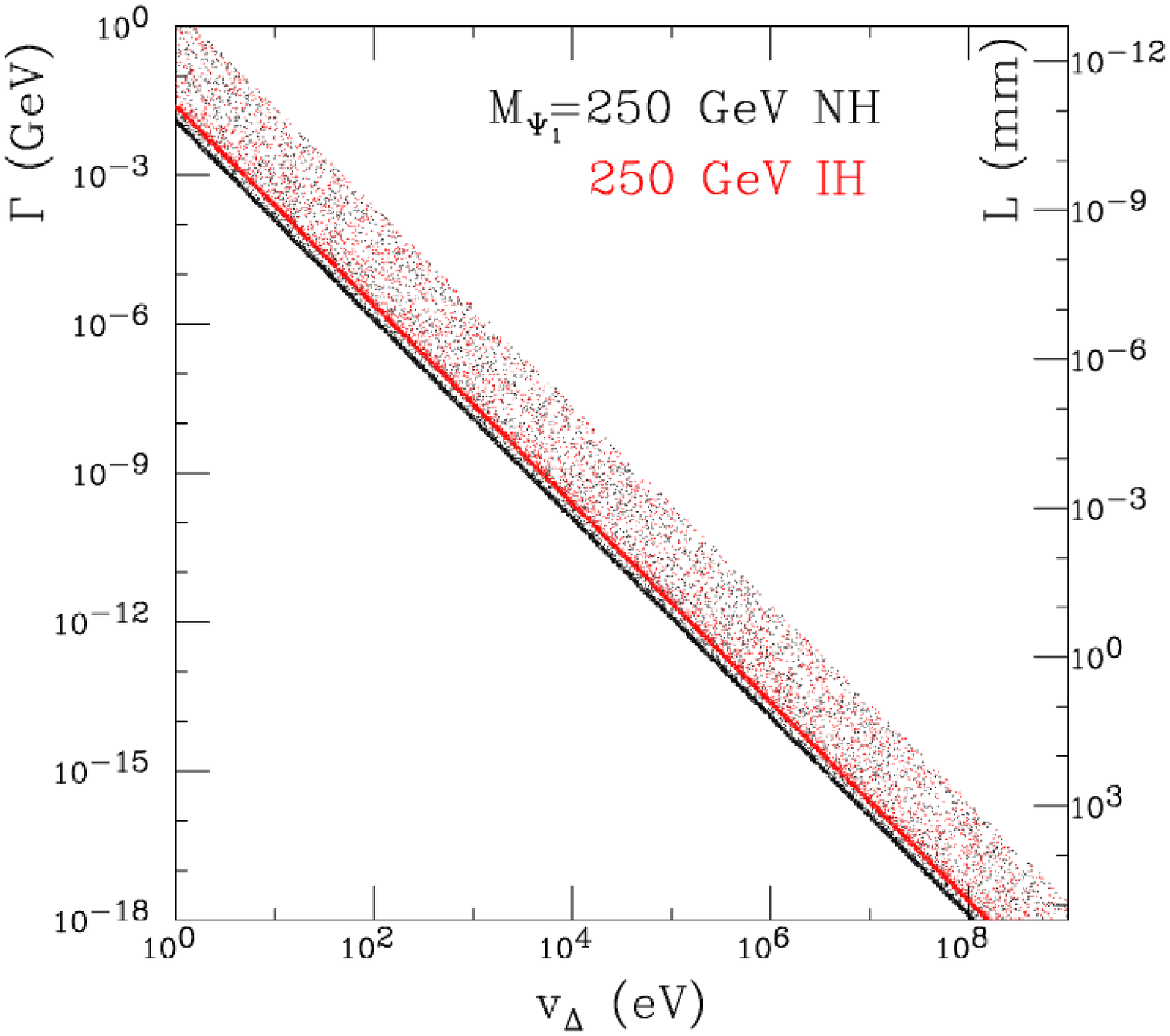}
\end{tabular}
\end{center}
\caption{Total decay width and length of $\Psi_1$
versus $v_{\Delta}$.}
\label{wilp1}
\end{figure}

To complete our study about the LQ properties,
in Fig.~\ref{wilp1} we plot the total width (left axis) and
decay length (right axis) for $\Psi_1$ versus $v_\Delta$
for $M_{\Psi_1}=250$ GeV in NH and IH. The total decay width
is proportional to $M_{\Psi_1}/v_\Delta^2$. At the values
$v_\Delta < 10^4$ eV, its decay is prompt. This is the standard scenario
for collider searches to be discussed in the next section. For larger
values of $v_\Delta$, the leptoquark can be long-lived in the
detector's scale, making the collider signatures rather different. One may need
to search for exotic hadrons or heavy charged tracks. 
Since these very large values of $v_\Delta$ are much higher than the 
light neutrino mass scale 
for our consideration of neutrino mass generation, we will not
pursue this parameter region further.
\section{Search for Leptoquarks  at the LHC}
In this section we study the main production mechanisms of leptoquarks and their experimental
signatures at the LHC. There exists extensive literature on this topic, treating 
both theoretical and phenomenological considerations~\cite{LQthe,nlopair,prod1,prod2}
and as well as experimental searches~\cite{Affolder:2000ny,LQexp}. 
Our approach here goes beyond the existing studies in two respects: first we predict specific correlations between observables involving different final state lepton flavors, and second we delineate the connections between these correlations and the light neutrino mixing angles and possible mass spectra.

\begin{figure}[tb]
\begin{center}
\begin{tabular}{cc}
\includegraphics[scale=1,width=8.5cm]{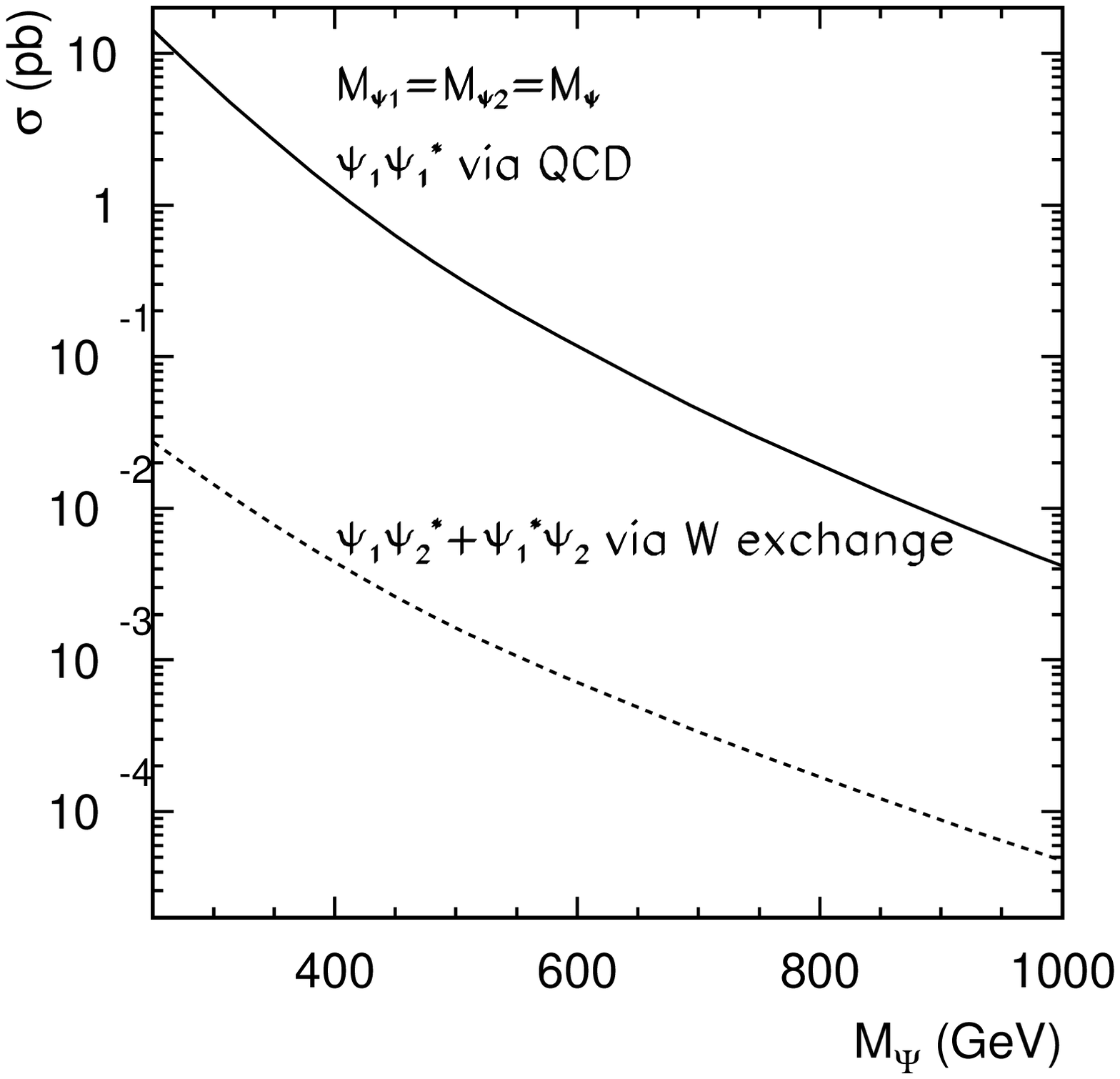}
\end{tabular}
\end{center}
\caption{Leptoquark pair production total cross section at the LHC versus leptoquark mass.
The solid curve is for $pp\to LQ + \overline{LQ}$ via QCD, assuming
$\mu_F=\mu_R=M_\Psi=M_{\Psi_1}$. The dotted curve is for $pp\to \Psi_1 \Psi_2^\ast +
 \Psi_1^* \Psi_2$ via $W^\pm$ exchange, assuming $M_{\Psi_1}=M_{\Psi_2}=M_\Psi$.}
\label{cspair}
\end{figure}

As is well known in the case of the leptoquarks, 
the leading production channel is via the QCD 
interaction
\begin{eqnarray}
q \ + \ \bar{q} \ & \to & \ LQ \ + \ \overline{LQ} \\
g \ + \ {g} \ & \to & \ LQ \ + \ \overline{LQ}.
\end{eqnarray}
The pair production total cross section  versus its mass at the LHC
is plotted in Fig.~\ref{cspair} (the solid curve).
The other unique channel  is the single production via the Yukawa interaction
\begin{equation}
g \ + \ q(\bar{q}) \ \to \ LQ\ (\overline{LQ}) \ + \ \bar \ell\ (\ell).
\label{eq:single}
\end{equation}
This cross section is rather small in our model due to the constraints from
the neutrino masses, as we will discuss later.
More importantly , the nature of the $SU(2)_L$ doublet $\Phi_b$ allows for associated production of $\Psi_1$ and $\Psi_2$
via $W$ exchange, 
\begin{eqnarray}
q(p_1) \ + \ \bar{q}'(p_2) \ & \to & \ \Psi_1(k_1) \ + \ \Psi_2^\ast(k_2).
\end{eqnarray}
In terms of the polar angle variable $y=\hat{p}_1\cdot \hat{k}_1$ in the parton c.m.~frame with energy $\sqrt{s}$, the parton level cross section for the associated production is
\begin{eqnarray}
{d\sigma\over dy}(q\bar{q}'\to \Psi_1\Psi_2^\ast)&=&{\pi\alpha^2\beta^3(1-y^2)\over 32N_c\sin^4\theta_W}{s\over (s-M_W^2)^2},
\end{eqnarray}
where $\beta=\sqrt{(1-(m_i+m_j)^2/s)(1-(m_i-m_j)^2/s)}$ is the speed factor of $\Psi_1$ and $\Psi_2$ in the c.m.~frame.
The total cross section  versus its mass at the LHC is plotted in
Fig.~\ref{cspair} (the dotted curve), assuming $M_{\Psi_1}=M_{\Psi_2}$.
As expected, the cross section for $\Psi_1 \Psi_2^\ast$ production is smaller than the QCD
pair production $\Psi_1 \Psi_1^\ast$ or $\Psi_2 \Psi_2^\ast$ by about three orders of magnitude.

We now turn to the signal observability at the LHC. In Sec.~A and B, we are mainly
concerned with the kinematical features for the signal and backgrounds. We will take
the decay branching fractions of  $\Psi_1,\  \Psi_2$ to be $100\%$ to the corresponding
channels under discussion. In Sec.~C, we will devote ourself to the determination for
the branching fractions.
\subsection{Pair Production of $\Psi_1$}
The QCD corrections to leptoquark pair production at the LHC
have been studied~\cite{nlopair} and a NLO K-factor of
order $1.5 - 1.9$ for leptoquark mass range
from $200~{\rm GeV}$ to $1500~{\rm GeV}$ is predicted.
We apply the K-factor $1.5$ to the processes in our numerical analysis.
Assuming that $\Psi_1$ is the lighter leptoquark, its decay modes are
\begin{eqnarray}
\Psi_1\to d_i e^+_j \ \ \ (d_i=d,s,b;\ \  e_j=e,\mu,\tau).
\end{eqnarray}
We now explore the signal observability according to the different lepton flavors.
\subsubsection{$\Psi_1\Psi_1^\ast\to \ell^+\ell^-jj \ \ \ (\ell=e,\mu)$}
\begin{figure}[tb]
\begin{center}
\begin{tabular}{cc}
\includegraphics[scale=1,width=8.5cm]{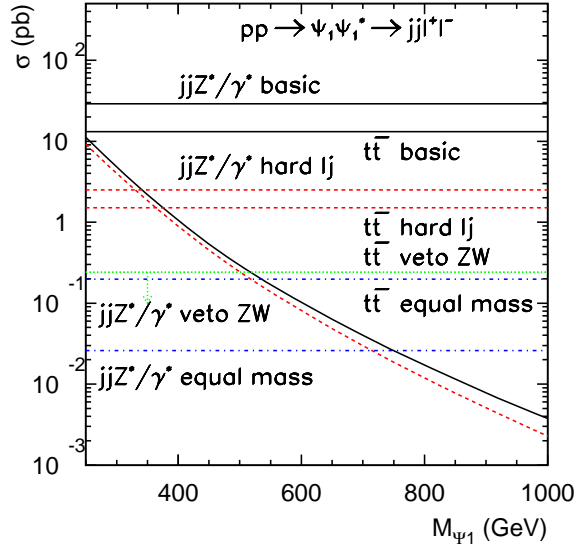}
\end{tabular}
\end{center}
\caption{Production cross section of $\Psi_1\Psi_1^\ast$ with basic cuts and hard final states cut. Branching fractions for leptoquark decay are not included in this plot. For comparison, the background processes are also included with the sequential cuts as indicated.}
\label{cswbkg}
\end{figure}

We start from the cleanest channels with $e,\mu$ in the final state from
$\Psi_1$ decay. The signal consists of one pair of opposite-sign
leptons of arbitrary $e, \mu$ flavor combinations plus two jets of $d,s,b$ quarks.
We employ the following basic acceptance cuts for the event selection~\cite{cms}
\begin{eqnarray}
&&p_T(\ell)\geq15~{\rm GeV}, \ |\eta(\ell)|<2.5,\\
&&p_T(j)\geq25~{\rm GeV}, \ |\eta(j)|<3.0, \\
&&\Delta R_{jj},\ \Delta R_{j\ell},\ \Delta R_{\ell\ell}\geq 1.0
\end{eqnarray}
where we require a large $\Delta R$ between the jet and lepton to resolve them because both $\Psi_1$s are produced nearly at rest implying that their two-body decay 
products are back-to-back.
To simulate the detector effects on the energy-momentum measurements,
we smear the electromagnetic energy, the electromagnetic energy and jet energy
by a Gaussian distribution whose width is parameterized as~\cite{cms}
\begin{eqnarray}
{ \Delta E\over E} &=& {a_{cal} \over \sqrt{E/{\rm GeV}} } \oplus b_{cal}, \quad
a_{cal}=10\%,\  b_{cal}=0.7\% ,
\label{ecal}\\
{ \Delta E\over E} &=& {a_{had} \over \sqrt{E/{\rm GeV}} } \oplus b_{had}, \quad
a_{had}=50\%,\  b_{had}=3\%.
\end{eqnarray}
The leading SM backgrounds to this channel are
\begin{eqnarray}
Z^\ast/\gamma^\ast jj\to \ell^+\ell^- jj,\qquad t\bar{t}\to \ell^+\ell^-jj+\cancel{E}_T.
\label{pairbkg}
\end{eqnarray}
Although the background rates are very large to begin with, the signal and background kinematics are  quite different. We outline the characteristics
and propose some judicious cuts as follows.
\begin{itemize}
\item For a few hundred ${\rm GeV}$ leptoquark decay, the leptons and
jets in final states are very energetic. We tighten up the kinematical cuts
\begin{eqnarray}
p_T^{\rm max}(\ell)>M_{\Psi_1}/4,\qquad p_T^{\rm max}(j)>M_{\Psi_1}/4.
\end{eqnarray}
\item To remove the $jjZ^\ast$ background, we veto the lepton pairs
with opposite charges in the $Z$-mass window $|M_{\ell^+\ell^-}-M_Z|>15~{\rm GeV}$.
This is a standard cut to remove the on-shell $Z$ contribution.
\item To remove the $t\bar t$ background, we veto the events with large missing
energy from $W$ decay: $\cancel{E}_T<25~{\rm GeV}$.
\item In order to select the correct lepton and jet combination and reconstruct
the leptoquark, we take advantage of the feature that the two leptoquarks have equal
masses $M_{\ell_1j_1}=M_{\ell_2j_2}$. In practice, we take
$|M_{\ell_1j_1}-M_{\ell_2j_2}|<M_{\Psi_1}/10$.
This helps for the background reduction, in particular for $jj \gamma^*/Z^*$.
\end{itemize}
The production cross section for the $\Psi_1\Psi_1^\ast$ signal with the basic cuts (solid curve)
and all of the cuts above (dotted curve) are plotted in Fig.~\ref{cswbkg}.  For comparison,
the background processes of $jj\gamma^\ast/Z^\ast$ and $t\bar{t}$
are also included with the sequential cuts as indicated. Incidentally, the two background
curves after the $Z/W$ veto cuts coincide with each other.
The backgrounds are suppressed substantially.
There are also other sub-leading backgrounds, such as $WZ,\ ZZ$ and  $W^+W^-jj$.
Their production rates are much smaller compared to those mentioned above. For instance,
including the decay branching fractions, we have~\cite{smwzzz}
\begin{eqnarray}
\nonumber
&& \sigma(WZ)\sim 17.25~{\rm pb}\times {2\over 3}\times 6.72\%=0.77~{\rm pb},\\
\nonumber
&& \sigma(ZZ)\sim 6.7~{\rm pb}\times 69.9\%\times 6.72\%=0.3~{\rm pb},\\
\nonumber
&& \sigma(W^+W^-jj) \sim 0.2 \ {\rm pb\quad with\ basic\ cuts.}
\end{eqnarray}
Therefore, these backgrounds become negligible after our selective cuts as discussed
above.

\begin{figure}[tb]
\begin{center}
\begin{tabular}{cc}
\includegraphics[scale=1,width=8cm]{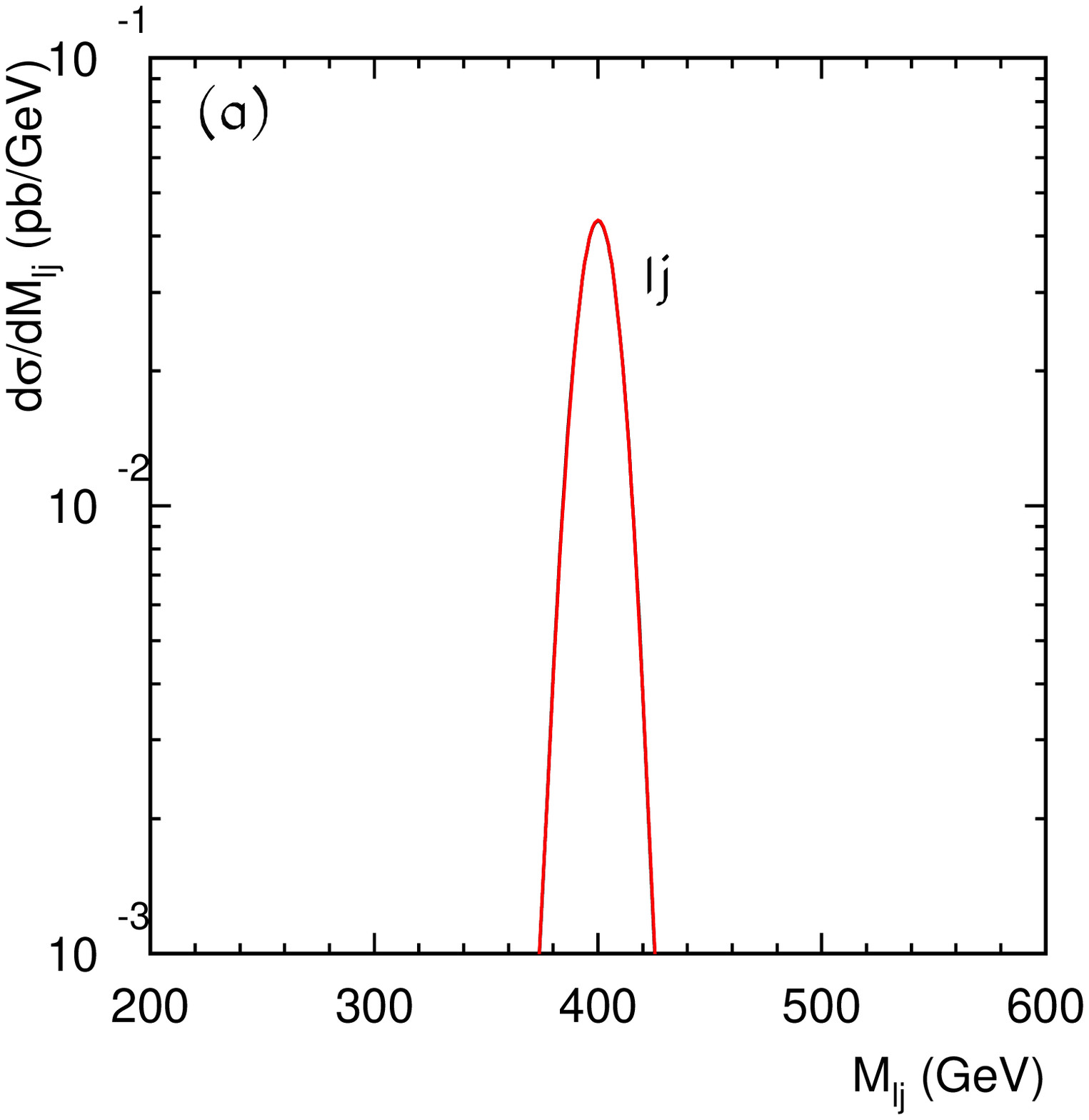}
\includegraphics[scale=1,width=8cm]{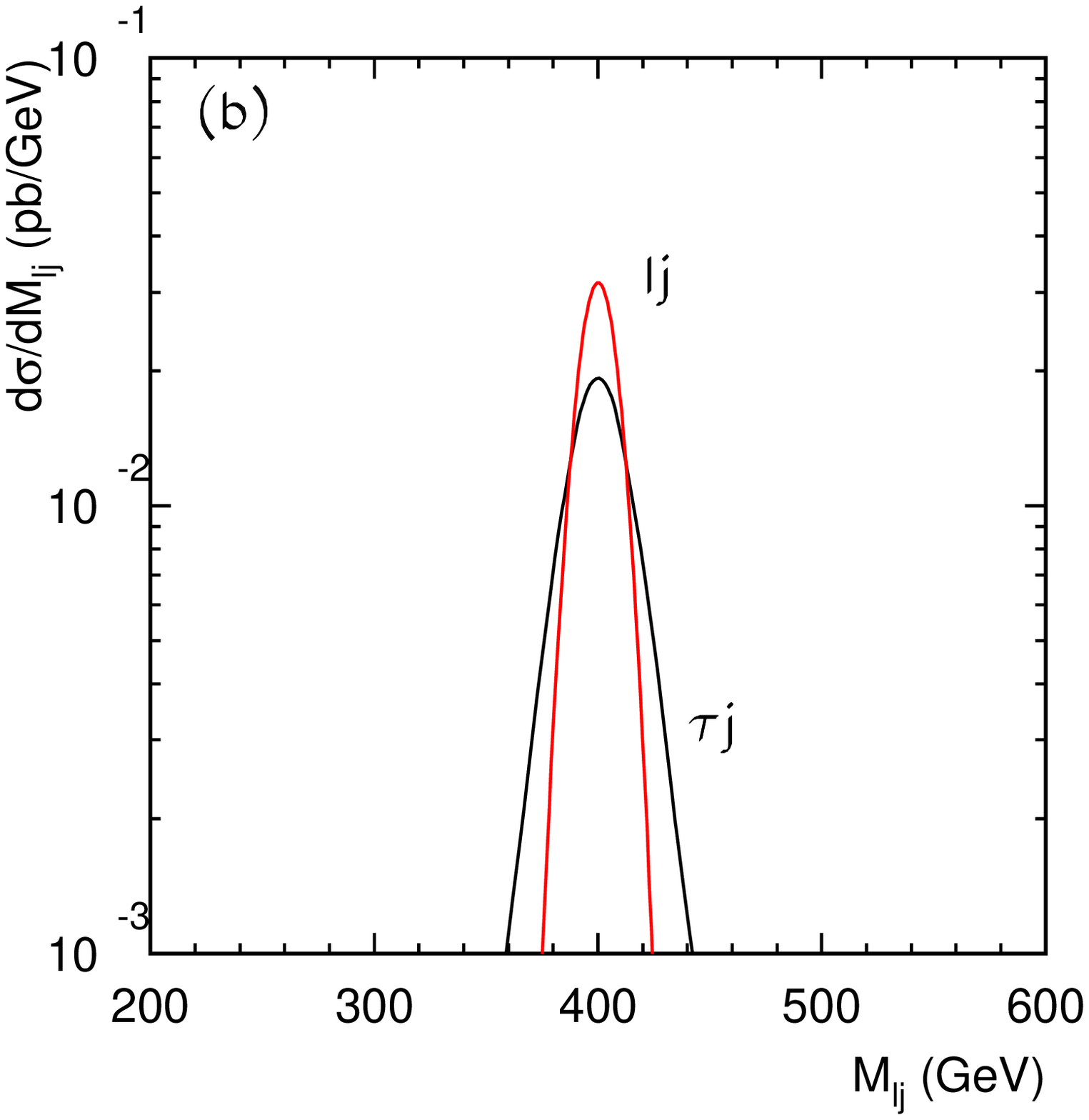}
\end{tabular}
\end{center}
\caption{Reconstructed invariant mass of $M(\ell j)$ for (a) $\ell^+\ell^- jj$ production
and (b) $\ell^\pm \tau^\mp jj$ production, with a leptoquark mass of $400~{\rm GeV}$.}
\label{inmass}
\end{figure}

Finally, when we perform a signal significance analysis, we look
for a resonance in the mass distribution of the $\ell j$ pair. 
The invariant mass of $\ell+j$ is plotted in Fig.~\ref{inmass}(a) 
for $400~{\rm GeV}$ leptoquark production.
The width of the distribution $\sim\pm20$ GeV
is governed by the detector resolution in our simulation.
If we look at a mass window of $|M_{\ell_1j_1,\ell_2j_2}-M_{\Psi_1}|<M_{\Psi_1}/20$,
the backgrounds will be at a negligible level.

We would like to comment on the $b\bar{b}\ell^+\ell^-$ signal 
because the $\Psi_1\to e^+b$ channel plays an important role in distinguishing 
different neutrino mass spectra. The b-tagging rate is 
about $50\%$~\cite{cms}. The signal is still sizable with two b jets tagged.
\subsubsection{$\Psi_1\Psi_1^\ast\to \tau^+\ell^-jj,\quad \tau^+\tau^-jj$}
Together with the $e$ and $\mu$ produced in the $\Psi_1$ decay, 
the $\tau$ lepton final state from leptoquark decay can play an important
role in distinguishing different neutrino mass patterns. Its identification
and reconstruction are different from $e,\mu$ final states because a $\tau$ decays
promptly and there will always be missing neutrinos in $\tau$ decay products.

In order to reconstruct the events with $\tau$s we note that all
the $\tau$s produced in the decay of  few hundred GeV LQ are highly energetic. 
The missing momentum will be along the direction of the charged track. 
We thus assume the momentum of  the missing
neutrinos to be reconstructed by
\begin{eqnarray}
\overrightarrow{p}({\rm invisible}) =\kappa\overrightarrow{p}({\rm track}).
\end{eqnarray}
Identifying $\overrightarrow{p_T}({\rm invisible})$ with the measured $\cancel{E}_T$,
we thus obtain the $\tau$ momentum by
\begin{eqnarray}
\nonumber
\overrightarrow{p}_T(\tau)=\overrightarrow{p}_T(\ell) +\overrightarrow{\cancel{E}}_T,\quad
p_L^{}(\tau)=p_L^{}(\ell) + { \cancel{E}_T \over p_T^{}(\ell) } p_L^{}(\ell).
\end{eqnarray}
The leptoquark pair kinematics are, thus, fully reconstructed. The reconstructed
invariant masses of $M(\ell j)$  and $M(\tau j)$ are plotted in Fig.~\ref{inmass}(b).
We see that $M(\tau j)$ distribution is slightly broader as anticipated. The rather
narrow mass peak of the $\ell j$ system nevertheless serves as the most distinctive
kinematical feature for the signal identification.

For $\tau\tau jj$ events with two $\tau$'s, we generalize the momenta reconstruction to
\begin{eqnarray}
\overrightarrow{p}({\rm invisible})=\kappa_1\overrightarrow{p}({\rm track}_1)+\kappa_2\overrightarrow{p}({\rm track}_2).
\end{eqnarray}
The proportionality constants $\kappa_1,\kappa_2$ can be determined
from the missing energy measurement as long as the two charge tracks are
linearly independent.

\begin{figure}[tb]
\begin{center}
\begin{tabular}{cc}
\includegraphics[scale=1,width=8.5cm]{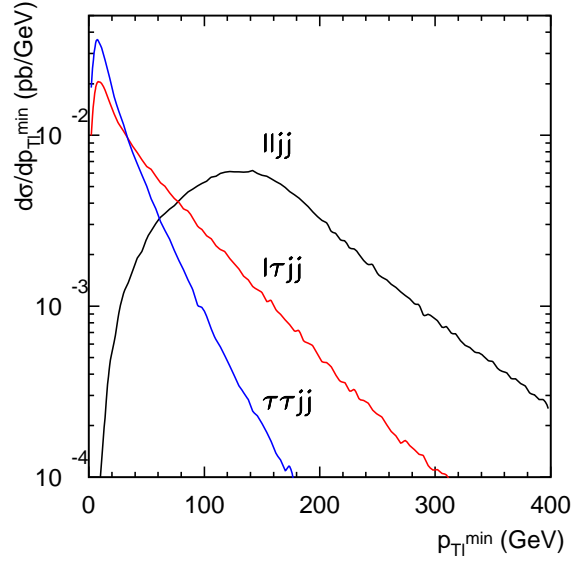}
\end{tabular}
\end{center}
\caption{$p_T$ distribution of the softer lepton from the  leptoquark  and
$\tau$  decays in the events of $\ell\ell jj$,  $\ell\tau jj$ and  $\tau\tau jj$,
for a leptoquark mass $400~{\rm GeV}$.}
\label{ptl}
\end{figure}

In practice when we wish to identify the events with $\tau$'s, we require a minimal
missing transverse energy
\begin{equation}
\cancel{E}_T > 20\ {\rm GeV}.
\end{equation}
This will effectively separate them from the $\ell\ell jj$ events.
Another important difference between the leptons from the primary leptoquark
decay and from the $\tau$ decay is that latter is much softer. In Fig.~\ref{ptl}
we show the $p_T$ distribution of the softer lepton
from the leptoquark and
$\tau$  decays in the events of $\ell\ell jj$,  $\ell\tau jj$ and  $\tau\tau jj$,
\subsection{Associated Production of $\Psi_1\Psi_2^\ast(\Psi_1^\ast\Psi_2)$}
The total cross section of $\Psi_1\Psi_2^\ast+\Psi_1^\ast\Psi_2$ is shown in Fig.~\ref{cspair}
by the dotted curve. As discussed earlier, we focus on the fermionic decay $\Psi_2\to d_i \nu$. Correspondingly,
we only consider the decays $\Psi_1\to d_i e$ and $d_i \mu$ in order to effectively reconstruct
the final state kinematics. The signal is thus
\begin{equation}
\Psi_1\Psi_2^\ast,\ \Psi_1^\ast\Psi_2 \to \ell^\pm j, \ \ \nu j \quad  (\ell=e,\mu),
\end{equation}
consisting of one charged lepton and two jets plus missing energy. Once again, we consider
the corresponding decay branching fractions to be $100\%$, and will leave the branching
fraction determination to the next section.

\begin{figure}[tb]
\begin{center}
\begin{tabular}{cc}
\includegraphics[scale=1,width=8.5cm]{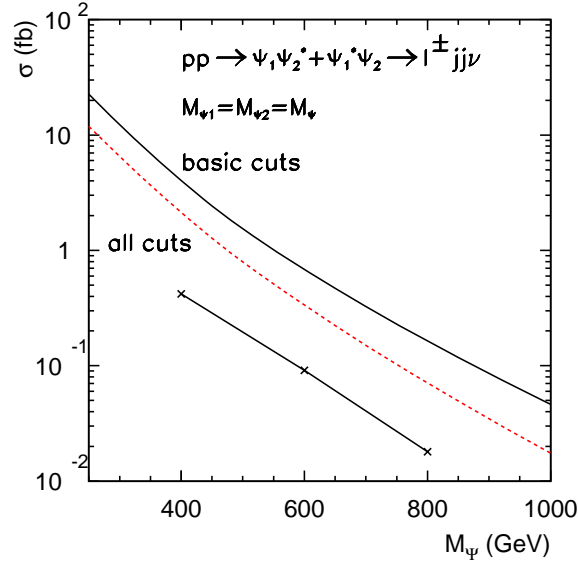}
\end{tabular}
\end{center}
\caption{Total cross section for $\Psi_1\Psi_2^\ast+\Psi_1^\ast\Psi_2$
production after the basic cuts (solid curve) and all cuts
(dotted curves) and the leading $Wjj$ background after all cuts (''$\times$'' marks).}
\label{cs12wbkg}
\end{figure}

We employ the following basic acceptance cuts for the event selection
\begin{eqnarray}
&&p_T(\ell)\geq15~{\rm GeV}, \ |\eta(\ell)|<2.5, \ \cancel{E}_T>40~{\rm GeV}\\
&&p_T(j)\geq25~{\rm GeV}, \ |\eta(j)|<3.0, \\
&&\Delta R_{jj},\ \Delta R_{j\ell}\geq 0.4.
\end{eqnarray}
To simulate the detector effects on the energy-momentum measurements,
we use the same smearing parameters as in the previous section.
The irreducible SM backgrounds to this channel are
\begin{eqnarray}
W^\pm jj,\ \  W^+ W^-,\ \ W^\pm Z \to \ell^\pm jj+\cancel{E}_T.
\end{eqnarray}
The $Wjj$ background is by far the largest. To further optimize the signal observability,
\begin{itemize}
\item we set additional cuts for hard lepton and jet
\begin{eqnarray}
p_T^{\rm max}(\ell)>M_\Psi/4,\quad p_T^{\rm max}(j)>M_\Psi/4.
\end{eqnarray}
\item The SM background events of $\ell^\pm \cancel{E}_T$ typically have
the origin from $W^\pm$. We thus veto the $W$ boson with
the help of a transverse mass cut $M_T(\ell\nu)>150~{\rm GeV}$.
After this cut, the $WW,WZ$ backgrounds are at a negligible level.
\item For the associated production of heavy particles like the
two leptoquarks of several hundred ${\rm GeV}$, the cluster mass
of the system indicates the large threshold. We define
\begin{eqnarray}
M_{cluster}=\sqrt{M_{2j}^2+(\sum\vec{p}_T^j)^2}+\sqrt{(\sum\vec{p}_T^\ell)^2}+\cancel{E}_T
\end{eqnarray}
and will impose a high mass cut to select the signal events $M_{cluster}>2M_\Psi$.
\item The mass reconstruction for $\ell j$ and $\nu j$ can be
very indicative. In order to select the correct
lepton and jet combination to reconstruct $\Psi_1$ we define a transverse mass $M_T$
by one jet and the missing transverse energy
\begin{eqnarray}
M_T(j\nu)= \sqrt{(E_T(j)+\cancel{E}_T)^2-(\vec{p}_T(j)+\vec{\cancel{p}}_T)^2}.
\end{eqnarray}
This variable has an upper bound $M_\Psi$, and is typically smaller than
$M_{\ell j}$. We thus accept the momentum combination if
$M(\ell j_1)>M_T(j_2\nu),\ M(\ell j_2)<M_T(j_1\nu)$,  or
$M(\ell j_2)>M_T(j_1\nu),\ M(\ell j_1)<M_T(j_2\nu)$.
However, there are two other possibilities that both transverse masses of the two
combinations are smaller or larger than their invariant masses. In these cases,
we select the combination with smaller $|M(\ell j)-M_T(j\nu)|$ as the right one.
In Fig.~\ref{re} we show the reconstructed invariant
mass and transverse mass distributions
for a leptoquark mass $400~{\rm GeV}$ for illustration.
\begin{figure}[tb]
\begin{center}
\begin{tabular}{cc}
\includegraphics[scale=1,width=8.5cm]{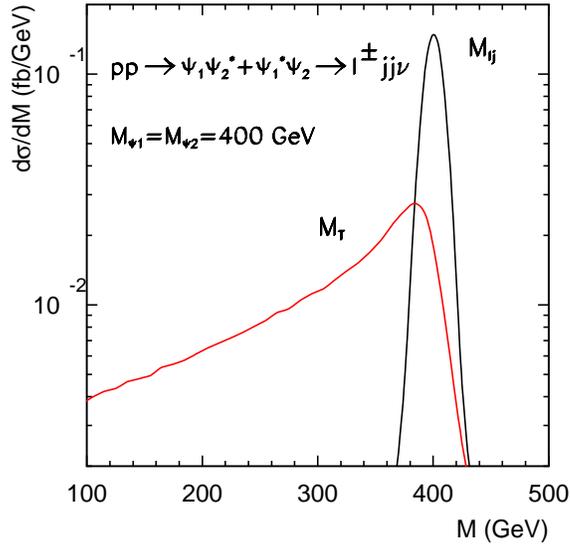}
\end{tabular}
\end{center}
\caption{Reconstructed invariant mass of $M(\ell^\pm j)$ and transverse mass $M_T(j\nu)$ for a leptoquark mass $400~{\rm GeV}$.}
\label{re}
\end{figure}
After selecting correct resonance, we impose transverse mass and
invariant mass cuts for leptoquarks
\begin{eqnarray}
M_T(j\nu)<M_\Psi, \ \ \ |M(\ell j)-M_\Psi|<20.
\end{eqnarray}
\end{itemize}
We list the signal and background cross sections with the consecutive cuts in Table.~\ref{TabIV},
for the case of $M_\Psi=400~{\rm GeV}$.
We see that  the $W^\pm jj$ background is at the $O(0.1~{\rm fb})$ level.
The production cross section of $\Psi_1\Psi_2^\ast + \Psi_1^\ast\Psi_2$
versus its mass with basic cuts and
all cuts are plotted in Fig.~\ref{cs12wbkg}. We see that the cuts are highly
efficient for the signal and the background processes of $Wjj$ is
under control up to $M_{\Psi}\sim 1$ TeV.

We would now like to consider the other potentially large backgrounds, namely,
$t\bar{t}\to bb\ell^+\ell^-+\cancel{E}_T$ and leptoquark pair production via QCD
$\Psi_1\Psi_1^\ast\to jj\ell^+\ell^-$. Demanding one isolated lepton with the basic
cuts and vetoing the extra leptons in the range
\begin{eqnarray}
p_T(\ell)>10~{\rm GeV}, \ \ \ |\eta(\ell)|<3.0,
\end{eqnarray}
the $t\bar t$ background is reduced by more than two orders of magnitude and
the $\Psi_1\Psi_1^\ast$ background by about  five orders of magnitude.
The lepton vetoing is considered part of the basic cuts.
With the additional $M_T$ and $M(\ell j)$ cuts, the backgrounds would be under control,
as listed in Table.~\ref{TabIV}.
\begin{table}[tb]
\begin{tabular}{| c || c | c |c | c| c|c|}
    \hline
     $\sigma$(fb) & Basic & $p_T^\ell,p_T^j$ cut & $M_W$ veto & $M_{\rm Cluster}$ & $M_T$ & $M_{\ell j}$  \\
cuts & Cuts & $>M_\Psi/4,M_\Psi/4$ & $M_T(\ell\nu)>150~{\rm GeV}$  & $>2M_\Psi$ & $<M_\Psi$ & $M_\Psi\pm 20~{\rm GeV}$  \\
     \hline
$\Psi_1\Psi_2^\ast+\Psi_1^\ast\Psi_2$ & & & & & & \\
$M_\Psi=400~{\rm GeV}$ & 3.8 & 3.4 & 2.9 & 2.7 & 2.4 & 2.1\\
\hline
$jjW^\pm$ & & & & & & \\
$M_\Psi=400~{\rm GeV}$ & $4.8\times 10^5$ & $2.3\times 10^4$ & 9.8 & 9.5 & 8.2 & 0.42\\
$M_\Psi=600~{\rm GeV}$ & $4.8\times 10^5$ & 7332 & 6.7 & 5.1 & 4.7 & 0.091\\
$M_\Psi=800~{\rm GeV}$ & $4.8\times 10^5$ & 2945 & 4.2 & 2.0 & 1.9 & 0.018\\
\hline
\hline
$t\bar{t}$ & & & & & & \\
$M_\Psi=400~{\rm GeV}$ & 186 & 11 & 7.5 & 0.6 & 0.58 & $-$\\
\hline
$\Psi_1\Psi_1^\ast$ & & & & & & \\
$M_\Psi=400~{\rm GeV}$ & $3.2\times 10^{-2}$ & $2.4\times 10^{-2}$ & $7.2\times 10^{-3}$ & $7.2\times 10^{-3}$ & $7.1\times 10^{-3}$ & $5.4\times 10^{-3}$
\\
\hline
\end{tabular}
\caption{$\Psi_1\Psi_2^\ast+\Psi_1^\ast\Psi_2$ signal and irreducible background $jjW^\pm$ and reducible backgrounds $t\bar{t}$ and $\Psi_1\Psi_1^\ast$ for $400~{\rm GeV}$ leptoquark mass. }
\label{TabIV}
\end{table}
\subsection{Measuring Branching Fractions and Probing the Neutrino Mass Patterns}
So far, we have only studied the characteristic features of the signal and backgrounds
for the leading channels and have taken the decay branching fractions of the  leptoquarks
to be $100\%$. As we presented earlier,
the $\Psi_1,\ \Psi_2$ decay branching fractions and the  light neutrino mass
matrix are directly correlated. Measuring the BR's of different flavor
combinations becomes crucial in understanding the neutrino mass pattern and thus
the mass generation mechanism.
In contrast, it has been customarily assumed~\cite{Affolder:2000ny,LQexp} 
that the leptoquark is flavor diagonal in each fermion generation.
We list the leading reconstructible leptonic channels along with the predicted
branching fractions in Table.~\ref{TabV}, where the light quarks ($d,s$) can be identified 
as  jets, and the $b$ quark may be flavor-tagged. 
We also associate these channels with predictions of the neutrino mass patterns.
We reiterate some of the key observations in connecting the LHC signals of the leptoquarks
to the neutrino parameters.

\begin{itemize}
\item $\Psi_2$ decays are independent of any phases. They are thus 
robust to test the mass pattern.
\item $\Psi_1$ decays to a lepton plus inclusive jets 
are also independent of any phases.  They are  robust to test the mass pattern as well.
\item These channels are not sensitive to Majorana phase $\Phi_2$.
\item The sensitivity to $\Phi_1$ can be significant if $b$-flavor tagging can be exploited 
in the cases of the IH and QD.
\end{itemize}

\begin{figure}[tb]
\begin{center}
\begin{tabular}{cc}
\includegraphics[scale=1,width=8cm]{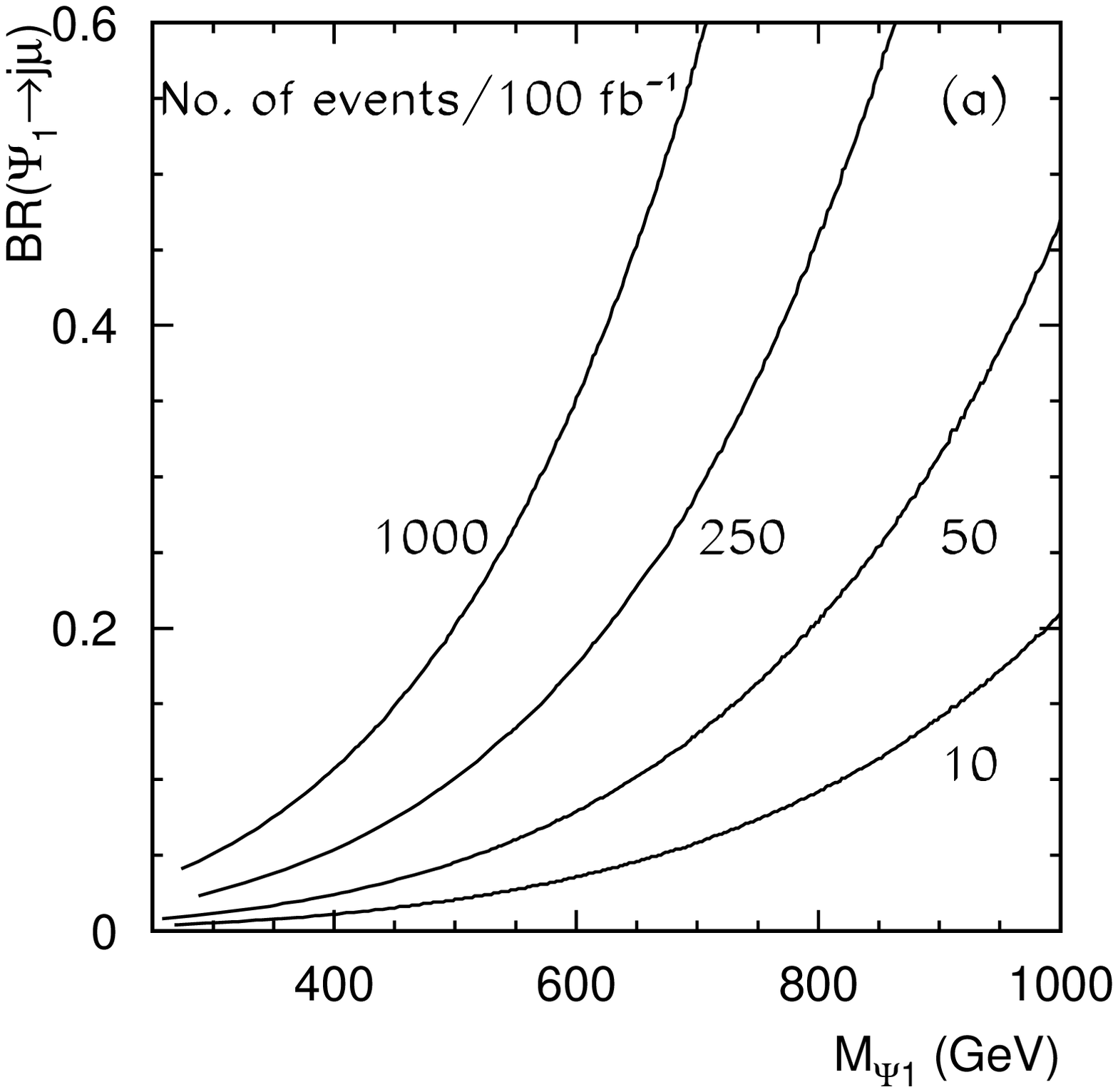}
\includegraphics[scale=1,width=8cm]{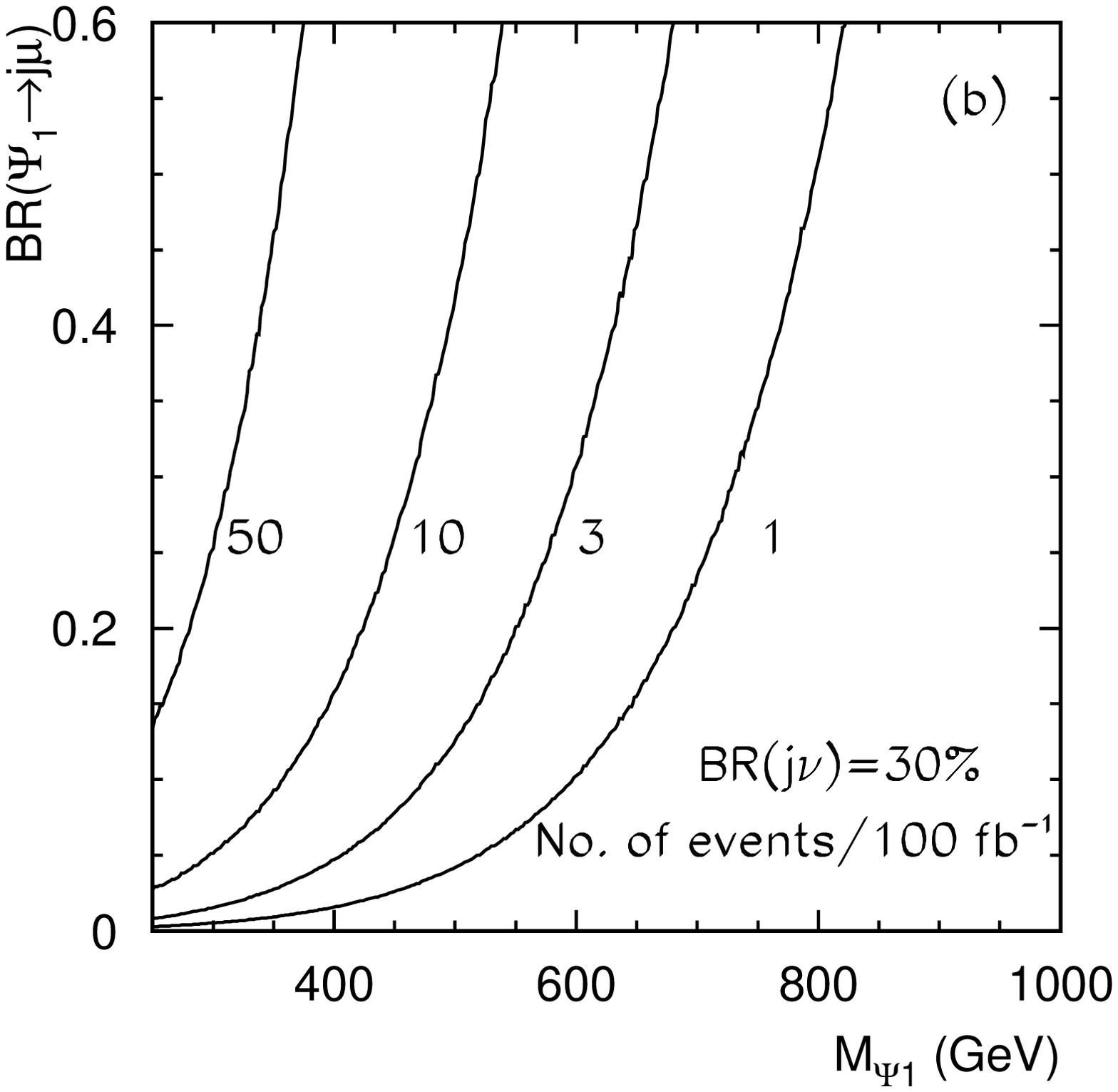}
\end{tabular}
\end{center}
\caption{Event contours in the BR$-M_{\Psi_1}$ plane at the LHC with an integrated luminosity $100~{\rm fb}^{-1}$ for (a) $\Psi_1 \Psi_1^*\to \mu^+ j\ \mu^-j $,
and (b)  $\Psi_1 \Psi_2^*+\Psi_1^* \Psi_2\to \mu^+ j\ j \nu$,
including all the judicious cuts describe in the early sections.}
\label{eve}
\end{figure}

\begin{center}
\begin{table}[tb]
\begin{tabular}[t]{|c|c|c|c|}
\hline
Signal channels & Leading modes and BR & Leading modes and BR & Leading modes and BR
\\
& Normal Hierarchy & Inverted Hierarchy & Quasi degenerate \\
\hline \hline
$\Psi_1\Psi_1^\ast$ &  & & \\
$\Phi_1=\Phi_2=0$ & & $e^+be^-\bar{b}: \ \ \ (40-50\%)^2$ &  \\
\hline
& $\mu^+ j\approx \tau^+ j:$ &  $\mu^+ j\approx \tau^+ j:$ &  $e^+b \approx \mu^+s\approx \tau^+ d:$ \\
 & any $\ell$ pairs: \ $(40-60\%)^2$ & any $\ell$ pairs: \ $(20-30\%)^2$& any $\ell$ pairs:\ $(30\%)^2$\\
    &  & $\mu^+j\ e^-\bar{b} \approx \tau^+j\ e^-\bar{b}:$ &  \\
        &  & $ (20-30\%)\times(40-50\%)$ &  \\
         \hline 
$\Phi_1\approx \pi,\Phi_2=0$ & independent of $\Phi_1$ & 
$ e b \leftrightarrow e j$,\ $\mu j \leftrightarrow \mu b$,  $\tau j \leftrightarrow \tau b$ & 
$ e b \leftrightarrow e j$,\ $\mu s \leftrightarrow \mu b$,  $\tau d \leftrightarrow \tau b$  \\
&  & 
$ e b + e j$,\ $\mu j + \mu b$,  $\tau j + \tau b$ & 
$ e b + e j$,\ $\mu j + \mu b$,  $\tau j + \tau b$  \\
&  & 
 unchanged & 
 unchanged \\
 \hline
$\Phi_1=0,\Phi_2\approx \pi$ & 
$s\mu \leftrightarrow d\mu$,\ \  $d\tau \leftrightarrow s\tau$ 
& independent of $\Phi_2$  &  
$s\mu \leftrightarrow d\mu$,\ \  $d\tau \leftrightarrow s\tau$ \\
& 
$\mu j,\ \tau j$ unchanged 
 & & $eb,\ \mu j,\ \tau j$ unchanged  \\
\hline \hline
$\Psi_1\Psi_2^\ast(\Psi_1^\ast\Psi_2)$ & & & \\
$\Phi_1=\Phi_2=0$ & $\mu^+ j\ j\bar{\nu}:$ 
& $e^+bb\bar{\nu}: \ \ (40-50\%)\times 50\%$ & $\mu^+s j\bar{\nu},e^+b j\bar{\nu} \ \ \ 2(30\%)^2$\\
&  $ (40-60\%)\times 100\%$ &  $\mu^+j b\bar{\nu}: \ \ (20-30\%)\times 50\%$ 
& $\mu^+s b\bar{\nu},e^+b b\bar{\nu} \ \ \ (30\%)^2$\\
 \hline
$\Phi_1\approx \pi,\Phi_2=0$ & independent of $\Phi_1$ &
 $eb\leftrightarrow ej,\ \mu j \leftrightarrow \mu b $ 
&  $eb\leftrightarrow ej,\ \mu s \leftrightarrow \mu b $  \\
 \hline
$\Phi_1=0,\Phi_2\approx \pi$ & $\mu^\pm j\ j\bar{\nu}$ unchanged  
 & independent of $\Phi_2$  & $eb,\ \mu s $ unchanged  \\ 
\hline
\end{tabular}
\caption{Leading fully reconstructible channels and the indicative ranges
of their branching fractions.
The light neutrino mass patterns of the NH and IH with the
lightest neutrino mass $m_0\lesssim 10^{-2}~{\rm eV}$ are shown in the first two
columns, including the effects of vanishing and large Majorana phases.
The last column shows the Quasi degenerate neutrino mass pattern when
$m_0 > 0.1~{\rm eV}$. A jet of a light quark $(d,s)$ is denoted by $j$.}
\label{TabV}
\end{table}
\end{center}
For illustration, consider first the cleanest
channel, $\Psi_1\Psi_1^\ast\to \mu^+j \ \mu^- j\ (j=d,s)$. The number of events is written as
\begin{eqnarray}
N=L\times \sigma(pp\to \Psi_1\Psi_1^\ast)\times {\rm BR}^2(\Psi_1\to \mu^+j),\label{event}
\end{eqnarray}
where $L$ is the integrated luminosity. Given a sufficient number
of events $N$, the mass of leptoquark is determined by the invariant
mass of lepton and jet $M_{\ell j}$. We thus predict the corresponding
production rate $\sigma(pp\to \Psi_1\Psi_1^\ast)$ for this given mass.
The only unknown in the Eq.~(\ref{event}) is the decay branching fraction.
We present the event contours in the BR$-M_{\Psi_1}$ plane in Fig.~\ref{eve}(a)
for $100~{\rm fb}^{-1}$ luminosity including all the judicious cuts described earlier, with which the backgrounds are insignificant.
We see that the LHC has tremendous sensitivity to probe the channel
$\Psi_1\to \mu +$jet  within and beyond the predicted branching fraction 
($35\%-60\%$ in NH, $20\%-30\%$ in IH, and $30\%$ in QD) 
up to and  beyond $M_{\Psi_1}\sim 1$ TeV.

Since the $\Psi_2$ decay is independent of the unknown Majorana phases, it is thus
important to search for the $\Psi_1 \Psi_2^\ast(\Psi_1^\ast\Psi_2)$ signal. In Fig.~\ref{eve}(b),
we show the event contours in the BR$-M_{\Psi_1}$ plane, for 
$100~{\rm fb}^{-1}$ luminosity including all the judicious cuts described earlier.
We see that with the estimated branching fraction for $\nu+$jet, one can reach the
coverage of about $M_{\Psi_1}\sim 0.6$ TeV or higher.
\subsection{Single Production of $\Psi_1$}
The leptoquark pair production is
well predicted by its SU$(3)_c$ and SU$(2)_L$ gauge interactions, and
their decays studied in the previous sections are governed by the neutrino
oscillation parameters, and all are model-independent.
The drawback for this nice feature, however, is that it does not
provide the model-dependent details, such as the Yukawa couplings or equivalently
the vev $v^{}_\Delta$. In order to further test the theory, one needs to involve the
Yukawa couplings in the production process. This is the single production of a leptoquark,
as in Eq.~(\ref{eq:single}).

\begin{figure}[tb]
\begin{center}
\begin{tabular}{cc}
\includegraphics[scale=1,width=8cm]{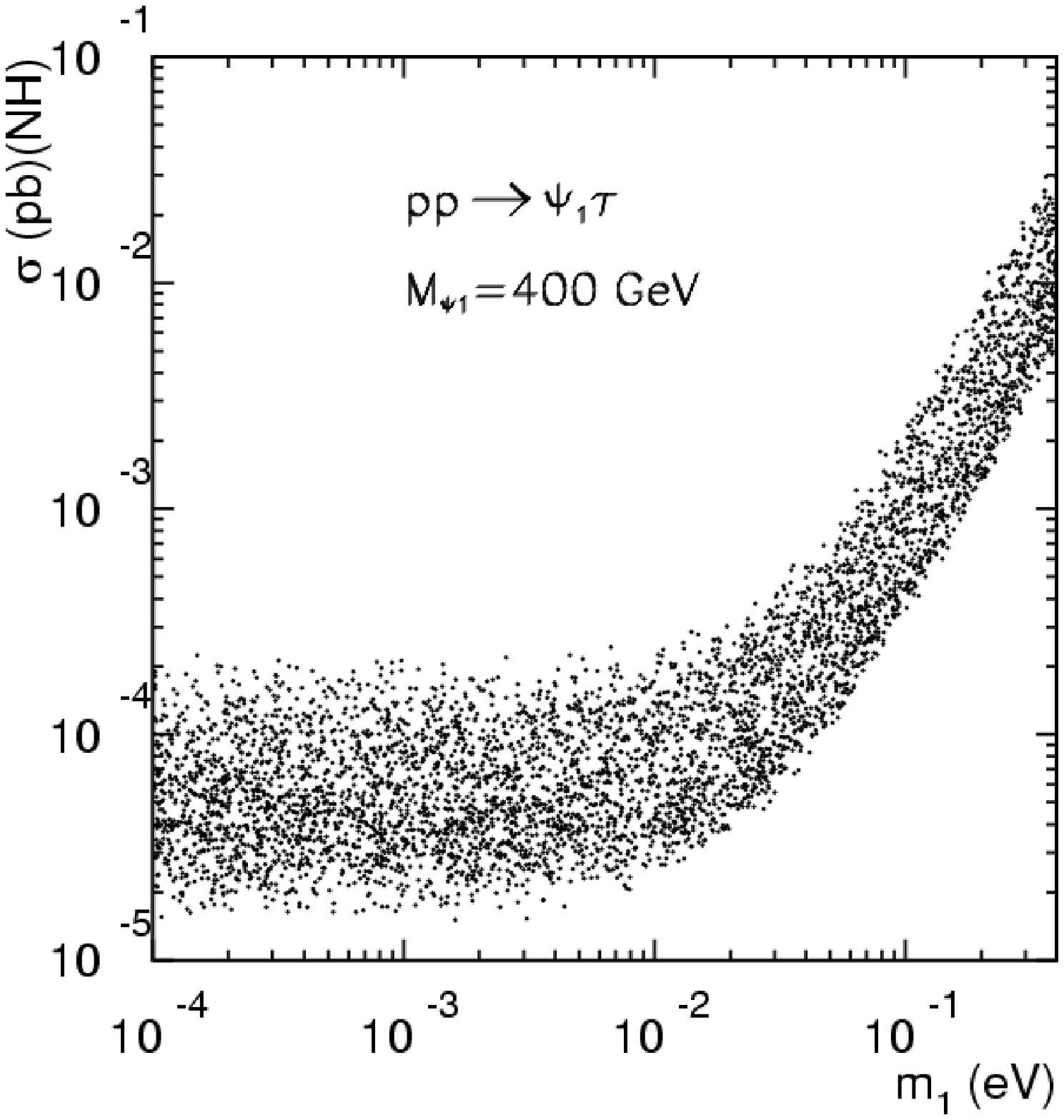}
\includegraphics[scale=1,width=8cm]{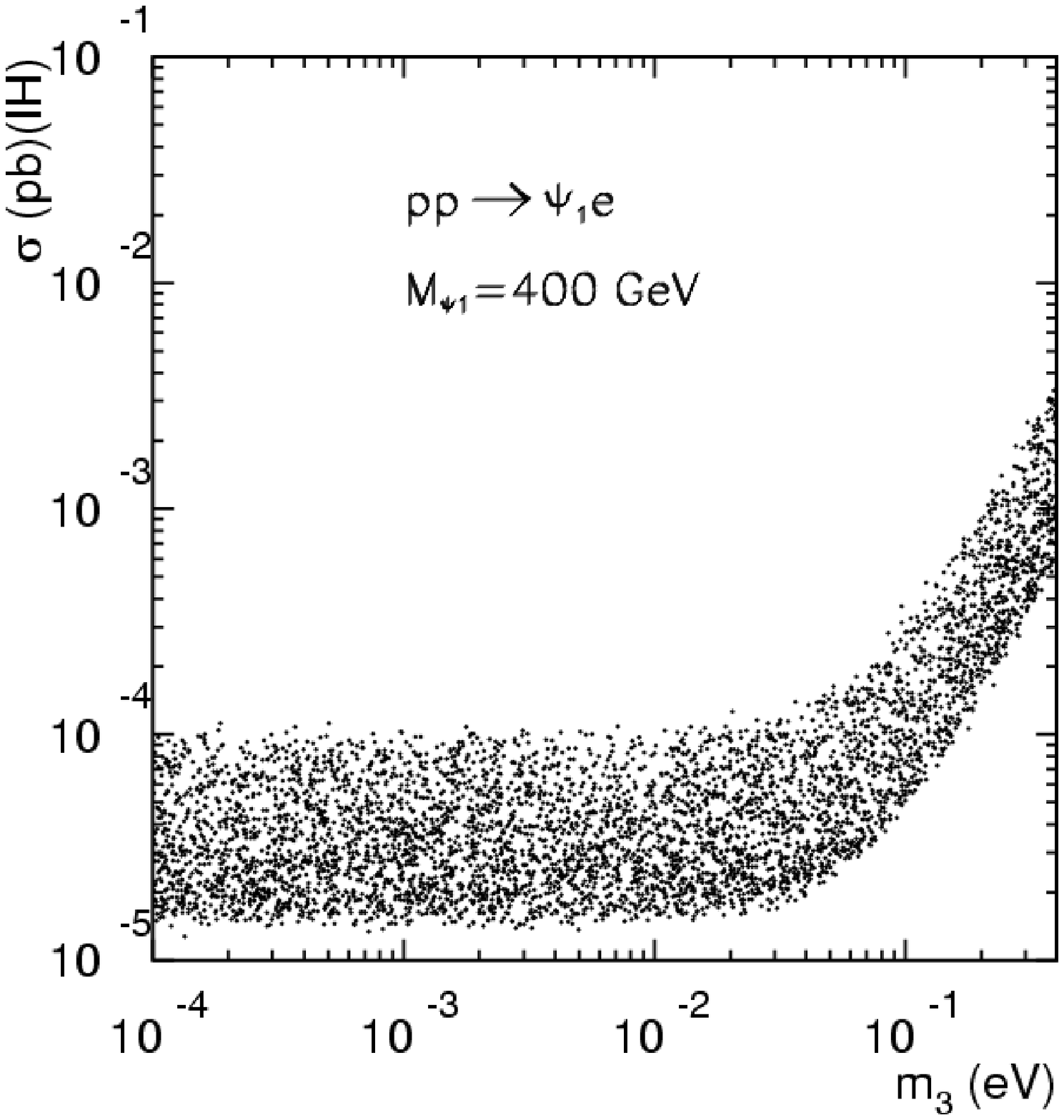}
\end{tabular}
\end{center}
\caption{Total cross section for single $\Psi_1$ production with $4~{\rm eV}\leq v_\Delta\leq 10~{\rm eV}$.}
\label{cssin}
\end{figure}

The parton level cross section for this process is
\begin{eqnarray}
{d\sigma\over d\hat{t}}(gd_i(\bar{d}_i)\to \Psi_1(\Psi_1^\ast)\ell^-_j(\ell^+_j))&=&{|\Gamma_1^{ij}|^2\alpha_s\over 48\hat{s}^2}\left[{\hat{s}+\hat{t}-M_{\Psi_1}^2\over \hat{s}}+{\hat{t}(\hat{t}+M_{\Psi_1}^2)\over (\hat{t}-M_{\Psi_1}^2)^2}+{\hat{t}(2M_{\Psi_1}^2-\hat{s})\over \hat{s}(\hat{t}-M_{\Psi_1}^2)}\right]
\end{eqnarray}
where $\hat{t}=(p_{d_i}-p_{\ell_j})^2$.
Taking into account the neutrino oscillation constraints, in Fig.~\ref{cssin} we show the scatter
plots for the total cross section of single $\Psi_1$ production versus the lightest neutrino mass,
for the NH case (left) and the IH case (right).
The total cross sections for both cases are  less than $1~{\rm fb}$, although it can become
larger when reaching the quasi degenerate case for $m_\nu > 0.1$ eV.
To estimate the backgrounds, we start from the cleanest channels with $e,\mu$ in the final state
from $\Psi_1$ decay. The signal consists of one pair of opposite-sign
same-flavor leptons and one light jet. We employ the same basic
acceptance cuts and smearing parameters as the pair production.

\begin{figure}[tb]
\begin{center}
\begin{tabular}{cc}
\includegraphics[scale=1,width=8cm]{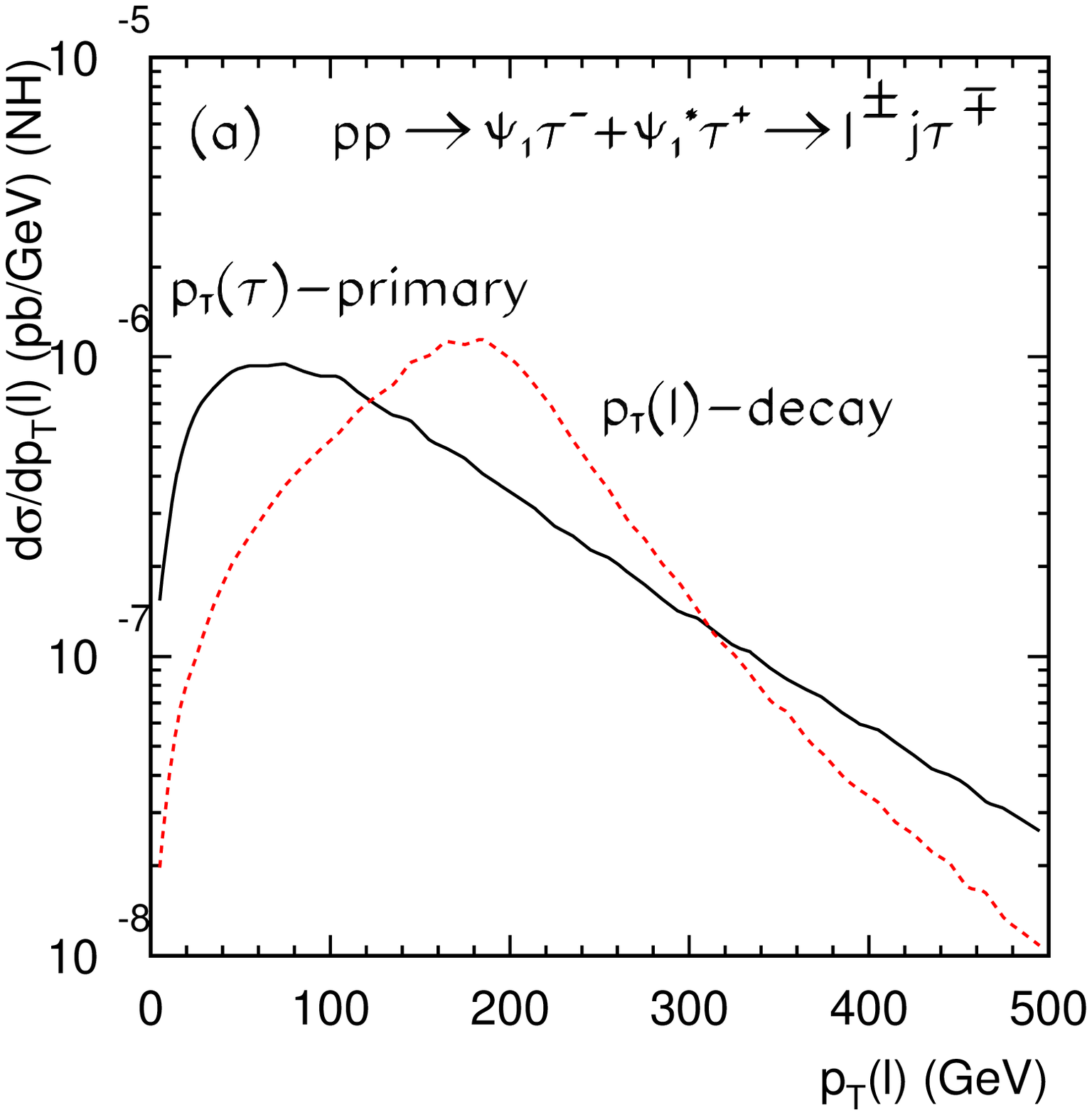}
\includegraphics[scale=1,width=8cm]{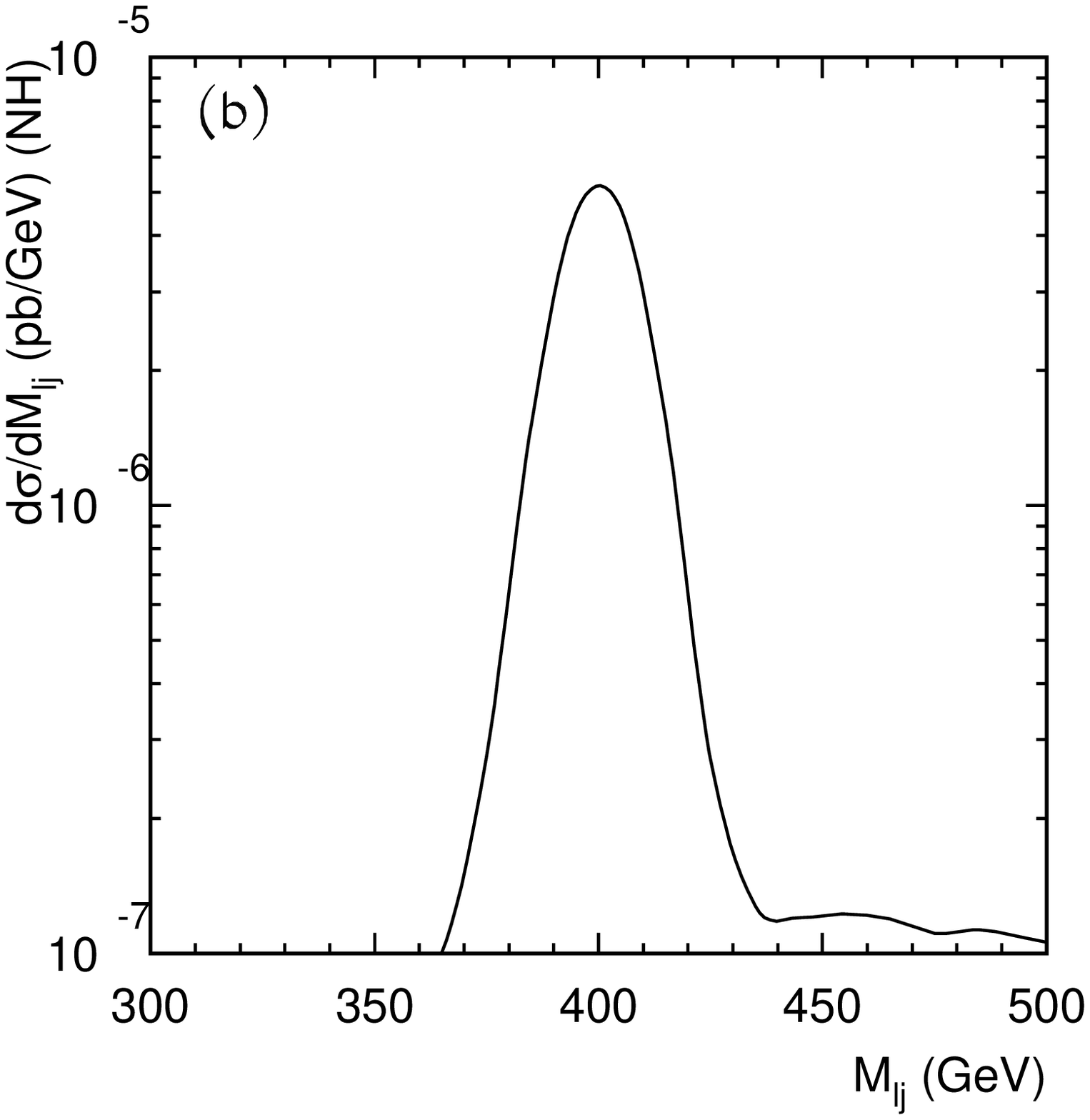}
\end{tabular}
\end{center}
\caption{(a) $p_T(\ell)$ distributions for the primary lepton and that from $\Psi_1$ decay,
and (b) invariant mass $M(\ell j)$ for a $400~{\rm GeV}$ single leptoquark production.}
\label{sinptlmass}
\end{figure}

The irreducible SM backgrounds to this channel are
\begin{eqnarray}
Z^\ast/\gamma^\ast j\to \ell^+\ell^- j, \quad W^+W^-j\to \ell^+\ell^-j+\cancel{E}_T.
\label{bkg}
\end{eqnarray}
Once again, the signal and background kinematics are quite. We outline the characteristics and propose
some judicious cuts as follows.
\begin{itemize}
\item The leptons and jets from the leptoquark decay are very hard. We tighten up the $p_T$
cuts
\begin{eqnarray}
p_T^{\rm max}(\ell)>M_{\Psi_1}/4,\quad p_T^{\rm max}(j)>M_{\Psi_1}/4.
\end{eqnarray}
\item Because the two leptons are largely back-to-back
for $jZ^\ast/\gamma^\ast$ background, we can set a cut for the angle between
the two leptons  in the transverse plane $\cos\phi_{\ell\ell}>-0.5$.
\item To remove the $Z$ background, we veto the lepton pairs
with opposite charges in the ${Z}$-mass window $|M_{\ell^+\ell^-}-M_Z|>15~{\rm GeV}$.
\item To remove the $jWW$ background, we veto the missing energy from $W$ decay and
require $\cancel{E}_T<25~{\rm GeV}$.
\item In order to select the correct lepton and jet combination
and reconstruct the leptoquark, we make use of  the feature that the lepton
from leptoquark decay is harder than that produced with $\Psi_1$, as seen
 in Fig.~\ref{sinptlmass}(a).
\item Finally, when we perform a signal significance analysis,
we look for a resonance in the mass distribution of $\ell j$.
The invariant mass of $\ell+j$ is plotted in Fig.~\ref{sinptlmass}(b) with
the event selection for $400~{\rm GeV}$ leptoquark mass. The long tail in the
high mass region is due to the wrong choice of the lepton.
We propose to look at a mass window of $|M_{\ell j}-M_{\Psi_1}|<M_{\Psi_1}/20$.
\end{itemize}
We summarize the  background rates with the above consecutive cuts, optimized
for a $M_{\Psi_1}=400~{\rm GeV}$ signal in Table.~\ref{TabSin}.
Unfortunately, even after all these optimal cuts, the $\gamma^*/Z j$ background
is still larger than the signal by more than an order of magnitude for both the NH
and IH cases. Unless we are deep in the quasi degenerate situation when  the
Yukawa coupling is substantially larger, the single LQ production would not be
observable above the huge SM background.
We will not pursue this channel further.
\begin{table}[tb]
\begin{tabular}{| c || c | c |c | c| c|c|}
    \hline
     $\sigma$(pb) & Basic & $\cos\phi_{\ell\ell}$ & $p_T^\ell,p_T^j$ cut & $M_Z$ veto & $M_W$ veto & $M_{\ell j}$  \\
cuts & Cuts & $>-0.5$ & $>100,100~{\rm GeV}$ & $\pm 15~{\rm GeV}$  & $\cancel{E}_T<25~{\rm GeV}$ & $\pm 20~{\rm GeV}$  \\
     \hline
     $j\gamma^\ast/Z^\ast$ & 309 & 113 & 9.3 & 0.82 & 0.82 & $6.5\times 10^{-2}$\\
\hline
      $jW^\pm W^\mp$ & 0.87 & 0.48 & $5.8\times 10^{-2}$ & $4.7\times 10^{-2}$ & $5.6\times 10^{-3}$ & $4.3\times 10^{-4}$\\
\hline
\end{tabular}
\caption{Backgrounds for a $400~{\rm GeV}$ leptoquark single production.}
\label{TabSin}
\end{table}
%
\section{Summary and Conclusions}
We have studied the properties of light scalar leptoquarks,
$\Phi_b^T = (\Psi_1, \Psi_2) \sim (3,2,1/6)$, in the context of
a simple grand unified theory where the neutrino masses are
generated through the Type II seesaw mechanism.
The $SU(5)$ symmetry of this theory  implies that the coupling of the
leptoquarks to matter are governed by the neutrino mass matrix.
We considered the low energy constraints on the model parameters
and explored the feasibility of testing the theory at the LHC with 
detailed parton-level simulations for the signal and SM backgrounds. 
We focus mainly on the different scenarios where the semileptonic 
decays of leptoquarks are the dominant decay channels.
We found the following interesting results:
\begin{itemize}
\item Among the low-energy constraints coming from meson decays,
lepton flavour violation, and electroweak precision data, the strongest bound
is from the $K_L\to \mu^+\mu^-$ decay, that leads to 
$M_{\Psi_1} v^{}_\Delta >1600$ GeV$\cdot$eV.
\item The decays of the scalar leptoquarks are quite
different in each spectrum for neutrino masses. Consider the most important
 $\Psi_1$ decay. 
In the NH scenario, the leading decay channels
 are $\mu+j$ and $\tau+j$, while the $e+j$ channel is absent.
In the IH scenario, the $e+b$ channel is the leading one, while
$\mu+j$ and $\tau+j$ are next. In the Quasi-degenerate case,
the leading channels are shared by $eb\approx \mu s \approx \tau d$.
\item The decays of $\Psi_2$ are independent
of the unknown phases in the theory. One can use these
decays  to test the neutrino mass pattern without additional ambiguity.
\item The signal dependence on the Majorana phase $\Phi_2$
is very weak. However, the dependence on $\Phi_1$
in the case of IH  and QD hierarchy can be strong,  and determining the $b\ell$ contributions
could help probe $\Phi_1$ if we can effectively exploit $b$-flavor tagging.
Specifically, we have, for the IH or QD hierarchy, 
\begin{eqnarray}
{\rm BR}(be) \left\{
\begin{array}{lll}
\gg {\rm BR}(b\mu),\ {\rm BR}(b\tau)  &  & {\rm for}\ \ \Phi_1 \approx 0; \\
\ll {\rm BR}(b\mu),\ {\rm BR}(b\tau)  &  & {\rm for}\ \ \Phi_1 \approx \pi .
\end{array}
\right.
\end{eqnarray} 
We reiterate that due to the symmetry in the neutrino mass matrix, the
theory has the interesting prediction for the correlations 
${\rm BR}(b\mu)={\rm BR}(se)$ and ${\rm BR}(b\tau)={\rm BR}(de)$.
We also note that the difference between the IH and  QD cases is rather small,
with $\ell b$ branching fraction to be $30\%$ in the QD case, compared to
$40-50\%$ in the IH case. 
\item The leading production of the leptoquarks is via the QCD process, 
$pp \to \Psi_1 \Psi_1^\ast$. We demonstrated that besides the clean di-lepton
channels from $e,\ \mu$, the $\tau$ final state can be effectively reconstructed
as well. Even with only the clean channel
 $\mu^+\mu^- +$jets, the signal observability can extend to 
$M_{\Psi_1} \sim 1$ TeV and beyond.
\item Although the rate is smaller than for QCD pair production, the electroweak process of associated production 
$\Psi_1 \Psi_2^\ast(\Psi_1^\ast \Psi_2)$ can be very useful by making use of
both $\Psi_1,\ \Psi_2$ decays simultaneously 
to identify the quantum numbers of the leptoquarks and
to distinguish between the neutrino mass spectra. Even with only the clean
channel of $\mu+\cancel{E}_T+$jets, the signal observability can reach 
about $M_{\Psi_1}\sim 0.6$ TeV or higher.
\item We have found that the single leptoquark production via its
Yukawa couplings is very small in both NH and IH spectrum.
\end{itemize}

The discovery of these leptoquarks at the LHC could test this appealing scenario
for neutrino mass generation at a fundamental level, 
and give us a hint about a possible candidate theory for grand unification.

\subsection*{Acknowledgments}
The work of P. F. P. and M.R-M. was supported in part by the U.S. Department of Energy
contract No. DE-FG02-08ER41531 and in part by the Wisconsin Alumni
Research Foundation. P. F. P would like to thank I. Dorsner,
R. Gonzalez Felipe, G. Rodrigo and G. Senjanovi\'c for discussions.
The work of T. H. is supported in part by the U.S. Department of Energy
under grant No. DE-FG02-95ER40896, and by the Wisconsin Alumni Research
Foundation. T. L. would like to thank Kai Wang for discussions and
the Ministry of Education of China for support and would also
like to acknowledge the hospitality of the Phenomenology Institute,
University of Wisconsin-Madison while the work was carried out.
\appendix
\section{EXPLICIT EXPRESSIONS FOR $\Gamma_1$ AND $\Gamma_2$}
\label{app:gammas}
As seen from Eq.~(\ref{G12}), we have
\begin{eqnarray}
&&\Gamma_1=\Gamma_2V^\dagger_{PMNS}K_3^* \ \ {\rm and} \ \
\Gamma_2=BK_3^*V^*_{PMNS}{m_\nu^{diag}\over v_\Delta}
\end{eqnarray}
and define
\begin{eqnarray}
&&Y_1^{ij}=\Gamma_1^{ij}\times v_\Delta, \ \ Y_2^{i}=\sum^3_{j=1}|\Gamma_2^{ij}|^2\times v_\Delta^2,\\
&&B=\left(
\begin{array}{lll}
 0 & 0 & e^{ i \beta_1} \\
 0 & e^{i \beta_2} & 0 \\
 e^{i \beta_3} & 0 & 0
\end{array}
\right), \ \ K_3=\left(
\begin{array}{lll}
 e^{ i \alpha_1} & 0 & 0 \\
 0 & e^{i \alpha_2} & 0 \\
 0 & 0 & e^{ i \alpha_3}
\end{array}
\right)
\end{eqnarray}
\begin{eqnarray}
Y_2^{1}&=&c_{13}^2c_{23}^2m_3^2+c_{12}^2c_{23}^2s_{13}^2m_1^2+c_{23}^2s_{12}^2s_{13}^2m_2^2+c_{12}^2s_{23}^2m_2^2+
s_{12}^2s_{23}^2m_1^2\nonumber \\
&-&2c_{12}c_{23}s_{12}s_{13}s_{23}\cos(\delta)(m_1^2-m_2^2),\\
Y_2^{2}&=&c_{23}^2s_{12}^2m_1^2+c_{13}^2s_{23}^2m_3^2+s_{12}^2s_{13}^2s_{23}^2m_2^2+c_{12}^2(c_{23}^2m_2^2+
s_{13}^2s_{23}^2m_1^2)\nonumber \\
&+&2c_{12}c_{23}s_{12}s_{13}s_{23}\cos(\delta)(m_1^2-m_2^2),\\
Y_2^{3}&=&c_{12}^2c_{13}^2m_1^2+c_{13}^2s_{12}^2m_2^2+s_{13}^2m_3^2.
\end{eqnarray}
and
\begin{eqnarray}
e^{-i\beta_3}Y_1^{31}&=&e^{-2i\alpha_1}(c_{13}^2s_{12}^2m_2+c_{12}^2c_{13}^2e^{-i\Phi_1}m_1+s_{13}^2e^{2i\delta}
e^{-i\Phi_2}m_3),\\
e^{-i\beta_2}Y_1^{21}&=&e^{-i\beta_3}Y_1^{32}=e^{-i(\alpha_1+\alpha_2)}c_{13}(s_{12}(c_{12}c_{23}-s_{12}s_{13}s_{23}e^{-i\delta})m_2\nonumber \\
&+&
c_{12}(-c_{23}s_{12}-c_{12}s_{13}s_{23}e^{-i\delta})e^{-i\Phi_1}m_1+s_{13}s_{23}e^{i\delta}e^{-i\Phi_2}m_3),
\end{eqnarray}
\begin{eqnarray}
e^{-i\beta_1}Y_1^{11}&=&e^{-i\beta_3}Y_1^{33}=e^{-i(\alpha_1+\alpha_3)}c_{13}(s_{12}(-c_{12}s_{23}-s_{12}s_{13}c_{23}e^{-i\delta})m_2\nonumber \\
&+&
c_{12}(s_{23}s_{12}-c_{12}s_{13}c_{23}e^{-i\delta})e^{-i\Phi_1}m_1+s_{13}c_{23}e^{i\delta}e^{-i\Phi_2}m_3),\\
e^{-i\beta_2}Y_1^{22}&=&e^{-2i\alpha_2}((c_{12}c_{23}-s_{12}s_{13}s_{23}e^{-i\delta})^2m_2+(c_{23}s_{12}+
c_{12}s_{13}s_{23}e^{-i\delta})^2e^{-i\Phi_1}m_1\nonumber \\
&+&c_{13}^2s_{23}^2e^{-i\Phi_2}m_3),\\
e^{-i\beta_1}Y_1^{12}&=&e^{-i\beta_2}Y_1^{23}=e^{-i(\alpha_2+\alpha_3)}(-(c_{12}s_{23}+c_{23}s_{12}s_{13}e^{-i\delta})(c_{12}c_{23}
-s_{12}s_{13}s_{23}e^{-i\delta})m_2\nonumber \\
&+&(s_{12}s_{23}-c_{12}c_{23}s_{13}e^{-i\delta})(-c_{23}s_{12}-
c_{12}s_{13}s_{23}e^{-i\delta})e^{-i\Phi_1}m_1\nonumber \\
&+&c_{13}^2c_{23}s_{23}e^{-i\Phi_2}m_3),\\
e^{-i\beta_1}Y_1^{13}&=&e^{-2i\alpha_3}((c_{12}s_{23}+c_{23}s_{12}s_{13}e^{-i\delta})^2m_2+(s_{12}s_{23}-
c_{12}c_{23}s_{13}e^{-i\delta})^2e^{-i\Phi_1}m_1\nonumber \\
&+&c_{13}^2c_{23}^2e^{-i\Phi_2}m_3).
\end{eqnarray}
\section{THE CONTRIBUTIONS OF LEPTOQUARKS TO THE OBLIQUE PARAMETERS}
\label{app:oblique}
We have computed the renormalized self-energy functions ${\hat\Pi_{VV'}}(q^2)$ in the $\overline{\mathrm{MS}}$ scheme and find that ${\hat\Pi_{VV'}}(0)=0$ for $V,V'=Z,\gamma$. Thus, the $T$ parameter is determined entirely by ${\hat\Pi_{WW}}(0)$ and reads
\begin{equation}
{\hat\alpha}(M_Z)\,  T =  \frac{{\hat\alpha}}{8\pi{\hat s}^2}\frac{N_C}{ M_W^2}\, \left[ \frac{1}{2}(M_1^2+M_2^2)+\frac{M_1^2 M_2^2}{M_1^2-M_2^2}\, \ln \frac{M_2^2}{M_1^2}\right]\ \ \ ,
\end{equation}
where $M_k\equiv M_{\Psi_k}$ ($k=1,2$) and where the running parameters ${\hat\alpha}$ and ${\hat s}$ on the RHS have been evaluated at the scale $\mu=M_Z$. The $S$ parameter, in contrast, requires inclusion of ${\hat\Pi_{VV'}}(M_Z^2)\not=0$ for $V,V'=Z,\gamma$, leading to
\begin{eqnarray}
\nonumber
{\hat\alpha}(M_Z)\,  S & = & \frac{{\hat\alpha} N_C Y_{\mathrm{LQ}}}{\pi M_Z^2}\, \sum_{k=1}^2\, \Bigl[
M_Z^2 F(M_k^2, M_k^2, M_Z^2) -\frac{M_Z^2}{6}\ln M_Z^2\\
&& + M_k^2\ln M_k^2-2 M_k^2 F_1(M_k^2, M_k^2, M_Z^2)\Bigr]\ \ \ ,
\end{eqnarray}
where the sum runs over the two LQ states having hypercharge $Y_{\mathrm{LQ}}$ and where
\begin{eqnarray}
F_n(a,b,c)&=&\int_0^1 dx\, x^n\, \ln[a(1-x)+bx-c x(1-x)-i\varepsilon]\\
\nonumber
F& = & F_1-F_2\ \ \ .
\end{eqnarray}

\end{document}